\documentclass[aps,prb,superscriptaddress,twocolumn,floatfix,letterpaper,nofootinbib]{revtex4-1}

\usepackage[all,arrow,matrix]{xy}

\newcommand{\w}{{\rm w}}

\newcommand{\gSq}{\mathbb{Sq}}
\newcommand{\Sq}{\text{Sq}}
\newcommand{\Bs}{\beta}
\newcommand{\RZ}{{\mathbb{R}/\mathbb{Z}}}
\newcommand\se[1]{\overset{\scriptscriptstyle #1}{=}}

\newcommand\hcup[1]{\underset{{\scriptscriptstyle #1}}{\smile}}
\newcommand\toZ[1]{\lfloor #1 \rceil}

\newcommand\ml[1]{$#1$}
\newcommand\ma[1]{
\begin{align}
#1
\end{align}
}

\usepackage[normalem]{ulem}

\begin{document}

\begin{titlepage}

\title{
Emergent (anomalous) higher symmetries from topological orders\\
and from dynamical electromagnetic field in condensed matter systems
}

\author{Xiao-Gang Wen}
\affiliation{Department of Physics, Massachusetts Institute of
Technology, Cambridge, MA 02139, USA}

\begin{abstract} 
Global symmetry (0-symmetry) acts on the whole space while higher $k$-symmetry
acts on all the codimension-$k$ closed subspaces.  The usual condensed matter
lattice theories do not include dynamical electromagnetic (EM) field and do not
have higher symmetries (unless we engineer fine-tuned toy models).  However,
for gapped systems, (anomalous) higher symmetries can emerge from the usual
condensed matter theories at low energies (usually in a spontaneously broken
form).  We pointed out that the emergent spontaneously broken higher symmetries
are nothing but a kind of topological orders. Thus the study of emergent
spontaneously broken higher symmetries is a study of topological order.  The
emergent (anomalous) higher symmetries can be used to constrain possible phase
transitions and possible phases induced by certain types of excitations in
topological orders.  (Anomalous) higher symmetry can also emerge in gapless
systems if the gapless excitations contain gapless gauge fields.  In
particular, EM condensed matter systems that include the dynamical EM field
have an emergent \emph{anomalous} $U(1)$-1-symmetry below  the energy gap of
the magnetic monopoles.  So  EM condensed matter systems can realize some
physical phenomena of anomalous higher symmetry.  In particular, any gapped
liquid phase of an EM condensed matter system (induced by arbitrary fluctuations
and condensations of electric charges and photons) must have a non-trivial
bosonic topological order.

\end{abstract}


\maketitle

\end{titlepage}

{\small \setcounter{tocdepth}{1} \tableofcontents }

\section{Introduction} 

A global symmetry acts on the whole space, and a local symmetry acts on all the
points (\ie all the sub manifolds of 0-dimension).  A higher symmetry, such as
a $k$-symmetry, acts on all the closed sub manifolds of codimension $k$.  Thus
a global symmetry is a 0-symmetry.  A lattice Hamiltonian system has a
$k$-symmetry, if the Hamiltonian is invariant under certain unitary
transformations defined on all the closed codimension-$k$ subspaces.  

Higher symmetry had been studied before in lattice systems, where exactly
soluble lattice Hamiltonians commuting with closed string and/or closed
membrane operators were constructed\cite{K032,W0303,LW0316,Y10074601,B11072707}
to realize topological orders\cite{W8987,WN9077,W9039}. A direct relation
between higher symmetries  and topological orders was pointed out in
\Ref{NOc0605316,NOc0702377}, under the name \emph{low-dimensional gauge-like
symmetries}.  

The term \emph{higher form symmetry} was first introduced in \Ref{GW14125148},
where it was stressed that higher symmetry can be viewed as a generalization of
the global symmetry (\ie 0-symmetry) and many results and intuitions for global
symmetry can be extended to higher symmetry.  For a Lagrangian field theory in
$(d+1)$-dimensional spacetime, a 0-symmetry is generated by constant fields
(closed $0$-forms) in spacetime, such as $\vphi \to \vphi+c$, while a
$k$-symmetry is generated by closed $k$-forms in spacetime.  From the field
theory definition of the $k$-symmetry generated by a closed $k$-form $\al_k$,
it appears that we require the field theory to have a $k$-form field $a_k$ so
that the $k$-symmetry transformation can be written as $a_k \to a_k+\al_k$.
However, from the lattice point of view, a $k$-symmetry does not require a
$k$-form field.

The emergence of higher symmetry from lattice model without higher symmetry was
also studied before in 2005\cite{HW0541}, under the name of emergent gauge
symmetry. A topological robustness of emergent higher symmetry against any
local perturbation was discovered. Such a topological robustness was used to
show the topological robustness of a Goldstone-like theorem for spontaneous
broken continuous higher symmetry: the gapless $U(1)$ gauge bosons from
spontaneous breaking of $U(1)$ higher symmetry remain to be gapless even after
we explicitly break the higher symmetry by arbitrary perturbations.  The result
of \Ref{HW0541} also suggests that every topological order containing Abelian
gauge theory has emergent higher symmetry.

Recently, higher symmetry and higher anomaly, as well as the related higher
symmetry protected trivial (SPT) orders and higher gauge theories, became an
active topic in field
theory.\cite{BSm0511710,GP07083051,BH10034485,KT13094721,GW14125148,S150804770,TK151102929,BM160606639,CT171104186,T171209542,P180201139,DT180210104,L180207747,CI180204790,HI180209512,BT180300529,BH180309336,NW180400677,ZW180809394,HO181005338,WW181211968,WW181211967,GW181211959,WW181211955}
\Ref{TK151102929,BM170200868,KR180505367} discussed higher symmetry and higher
anomaly in lattice systems.  \Ref{TK151102929} studied higher SPT phases and
higher anomaly in-flow.  \Ref{BM170200868} discussed how to construct a lattice
Hamiltonian of a higher gauge theory.  \Ref{KR180505367} obtained a
Lieb-Schultz-Mattis type theorem from higher anomalies.  

In this paper, we will
concentrate on applications of higher symmetry to condensed matter systems, by
studying the lattice aspect of higher symmetry, higher anomaly, and how higher
symmetry in field theory can emerge from lattice models without higher
symmetry.  The following is a summary of the results: 
\begin{enumerate} 
\item
The usual condensed matter theories (which do not include the dynamical
electromagnetic field) do not have higher symmetries. However, higher
symmetries can emerge in usual gapped condensed matter theories in a
spontaneously broken form, at energies below the energy gap. In fact, the
spontaneously broken finite higher symmetry is nothing but a special kind of
topological orders (see Section \ref{Z21symm}).  

\item
If certain topological excitations in topological orders have very large energy
gap $\Del_\text{top}$, while other topological excitations have small gaps of
order $\Del\ll \Del_\text{top}$, we may have emergent higher symmetry or
emergent anomalous higher symmetry at energies below $\Del_\text{top}$. Note
that such an emergent (anomalous) higher symmetry can appear at energies above
$\Del$, and is smaller than the emergent higher symmetry below $\Del$.  Thus
emergent (anomalous) higher symmetry in a topological order is characterized by
a set $\cC_\text{a}$ of low energy allowed topological excitations.  The set is closed
under the fusion and braiding of the topological excitations (see Section
\ref{Z21symm}).  

\item
For bosonic 2+1D Abelian topological orders, the emergent higher symmetry
characterized by $\cC_\text{a}$ is anomaly free iff the topological excitations in
$\cC_\text{t}$ are bosons with trivial mutual statistics.  Here $\cC_\text{t}$ is
the set of topological excitations that have trivial mutual statistics with all
the topological excitations in $\cC_\text{a}$ (see Section \ref{Z21symm}).

\item
The emergent (anomalous) higher symmetry will constrain the possible phase
transitions and phases induced by the low energy topological excitations (see
Section \ref{example}).  In particular, for a topological order with emergent
\emph{anomalous} higher symmetry characterized by $\cC_\text{a}$, the
topological order cannot change into a trivial phase with no topological order
no matter how we condense the topological excitations in $\cC_\text{a}$.

\item
Anomalous higher symmetries can be realized at boundary of higher symmetry
protected topological states in one higher dimension protected by the
corresponding anomaly-free higher symmetry.

\item
We find a spacetime lattice regularization of 3+1D $U^\ka(1)$ gauge theory with
$2\pi$-quantized topological term
\begin{align}
\ \ \ 
Z = \int D[a_I] & \ee^{\ii \int_{M^4} \frac{K_{IJ}}{4\pi}
f_I f_J
-\int_{M^4}\frac{|f_I|^2}{g} 
}
,
\end{align}
where $f_I=\dd a_I$ is the field-strength 2-form of the $U(1)$ gauge field
$a_I$.  We show that the lattice model is a local bosonic model with a $
Z_{k_1}\times Z_{k_2}\times \cdots$-1-symmetry which is determined by the even
integer matrix $K$, where $k_I$ are the diagonal elements of the Smith normal
form of $K$.  The lattice model \eq{K4DdadaC1} is exactly soluble on closed
spacetime in $g\to\infty$ limit, and realizes a higher SPT phase (see Section
\ref{Uk1}). 

\item
The Abelian higher symmetry and the non-Abelian 0-symmetry can have a
non-trivial mix  described by a higher group.\cite{KT13094721,CI180204790} In
this paper, we discuss a general way to construct lattice models with a
combined 0-symmetry, 1-symmetry, etc (see Section \ref{ST}).  

\item
We systematically construct lattice bosonic models that realize higher SPT
phases with higher symmetry described by higher group $\cB(G,\Pi_2,\cdots)$ in
any dimension.\cite{TK151102929}  Our construction suggests a (many-to-one)
classification of $(d+1)$D bosonic higher SPT phases in terms of the cohomology
of an extended higher group: $H^{d+1}[\cB(G\gext
SO_\infty,\Pi_2,\cdots),\R/\Z]$ (see Section \ref{SPTSO}).  Using fermion
worldline decoration, we also systematically construct lattice fermionic models
that realize higher SPT phases with higher symmetry $\cB(G_f,\Pi_2,\cdots)$  in
any dimension (see Section \ref{SPTSOF}).

\item 
The condensed matter theories that include the dynamical electromagnetic (EM)
field (which will be called the EM condensed matter theories) can be viewed as
local bosonic theories with an \emph{anomalous} $U(1)$-1-symmetry, since the
magnetic monopoles can be ignored at low energies.  Such an anomaly comes from
the property that all fermions carry odd electric charges  and all bosons carry
even electric charges.  Using the anomalous $U(1)$-1-symmetry, we show that any
gapped liquid phase\cite{ZW1490,SM1403} of any EM condensed matter system
must have a non-trivial bosonic topological order (see Section \ref{EMsys}).  

\end{enumerate}

Exactly soluble models have been constructed to realize various topological
orders \cite{K032,W0303,LW0316,Y10074601,B11072707}.  The toy model Hamiltonian
commute with closed string and/or membrane operators, and thus has higher
symmetries.  However, the real condensed matter systems (after ignoring the
dynamical electromagnetic field) do not have higher symmetries.  In this paper
we pointed out the topologically ordered phases and the states near the
topologically ordered phases has emergent higher symmetries bellow the energy
gap of \emph{some} topological excitations.  Thus if those topological
excitations remain to be gapped, we can use the emergent higher symmetries to
study the low energy dynamics and phase transition of the topologically ordered
phases.

\section{Notations and conventions}

In part of this paper, we will use extensively the notion of cochain, cocycle,
and coboundary, as well as their higher cup product $\hcup{k}$ and Steenrod
square $\gSq^k$.  A brief introduction can be found in Appendix \ref{cochain}.
We will abbreviate the cup product $a\smile b$ as $ab$ by dropping $\smile$.
We will use $\se{n}$ to mean equal up to a multiple of $n$, and use $\se{\dd }$
to mean equal up to $\dd f$ (\ie up to a coboundary).  We will use $\<l,m\>$ to
denote the greatest common divisor of $l$ and $m$ ($\<0,m\>\equiv m$).  We will
also use $\toZ{x}$ to denote the integer that is closest to $x$.  (If two
integers have the same distance to $x$, we will choose the smaller one, \eg
$\toZ{\frac12}=0$.)

In this paper, we will deal with many $\Z_n$-value quantities. We will denote
them as, for example, $a^{\Z_n}$. However, we will always lift the $\Z_n$-value
to $\Z$-value, so the value of $a^{\Z_n}$ has a range from $ -\toZ{\frac N2}$
to $ \toZ{\frac N2}$.  In this case, even the expression like $a^{\Z_n}+a^{\Z_m}$
makes sense.  

We introduced a symbol $\gext$ to construct fiber bundle $X$ from the fiber $F$
and the base space $B$:
\begin{align}
pt\to  F \to X=F\gext B \to B\to pt .
\end{align}
We will also use $\gext$ to construct group extension of $H$ by $N$
\cite{Mor97}:
\begin{align}
1 \to  N \to N\gext_{e_2,\al} H \to H\to 1 .
\end{align}
Here $e_2 \in H^2[H;Z(N)]$, $Z(N)$ is the center of $N$, and $\al: H \to
\text{Aut}(N)$ is a non-trivial action of $H$ on $Z(N)$.  Thus $e_2$ and $\al$
characterize different group extensions.

We will use $K(\Pi_1,\Pi_2,\cdots,\Pi_n)$ to denote a connected topological
space with homotopy group $\pi_i(K(\Pi_1,\Pi_2,\cdots,\Pi_n)) =\Pi_i$ for
$1\leq i\leq n$, and $\pi_i(K(\Pi_1,\Pi_2,\cdots,\Pi_n)) =0$ for $i>n$.  If
only one of the homotopy groups, say $\Pi_d$, is non-trivial, then
$K(\Pi_1,\Pi_2,\cdots,\Pi_n)$ is the Eilenberg-MacLane space, which is denoted
as $K(\Pi_d,d)$.  If only two of the homotopy groups, say $\Pi_d$, $\Pi_{d'}$,
is non-trivial, then we denote the space as $K(\Pi_d,d; \Pi_{d'},d')$, \etc.
We will use $\cB(\Pi_1;\Pi_2;\cdots;\Pi_n)$, $\cB(\Pi_d,d)$, and $\cB(\Pi_d,d;
\Pi_{d'},d')$ to denote the simplicial sets with only one vertex satisfying Kan
conditions that describe a special triangulation of
$K(\Pi_1,\Pi_2,\cdots,\Pi_n)$, $K(\Pi_d,d)$, and $K(\Pi_d,d; \Pi_{d'},d')$
respectively. Since simplicial sets satisfying Kan conditions are viewed as
higher groupoids in higher category theory,  the  simplicial sets
$\cB(\Pi_1;\Pi_2;\cdots;\Pi_n)$, $\cB(\Pi_d,d)$, and $\cB(\Pi_d,d;
\Pi_{d'},d')$, with only one vertex (unit),  can be viewed as higher groups.
In this paper, higher groups are treated therefore as this sort of special
simplicial sets, as in \Ref{ZW180809394}. 

\section{Symmetry and higher symmetry on lattice}

The notion of phase and phase transition play a major role in condensed matter
physics in our attempts to understand properties of various materials.
However, to mathematically define the concepts of  quantum phase and quantum
phase transition at zero temperature, we need to first introduce the notion
Hamiltonian class.  The notions of  phase and phase transition can only be
defined relative to a Hamiltonian class.  (For example, the notion of  phase is
not a property of a single Hamiltonian.)

For example, we can define a Hamiltonian class as a set of local Hamiltonians
for bosons on lattice.  Relative to such a Hamiltonian class, we can define the
notion bosonic topological orders.\cite{W8987,WN9077} Such a precise definition
of bosonic topological orders allows us to classify them in 1 spatial
dimension\cite{FNW9243,VCL0501} where there is no non-trivial bosonic
topological orders, as well as in 2 spatial
dimensions\cite{W9039,K062,RSW0777,W150605768,LW150704673} and in 3 spatial
dimensions\cite{LW170404221,LW180108530}, where very rich bosonic topological
orders exist.  

We may define another Hamiltonian class as a set of local Hamiltonians for
fermions on lattice.  Relative to such a Hamiltonian class, we can define the
notion fermionic topological orders.\cite{GWW1017} There is only one
non-trivial fermionic topological order in 1 spatial dimension -- the 1+1D
topological $p$-wave superconductor.\cite{K0131,LMc0108266} The fermionic
topological orders in 2 spatial dimensions are classified in \Ref{LW150704673},
and a proposal to classify fermionic topological orders in 3 spatial dimensions
is given in \Ref{LW180108530}.

We can also introduce a Hamiltonian class as a set of local Hamiltonians $H$ on
lattice with an on-site symmetry\cite{CGL1314,W1313} described by a group $G$:
\begin{align}
 HW_g = W_gH,\ \ \
W_g = \prod_i W_g(i),\ \ \ g\in G,
\end{align}
where $W_g(i)$ is a representation of the symmetry group $G$ acting on the
local Hilbert space on site-$i$.  Relative to such a Hamiltonian class, we can
define the notion of spontaneous symmetry breaking orders, symmetry protected
topological (SPT) orders\cite{GW0931,CLW1141} (also known as symmetry protected
trivial orders\cite{W1477,W161003911}), and symmetry enriched topological (SET)
orders.\cite{CGW1038,LV1334,CY14126589,Y161008645} The classification of
spontaneous symmetry breaking orders is given by a pair of groups
$G_\text{grnd} \subset G$, where $G_\text{grnd}$ is the symmetry group of the
ground state (the unbroken symmetry group).  The classification of SPT and SET
orders are given in
\Ref{CGW1107,SPC1139,CGL1314,CY14126589,LW160205946,LW170404221,LW180108530,KT170108264,WG170310937,WG181100536},
via group cohomology theory, cobordism theory, and (higher) category theory.

The on-site symmetry $G$ is also called global symmetry.  However, a global
symmetry in field theory or in lattice theory may not be an on-site symmetry
$G$.  In this case, we say\cite{W1313,KT1430} the global symmetry to have a t'
Hooft anomaly.\cite{H8035} We will also call the on-site symmetry $G$ as an
on-site 0-symmetry, where the symmetry transformation acts on codimension-0
space or spacetime.\cite{GW14125148,L180207747}

Now, we are ready to define higher symmetry on lattice, by introducing a
Hamiltonian class as a set of local Hamiltonians $H$ on a triangulation of
$d$-dimensional space (which is a more precise definition of lattice)
that satisfy
\begin{align}
 HW_a(C^{d-k})  = W_a(C^{d-k})H,
\end{align}
for any $d-k$-dimensional (\ie codimension-$k$) closed subcomplex $C^{d-k}$ of
the triangulated $d$-dimensional space.  Here $W_a(C^{d-k})$ is an operator
that acts on the degrees of freedom on the closed subcomplex $C^{d_k}$ with the
following ``on-site'' property 
\begin{align}
W_a = \prod_{i\in C^{d-k}}  W_a(i) .
\end{align}
 The operator
$W_a(C^{d-k})$ labeled by $a$ satisfy a so called pointed fusion rule
\begin{align}
 W_a(C^{d-k}) W_b(C^{d-k}) =  W_c(C^{d-k}),
\end{align}
and they commute with each other
\begin{align}
  [W_a(C^{d-k}),W_b(\t C^{d-k})]=0 .
\end{align}
Such a pointed fusion rule makes the index set $\{a\}$ into an Abelain group
$\Pi$.  The Hamiltonian class defined above is referred as having an on-site
$k$-symmetry since the symmetry acts on codimension-$k$ subcomplex of the
space.\cite{GW14125148,L180207747}

\section{An example of anomaly-free (on-site) $Z_2$-1-symmetry}
\label{Z21symm}

\subsection{The 3+1D bosonic lattice model}

\begin{figure}[t]
\begin{center}
\includegraphics[scale=0.5]{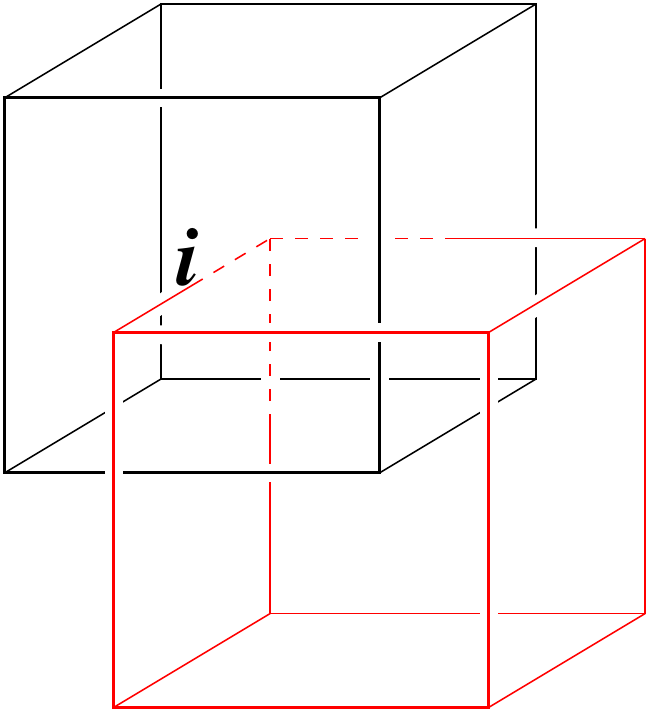} \end{center}
\caption{ ( Color online) 
A cubic lattice (black) and its dual cubic lattice (red).
The index $\v i$ labels the faces of the cubic lattice
and the links of the dual cubic lattice.
}
\label{cubeDcube}
\end{figure}

The simplest on-site 1-symmetry is a $Z_2$-1-symmetry.  Let us consider a qubit
model on a 3-dimensional cubic lattice (see Fig. \ref{cubeDcube}), where the qubits live on the square
faces of the cubic lattice. We choose the closed subcomplex $C^{2}$ to the
closed 2-dimensional surfaces formed by the square faces of the cubic lattice.
The generator of the $Z_2$-1-symmetry is given by
\begin{align}
\label{WC2}
  W(C^2)=\prod_{\v i \in C^{2}} \si^z_{\v i}
\end{align}
where $\v i$ labels the   square faces of the cubic lattice.  We can check that
$W(C^{2})$'s all commute with each other and their pointed fusion is
described by a $Z_2$ group.  So we say   $W(C^{2})$ generates an on-site $Z_2$
1-symmetry.

A Hamiltonian with the above $Z_2$
1-symmetry is given by\cite{K032,W0303,LW0316}
\begin{align}
\label{H3dZ2}
H &=
-U_1\sum_{\<\v i\v j\v k\v l\>}\si^x_{\v i}\si^x_{\v j}\si^x_{\v k}\si^x_{\v l}
\\
&\ \ \ \
-U_2 \sum_{\<\v i\v j\v k\v l\v m\v n\>}
\si^z_{\v i}\si^z_{\v j}\si^z_{\v k}\si^z_{\v l} \si^z_{\v m}\si^z_{\v n}
-B \sum_{\<\v i\>} \si^z_{\v i},
\nonumber 
\end{align}
where $\sum_{\<\v i\>}$ sums over all the square faces, $\sum_{\<\v i\v j\v k\v
L\v m\v n\>}$ sums over all cubes, and $\sum_{\<\v i\v j\v k\v l\>}$ sums over
all squares formed by the links in the dual cubic lattice.  (We note that $\v
I$'s also label the links of the dual cubic lattice. See Fig. \ref{cubeDcube})

When $|B|\ll |U_1|,|U_2|$, the Hamiltonian \eq{H3dZ2} has a topologically
ordered ground state.  The topological order is described by a $Z_2$ gauge
theory.  A pair of $Z_2$-charge $e$ is created by an open string operator
\begin{align} 
\label{string}
S_\text{str}(\t C^1) = \prod_{\v i \in \text{ string } \t C^1} \si^x_{\v i},  
\end{align} 
where the string is formed by the links of the dual cubic lattice.  Note that
the open string creation operators break the $Z_2$-1-symmetry.  Thus we cannot
even include the short open string operators in the Hamiltonian.  This implies
that in the presence of the $Z_2$-1-symmetry, the $Z_2$-charge is not mobile.
A $Z_2$-flux loop $s$ is created by an open membrane operator bounded by the
loop:
\begin{align} 
\label{memb}
M_\text{memb}(C^2) = \prod_{\v i \in \text{ membrane } C^2} \si^z_{\v i},  
\end{align}
where the membrane is formed by the square faces of the original cubic lattice.
The $Z_2$-1-symmetry allow the Hamiltonian to have such a open membrane
operator. Thus the $Z_2$-flux loop $s$ is mobile even in the presence of the
$Z_2$-1-symmetry.

We also note that the closed membrane operator happen to be the generator
\eq{WC2} of the  $Z_2$-1-symmetry. Thus we say that the $Z_2$-1-symmetry is
generated by the topological excitations of the $Z_2$-flux loops.

\subsection{$Z_2$-1-symmetry and unbreakable strings}

In this subsection, we discuss a meaning of 1-symmetry in our lattice model
\eqn{H3dZ2}.  Let $|\up\>$, $|\down\>$ be the eigenstates of $\si^z$.  We view
$\otimes_{\v i} |\up\>_{\v i}$ as a reference state.  We create a closed-string
state by changing $|\up\>_{\v i}$ to $|\down\>_{\v i}$ for $\v i$ on closed
strings.  Here strings are formed by links of the dual cubic lattice.

Because of the $Z_2$-1-symmetry, the ground state of \eqn{H3dZ2} is a
superposition of closed strings (assuming $U_2,B>0$).  When $B=0$, the ground
state is an equal weight superposition of all closed strings, and spontaneously
break the $Z_2$-1-symmetry (on space with non-trivial first homotopy group
$\pi_1$).  When $B\to +\infty$, the ground state has no strings and does not
break the $Z_2$-1-symmetry.

From this example, we see that the physical meaning of the $Z_2$-1-symmetry is
the appearance of unbreakable strings.  Even when we force a string breaking,
the $Z_2$-1-symmetry requires that the end of the string (\ie the $Z_2$-charge)
cannot move and have no dynamics.  

To summarize, the $Z_2$-1-symmetry is
generated by the $Z_2$-flux-line excitations.  Such a $Z_2$-1-symmetry forbid
the excitations that have non-trivial mutual statistics with the $Z_2$-flux
lines, such as the $Z_2$-charge excitations.  This can be achieved by including
the $Z_2$-1-symmetry generator \eq{WC2} in the Hamiltonian with a large
coefficient (the $U_2$ term in \eqn{H3dZ2}), which gives the $Z_2$-charge a
large energy gap.

\subsection{Emergence of generic  higher symmetry in topological orders}

The above physical understanding of $Z_2$-1-symmetry can also be generalized:
\frmbox{If a topological order contains a topological excitation $\eta$ of unit
quantum dimension, then the topological order can have an emergent higher
symmetry generated by $\eta$ below an energy gap $\La$, if all the topological
excitations with non-trivial mutual statistics respect to any combination of
$\eta$ have a large gap beyond $\La$.} We note that the emergent higher
symmetry allows the topological excitations with trivial mutual statistics
respect to $\eta$ to have a small gap, or even become gapless and drive a phase
transition via their condensation.  The emergent higher symmetry will be
present through the phase transition.

To understand the above result in more details, let us
consider a dimension-$n$ topological excitation in a topological order.  We
assume that the topological excitation can be created by a dimension-$n+1$
operator $W_1(C^{n+1})$ at its boundary $\prt C^{n+1}$.  Here $W_1(C^{n+1})$
is a operator acts on the $(n+1)$-dimensional subcomplex $C^{n+1}$ in the space
(not spacetime).  If the quantum dimension of the topological excitation
is 1, then $W_1(C^{n+1})$ on a closed subcomplex $C^{n+1}$ generates a $Z_N$
fusion
\begin{align}
 W_a(C^{n+1}) W_b(C^{n+1}) = W_{a+b}(C^{n+1}),\ \ \ 
W_N(C^{n+1}) = 1,
\end{align}
for a certain integer $N$.  Now, we require the lattice Hamiltonians to commute
with $W_1(C^{n+1})$ for all closed $C^{n+1}$.  This way we obtain a Hamiltonian
system that has a $Z_n$ $D-n-2$-symmetry where $D$ is the spacetime dimension.
If a Hamiltonian with such a higher symmetry realizes the above topological
order, then all the topological excitations with non-trivial mutual statistics
with the $n$-dimensional topological excitation are not mobile.  We can even
make those  topological excitations to have a large gap by adding the
$W_a(C^{n+1})$ terms to the Hamiltonian with a large coefficient.  Only
topological excitations having trivial mutual statistics with the
$n$-dimensional topological excitation are allowed to appear at low energies
and to have non-trivial dynamics.

We stress that, in our construction, the higher symmetry is a property of the
pair: the topological order plus the allowed low energy  topological
excitations: the higher symmetry is generated from the topological excitations
of unit quantum dimension that have trivial mutual statistics with the allowed
low energy  topological excitations.  Even in the same topological order,
allowing different types of  topological excitations to appear at low energy
will leads to different
(anomalous) higher symmetries.  Later, we will present several examples of this
phenomenon.

For the 3+1D $Z_2$ topological order discussed above,   if we only allow the
$Z_2$-flux line excitations at low energies (\ie the $Z_2$-charges associated
with the string ends all have very high energies), then we will have the $Z_2$
1-symmetry at low energies, generated by the $Z_2$-flux lines.  Later we will
show that if we only have the $Z_2$-charge excitations at low energies, we will
have a $Z_2$-2-symmetry at low energies (see \eqn{Z2tgauge}), generated by the
$Z_2$-charge excitations.  If we only allow the trivial excitations at low
energies, then we will have the $Z_2$-1-symmetry generated by the $Z_2$-flux
lines, and the $Z_2$-2-symmetry, generated by the $Z_2$-charge excitations.  If
we allow both $Z_2$-flux and $Z_2$-charge excitations at low energies, we will
explicitly break the $Z_2$-1-symmetry and the $Z_2$-2-symmetry at low energies.
In fact, there are no topological excitations that have trivial mutual
statistics with both $Z_2$-flux and $Z_2$-charge excitations, and there is no
higher symmetry.

We like to remark that the above $Z_2$-1-symmetry and $Z_2$-2-symmetry have
mixed anomaly between them (see Section \ref{}). A system with both of those higher symmetries
cannot realize a trivial topological order.

\subsection{Generalized higher symmetry}

In the above, we have constructed a higher symmetry from one topological
excitation of unit quantum dimension.  Certainly, we can construct more general
higher symmetry from several topological excitations of unit quantum dimension.
We can even construct something from a topological excitations with higher
quantum dimensions.  We call the ``something'' generalized higher symmetry (see
\Ref{KLTW}): \frmbox{ A topological order can have an emergent generalized
higher symmetry generated by a topological excitation $\eta$, if all the
topological excitations with non-trivial mutual statistics respect to any
combination of $\eta$ have a large gap.  } 

Since a topological order, by definition, always has a finite energy gap,  at
energies much below the energy gap, the topological order always has an
emergent generalized higher symmetry, and such a generalized higher symmetry is
spontaneously broken.  If the topological order contains excitations with unit
quantum dimension, then, part of the \emph{generalized higher symmetry} can be
viewed as the \emph{higher symmetry}.  If certain topological excitations have
a large gap and other topological excitations have a small gap, then below the
large gap, the topological order may have an emergent generalized higher
symmetry, which may be smaller then the emergent generalized higher symmetry
below the small gap.

\subsection{Spontaneous higher symmetry breaking and topological order}

When $|B|\gg |U_1|,|U_2|$, the ground state of the Hamiltonian \eq{H3dZ2} is a
product state without topological order.  When $|B|\ll |U_1|,|U_2|$, the ground
state has a non-trivial topological order.  The small $|B|$ topologically
ordered phase and the large $|B|$ trivial phase can also be distinguished by
\emph{spontaneous 1-symmetry breaking}.  The $Z_2$-1-symmetry is generated by
$W(C^2)$ in \eqn{WC2}.  

On space $S^1\times S^1 \times S^1$, the large $|B|$ trivial phase has an
unique ground state, which is is invariant under all the $Z_2$-1-symmetry
transformations.  The $Z_2$-1-symmetry is not broken.  While the topologically
ordered phase for small $|B|$ has 8 ground states on space $S^1\times S^1
\times S^1$.  Some the $Z_2$-1-symmetry transformations act non-trivially in
the 8-dimensional ground state subspace, \ie are not propotional to an identity
operator.  Thus, the $Z_2$-1-symmetry is spontaneously broken.

The spontaneous 1-symmetry broken state is nothing but a topologically ordered
state. Since the 1-symmetry is not spontaneously broken in the large $|B|$
trivial product state, the transition from the trivial product state to the
topologically ordered state can be viewed as a spontaneous breaking of the
1-symmetry.  This result is general: \frmbox{A spontaneous higher symmetry
broken state always corresponds to a topologically ordered state.} 

Here we have assumed that the higher symmetry is finite.  The spontaneous
breaking of continuous higher symmetry is discussed in
\Ref{GW14125148,L180207747}, and give rise to gapless states.  However, even
though the spontaneous breaking of continuous higher symmetry produce gapless
excitations, the gaplessness of the excitations do not need higher symmetry.
Even after we explicitly break the higher symmetry, the gapless excitations
remain gapless (see Section \ref{robust}).\cite{HW0541} This is very different
from the gapless excitations from the spontaneous breaking of continuous
0-symmetry.

We also like to point out that a topologically ordered state can be more
general and may not correspond to a spontaneous higher symmetry broken state.
For example, we can break the higher symmetry explicitly.  Even without higher
symmetry, we can still have topologically order.  Even though, some topological
orders, such as $Z_2$-gauge theory, can be viewed as spontaneous higher
symmetry broken states.  Some other topological orders, such as $S_3$-gauge
theory, cannot be viewed as spontaneous higher symmetry broken states, since
spontaneous 1-symmetry broken states only give rise to Abelian gauge theory.

\subsection{The usefulness of higher symmetry in condensed matter}

We see that some topological orders can be understood as spontaneous higher
symmetry breaking in the systems with higher symmetry.  But the usual condensed
matter theories on lattice never have higher symmetry.  (Here by ``usual
condensed matter theories'' we mean the theories that do not include the
dynamical electromagnetic fields.) So it appears that this way to understand
topological order may not be very useful.  Also, this way to understand
topological order misses a key feature of topological order: topological order
is robust against any local perturbation that can break all the symmetries and
higher symmetries.

However, the notion of higher symmetry and their spontaneous breaking can still
be useful in condensed matter in the following sense: A gapped liquid
state\cite{ZW1490,SM1403} may have many emergent symmetries and higher
symmetries at low energies.  If some of the emergent higher symmetries are
spontaneously broken, then the corresponding  gapped liquid state has
topological orders.  This allows us to use spontaneously broken emergent higher
symmetry to characterize a subclass of topological orders.  

The higher symmetries in low energy effective field theory may come from
lattice model with exact higher symmetry, or may emerge from a lattice model
that has no higher symmetry.  Therefore, even though the usual condensed matter
theories on lattice do not have higher symmetry, low energy effective theories
with higher symmetries can still be used to describe condensed matter systems,
since higher symmetry can be emergent.

Later, we will see that if we include the dynamical electromagnetic (EM) fields
in condensed matter theories, the resulting EM condensed matter theories will
have an approximate $U(1)$-1-symmetry if we ignore the magnetic monopoles.
(The appearance of magnetic monopoles in condensed matter energy scale will
break the $U(1)$ higher symmetry.) In this sense, $U(1)$-1-symmetry is
useful for real condensed matter systems.

The notion of higher symmetry can also be useful for condensed matter in
another way.  We can construct toy models with higher symmetries, and make
the ground states spontaneous break the higher symmetry.  This way, we
construct toy models that realize some topological orders.  There are already
many different ways to construct exactly soluble models to systematically
realize topological orders, SPT orders, and SET
orders.\cite{K032,LW0510,GWW1017,WW1132,CGL1314,BK160501640,WW160607144,ZW180809394,WG181100536}
But it does not hurt to have one more construction.  In this paper, we will
construct some simple toy models with higher symmetry that realize some
topological orders, SET orders, and SPT orders.

Last, higher symmetry can also be used to constrain possible phase transitions
and possible phases.\cite{KR180505367}  In certain topological orders, if we
only allow a certain type of topological excitations at low energies, the
topological orders plus the topological excitations may have an emergent
(anomalous) higher symmetry.  Then any phase transition \emph{induced by this
particular type of topological excitations} must preserve the anomaly of the
higher symmetry.  This puts constraint on possible phase transitions and
possible resulting phases.

\section{An example of anomalous $Z_2$-1-symmetry}

\subsection{The 2+1D bosonic lattice model}

\begin{figure}[t]
\begin{center}
\includegraphics[scale=0.6]{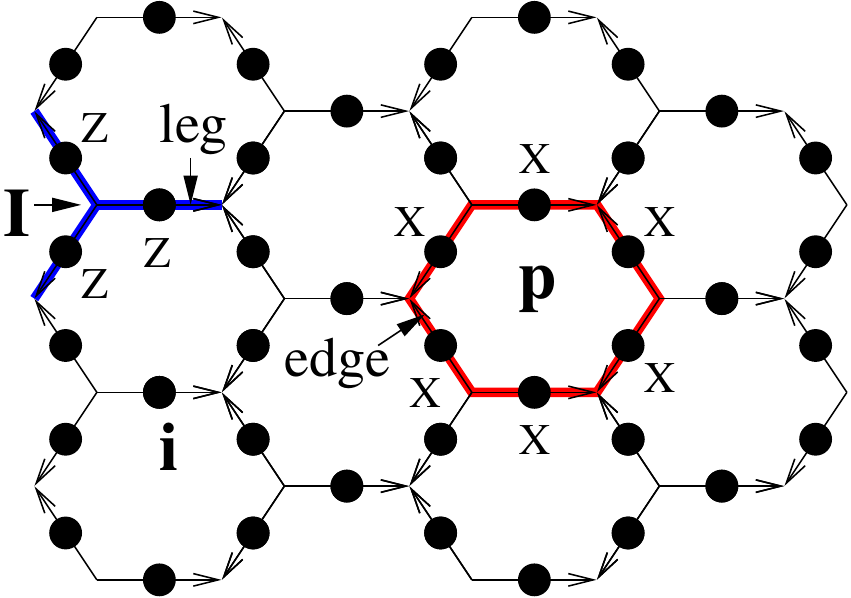}
\end{center}
\label{KlattZ2}
\caption{ ( Color online) 
A qubit model, where qubits live on the links of a honeycomb lattice.
$\v I$ labels the vertices 
and $\v i$ labels the links (the qubits) of the  honeycomb lattice.
}
\end{figure}

In this section, we are going discuss another simple lattice that have several
$Z_2$-1-symmetries and one of them is an anomalous  $Z_2$-1-symmetry.  In this
model the qubits live on the links of a honeycomb lattice (see Fig.
\ref{KlattZ2}), with a Hamiltonian:\cite{K032}
\begin{align}
\label{torH}
 H &= - U\sum_{\v I} Q_{\v I} -g\sum_{\v p} F_{\v p},
\nonumber\\
Q_{\v I} &= \prod_{\text{legs of }\v I} \si^z_{\v i},
\nonumber\\
F_{\v p} &=  \prod_{\text{edges of }\v p} \si^x_{\v i},
\end{align}
where $\sum_{\v I}$ sum over the vertices and where $\sum_{\v p}$ sum over the
hexagons of the honeycomb lattice (see Fig. \ref{KlattZ2}).
Notice that
\ml{H} is a sum of commuting operators
\ml{ [F_{\v p}, F_{\v p'}]=0},
\ml{ [Q_{\v I}, Q_{\v I'}]=0},
\ml{ [F_{\v p}, Q_{\v I}]=0}, and
\ml{F_{\v p}^2=Q_{\v I}^2 =1}
Thus the ground state \ml{|\Psi_\text{grnd}\>} is given by \ml{F_{\v
p}|\Psi_\text{grnd}\>=Q_{\v I}|\Psi_\text{grnd}\>=|\Psi_\text{grnd}\>}

\begin{figure}[t]
\begin{center}
\includegraphics[scale=0.5]{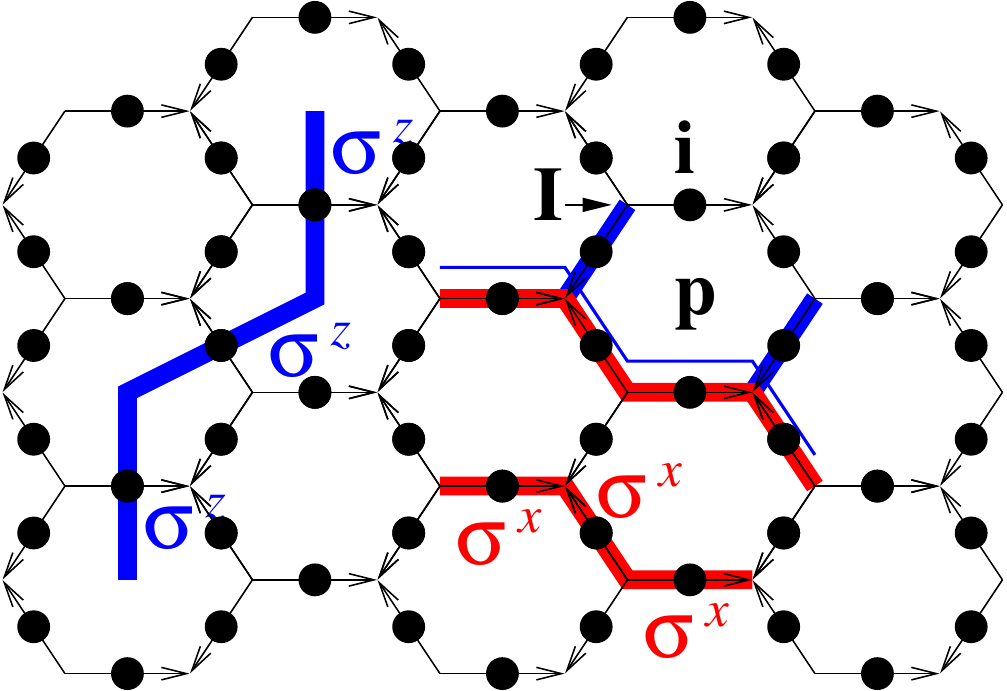}
\end{center}
\caption{ ( Color online) 
The string operators.
}
\label{KlattStrOp}
\end{figure}

There are two types of topological excitations above the ground state with
$Q_{\v I}=F_{\v p}=1$: \ml{e}-type with \ml{ Q_{\v I}=-1} and
\ml{m}-type with \ml{F_{\v p}=-1}.  Those excitations cannot be created
individually.  They can only be created in pairs by string operators.

We have type-\ml{e} string operator \ml{W_e(C_1)=\prod_{\v i\in C_1}
\si^x_{\v i}} where the string $C_1$ is formed by the links of the honeycomb
lattice (see Fig. \ref{KlattStrOp}).  An open $e$-string operator creates two
$e$-type topological excitations at its ends.  We also have type-\ml{m} string
operator \ml{W_m(\t C_1)=\prod_{\v i\in \t C_1} \si^z_{\v i}} where the
string $\t C_1$ is formed by the links of dual of the honeycomb lattice (see
Fig. \ref{KlattStrOp}).  An open $m$-string operator creates two $m$-type
topological excitations at its ends.

We can also fuse the $e$-string and $m$-string operators together to form a
type-\ml{f} string operator \ml{W_f(C_1\prod \t C_1)=\prod_{\v i \in_1 C_1}
\si^x_{\v i} \prod_{\v i \in \t C_1} \si^z_{\v i} } where the string $C_1$ in
the lattice and the string $\t C_1$ in the dual lattice closely follow each
other (see Fig. \ref{KlattStrOp}).  An open $f$-string operator creates two
$f$-type topological excitations at its ends.  It turns out that $e$ and $m$
are bosons, and $f$ is a fermion.  They all have a $\pi$ mutual statistics
respect to each other.

We find that $H$ in \eqn{torH} commutes with the above three types of string
operators if the strings are closed: 
\ma{ [H,W^\text{closed}_e]=
[H,W^\text{closed}_m]= [H,W^\text{closed}_f]=0. 
}  
Therefore, our lattice model has two $Z_2$-1-symmetries since
\begin{align}
 W_e^2= W_m^2= W_f^2=1, \ \ \ 
 W_e W_m= W_f . 
\end{align}

On a torus, the model \eq{torH} has four degenerate ground states, and the
closed string operators $W^\text{closed}_e$, $W^\text{closed}_m$, and
$W^\text{closed}_f$ act non-tivially in the ground state subspace when the
closed strings are not contractible.  Thus the ground state on \eqn{torH}
spontaneously breaks the two $Z_2$-1-symmetries, and has a $Z_2$ topological
order.

Now we consider the following model with
\ml{U \gg g,J}
\ma{
 H &= - U\sum_{\v I} Q_{\v I} -g\sum_{\v p} F_{\v p} 
- J \sum_{\v i} \si^z_{\v i}  , 
\nonumber\\
Q_{\v I} &= \prod_{\v i \in \v I} \si^z_{\v i}, \ \ \
F_{\v p} =  \prod_{\v i \in \v p} \si^x_{\v i}
}
As we go from \ml{g\gg J} to \ml{g \ll J} the ground state undergoes a phase
transition that change the \ml{Z_2} topological order to trivial product state,
driven by \ml{m} particle condensation. This is because that the \ml{J}-term is
the hopping for \ml{m} and can drive the \ml{m} excitations to have a negative
energy.

The above Hamiltonian and the transition has the \ml{Z_2^m}-1-symmetry
generated by the closed $m$-strings
\ml{ W_m(\t C_1)=\prod_{\v i \in \t C_1} \si^z_{\v i} }, 
but does not have the \ml{Z_2^e}-1-symmetry generated by the
closed $e$-strings \ml{W_e(C_1)=\prod_{\v i \in C_1}  \si^x_{\v i}} and the
\ml{Z_2^f}-1-symmetry generated by the closed $f$-strings \ml{W_f(C_1\prod \t
C_1)}.

The end of \ml{W_m} (the 1-symmetry generator) is the \ml{m} particle.  The low
energy allowed excitations of the above Hamiltoniam are the particles with
trivial mutual statistics with the \ml{m} particle. Thus the low energy allowed
excitations include the \ml{m} particles, but not include the \ml{e} particles
and the \ml{f} particles (the fermions).

To summerize,
the \ml{Z_2} topological order in 2+1D has three type of topological excitations:\\
1. the \ml{Z_2}-charge \ml{e} -- boson\\
2. the \ml{Z_2}-flux \ml{m} -- boson\\
3. the charge-flux bound state \ml{f=m\otimes e} -- fermion\\
The three particles have mutual \ml{\pi} statistics respect to each other.
Below the minimal gap of the three particles \ml{\Del_e,\Del_m,\Del_f}, we have
three \ml{Z_2}-1-symmetries generated by closed string operators \ml{W_e(C_1)},
\ml{W_m(\t C_1)}, and \ml{W_f(C_1\otimes \t C_1)}.

If \ml{\Del_m \ll \Del_e,\Del_f}, then below
\ml{\Del_e,\Del_f} (but may be above \ml{\Del_m}), we have a \ml{Z_2}-1-symmetry
generated by closed string operators \ml{W_m(\t C_1)}, but not the ones from
\ml{W_e(C_1)} and \ml{W_f(C_1\otimes \t C_1)}.
The low energy allowed particles are \ml{\cC_a=\{m\}}. The 1-symmetry is
generated by string operators for the particles \ml{\cC_\text{t}=\{m\}}. 
If we reduce \ml{\Del_m} to make it negative, we will induce a
Bose-condensation of the \ml{Z_2}-flux and a \ml{Z_2^m}-1-symmetric
confinement transition:
the \ml{Z_2} topological order is changed the trivial product state.

If \ml{\Del_f \ll \Del_e,\Del_m}, then below
\ml{\Del_e,\Del_m} (but may be above \ml{\Del_f}), we have a \ml{Z_2}-1-symmetry
generated by closed string operators \ml{W_f(C_1\otimes \t C_1)}, but not the ones from
\ml{W_e(C_1)} and \ml{W_m(\t C_1)}.
The low energy allowed particles is \ml{\cC_a=\{f\}}, and the 1-symmetry is
generated by string operators for the particles \ml{\cC_\text{t}=\{f\}}. 

If we reduce \ml{\Del_f} to make it negative, can we still induce confinement
transition to change the \ml{Z_2} topological order to a trivial product state
with no topological order?  Since \ml{f} is a fermion, it cannot Bose-condense.
But it can condense into some other topologically ordered state.  Can the new
topological order cancel the parent \ml{Z_2} topological order to produce a
trivial phase without topological order?

The condensation of \ml{f} is a \ml{Z_2^f}-1-symmetric phase transition.  We
will show later that the \ml{Z_2^f}-1-symmetry is anomalous, and the
\ml{Z_2^f}-1-symmetric phase transition cannot induce a trivial product state.

\subsection{On-site/non-on-site higher symmetry}

To understand the  anomalous higher symmetry, let us first review the
connection between non-on-site symmetry and anomalous symmetry.\cite{W1313}
An on-site symmetry (on-site 0-symmetry) of group \ml{G} is generated by a
transformation of the following form:
 \ma{ U(g) = \prod_i
U_i(g),\ \ \ U(g)U(h) = \prod_i U_i(gh)=U(gh),
} 
where $g,h \in G$,  
$i$ label the lattice site and $U_i(g)$ only acts on the degrees of
freedom on site-$i$.  The on-site symmetry can be gauged to get a local
\ml{G^{\otimes N_\text{site}}} symmetry
\ma{
 U(\{g_i\}) &= \prod_i U_i(g_i),
\nonumber\\
 U(\{g_i\}) U(\{h_i\}) &= \prod_i U_i(g_ih_i)= U(\{g_ih_i\}).  
} 
An on-site symmetry is also called anomaly-free symmetry.

Roughly speaking, an non-on-site symmetry of group \ml{G} 
does have the product form
\ma{ 
U(g) \neq [\prod_i U_{i}(g)] ,\ \ \ \ U(g)U(h) = U(gh).
} 
It cannot be gauged
to get a \ml{G^{\otimes N_\text{site}}} symmetry:
\ma{ 
U(\{g_i\}) &= \prod_i U_{i,i+1}(g_i),
\nonumber\\
 U(\{g_i\}) U(\{h_i\}) &\neq U(\{g_ih_i\}).  
} 
An non-on-site symmetry is also called anomalous symmetry.  For a more accurate
discussion of non-on-site and anomalous symmetry, see \Ref{W1313}.

Similarly, an on-site \ml{k}-symmetry of an Abelian group \ml{\Pi} in
$d$-dimensional space is given by
\ma{ 
U(C_{d-k},g) & = \prod_{i\in C_{d-k}} U_i(g),
\nonumber\\
 U(g)U(h) & = \prod_{i\in C_{d-k}} U_i(gh)=U(gh),\ g \in
\Pi_2.  }
Here we stress that we have assumed that the space is a complex (a lattice)
and there are independent degrees of freedom living on the $(d-k)$-cells of the
complex.  The operator  $U_i(g)$ only acts on the degrees of freedom on the
$(d-k)$-cell labeled by $i$.  $C_{d-k}$ is a collection of  $(d-k)$-cells and
$\prod_{i\in C_{d-k}}$ is a product over all the $(d-k)$-cells in $C_{d-k}$.

The on-site \ml{(d-1)}-symmetry can be gauged 
\ma{ 
& U(C_{d-k},\{g_i\}) = \prod_{i\in C_{d-k}} U_i(g_i),
\\
& U(C_{d-k},\{g_i\}) U(C_{d-k},\{h_i\}) =
U(C_{d-k},\{g_ih_i\}).  
\nonumber 
} 
An on-site higher symmetry is also called anomaly-free
higher symmetry.

Non-on-site \ml{(d-1)}-symmetry for a group \ml{\Pi} 
\ma{ 
U(C_{d-k},g) \neq \prod_{i\in C_{d-k}} U_{i}(g), 
\nonumber\\
U(C_{d-k},g)U(C_{d-k},h) = U(C_{d-k},gh),\ \ \ \ g \in \Pi.  
}
The non-on-site higher symmetry cannot be gauged 
\ma{ 
U(C_{d-k},\{g_i\}) U(C_{d-k},\{h_i\})\neq U(C_{d-k},\{g_ih_i\}).  
}
A non-on-site higher symmetry is called anomalous higher symmetry, if we cannot
make it on-site via some local unitary operations.\cite{CGW1038}
More precisely
\frmbox{
A higher symmetry (which may be generated by operators in several different
dimensions) is anomaly-free if it allows a symmetric ground state with trivial
topological order.  A higher symmetry is anomalous if it does not allow a
symmetric ground state with trivial topological order.
}

For example, the \ml{Z_2}-1-symmetry generated by \ml{W_e(C_1) = \prod_{i\in
C_1} \si^x_i} are on-site and anomaly-free.  We note that the open string
operator $W_e(C_1)$ creates two bosons at its ends.

Also, the \ml{Z_2}-1-symmetry generated by \ml{W_m(\t C_1)=\prod_{j\in\t C_1}
\si^z_j} are on-site and anomaly-free.  Again the open string operator $W_m(\t
C_1)$ creates two bosons at its ends.

The \ml{Z_2}-1-symmetry generated by \ml{W_f(C_1\otimes\t C_1) = \prod_{i\in
C_1} \si^x_i \prod_{j\in \t C_1} \si^z_j } is not on-site and maybe anomalous.
But how to determine if a higher symmetry is anomaly-free or anomalous?  Later,
we will show that an 1-symmetry generated by a string operator \ml{W(C_1)} is
anomaly-free if and only if the end of the string is a boson.  So the
\ml{Z_2}-1-symmetry generated by \ml{W_f(C_1\otimes\t C_1) = \prod_{i\in C_1}
\si^x_i \prod_{j\in \t C_1} \si^z_j } is anomalous, since the end of string
$W_f(C_1\otimes\t C_1)$ is a fermion.  In fact, one can show that the open
string operators $W_f(C_1\otimes\t C_1)$ satisfy the so-called fermion-hopping
algebra, which make the string end to be a fermion.  The string operators
satisfying fermion-hopping algebra cannot be made into on-site operators.

The \ml{Z_2\times Z_2}-1-symmetry generated by \ml{W_e(C_1) = \prod_{i\in
C_1} \si^x_i} and by \ml{W_m(\t C_1)=\otimes_{j\in\t C_1} \si^z_j} is also
anomalous. This is because it contains the \ml{Z_2}-1-symmetry generated by
\ml{W_f(C_1\otimes\t C_1) = \prod_{i\in C_1} \si^x_i \prod_{j\in \t C_1}
\si^z_j } which is anomalous.  We also note that the end of string $W_e(C_1)$
and the end of string $W_m(\t C_1)$ have a non-trivial mutual statistics
between them, which implies a mixed anomaly between the two
\ml{Z_2}-1-symmetries.

If we have a higher symmetry of generated by operators in several dimensions
defined on the same spacial complex, $U(C_{d-k_1},g_1)$, $U(C_{d-k_2},g_2)$,
$\cdots$ and if all those operators are on-site, then the higher symmetry a
anomaly-free.  We note that since all the operators are defined on the same
spacial complex the higher symmetry generators with different dimensions act on
different degrees of freedoms living on cells of different dimensions.  So the
higher symmetry generators with different dimensions always commute with each
other.  If some higher symmetry generators are defined on a complex while other
higher symmetry generators are defined on the dual complex, then the higher
symmetry generators may not commute and may be anomalous.

\begin{table*}[t]
\caption{The higher category theory is
actually a theory of pointlike, stringlike, $\cdots$, excitations in
physics.\cite{KW1458} This table lists the corresponding concepts in
mathematics and in physics.
}
  \label{tab:dic}
  \centering
  \begin{tabular}{|p{3in}|p{3.5in}|}
    \hline
    \textbf{~~Concepts in higher category} ~~~ &  ~~\textbf{Concepts in physics}\\
    \hline
     Unitary $D$-category $\cM$ & Topological excitations with their braiding fusion properties in a topologically ordered state in $D$-spacetime dimension\\
    \hline
    Objects (0-morphisms) & The ground states\\
    \hline
    Simple 1-morphisms & The codimension-1 topological excitations\\
    \hline
    Simple $D-2$-morphisms & The stringlike topological excitations\\
    \hline
    Simple $D-1$-morphisms & The pointlike topological excitations\\
    \hline
    Composite  morphisms & The topological excitations with accidental degeneracy\\
    \hline
    The collection of simple $D-1$-morphisms, simple $D-2$-morphisms, etc & Topological excitations\\
    \hline
    Trivial morphisms & The excitations that can be created by local operators (non-topological excitations)\\
    \hline
  \end{tabular}
\end{table*}

More generally, the boundary of higher symmetry generators can produce
pointlike, stringlike, $\cdots$, topological excitations.  We can use a higher
category $\cC_\text{t}$ with one object to describe their fusion and braiding (see
Table \ref{tab:dic}).  In fact the higher category $\cC_\text{t}$ characterizes the
higher symmetry completely.  We like to conjecture that   
\frmbox{ 
The higher symmetry is anomaly-free if and only if all the morphisms
in $\cC_\text{t}$ have a unit quantum dimension, and have no phase factor under
exchange, braiding and fusion.  
}
Here the statement ``have no phase factor under exchange, braiding and fusion''
need a more precise definition.  For pointlike excitations in 2-dimensional
space and higher, ``no phase factor under exchange and braiding'' means that
the pointlike excitations are all bosons with trivial mutual statistics.  If
the fusion of some excitations is described by a pointed fusion category, the
``no phase factor under fusion'' means the $F$-symbol of the fusion category is
equal to 1. For more details, see \Ref{KLTW}.

\section{Simple lattice examples that realize higher symmetry protected topological phases} 
\label{example}

One way to show a higher symmetry in a system is anomalous is to show that the
symmetric  system can be regarded as a boundary of higher symmetry protected
topological (hSPT) state in one-higher dimension, using the relation between
anomaly and SPT state in one-higher dimension.\cite{W1313}  In this section, we
will describe some examples of hSPT states.  Using those examples, we will show
that a higher symmetry generated by several types of closed string operator is
anomaly-free only if the ends of string are bosons with trivial mutual
statistics with each others.

\begin{figure}[t]
\begin{center}
\includegraphics[scale=0.8]{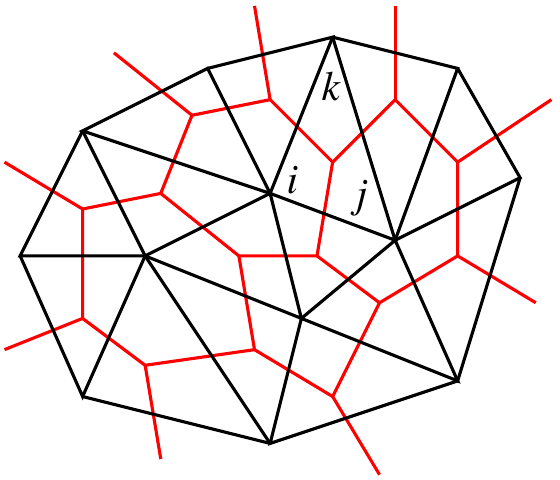}
\end{center}
\caption{(Color online) 
The black lines describe a 2-dimensional spacetime complex $\cM^2$.
The red lines describe the dual complex $\t \cM^2$.
}
\label{TriDTri}
\end{figure}

To construct lattice models with 0-symmetries and higher symmetries, it is more
convenient to do so in the spacetime Lagrangian formalism.  We construct a
spacetime lattice by first triangulating a $D$-dimensional spacetime manifold
$M^D$. (In this paper, we will use $D$ to denote spacetime dimensions and $d$
to denote space dimensions.) So a spacetime lattice is a $D$-complex $\cM^D$
with vertices labeled by $i$, links labeled by $ij$, triangles labeled by
$ijk$, \etc (see Fig. \ref{TriDTri}).  The $D$-complex $\cM^D$ also has a dual
complex denoted as $\t \cM^D$.  The  vertices of $\cM^D$ correspond to the
$D$-cells in $\t \cM^D$, The links of $\cM^D$ correspond to the $(D-1)$-cells
in $\t \cM^D$, \etc

Our spacetime lattice model may have a field living on the vertices, $g_i$.
Such a field is called a 0-cochain.  The model may also have a field living on
the links, $a_{ij}$. Such a field is called a 1-cochain, \etc.  To construct
spacetime lattice models, in particular, the topological spacetime lattice
models,\cite{KT13094721,TK151102929,W161201418,LW180901112} we will use
extensively the mathematical formalism of cochains, coboundaries, and cocycles
(see Appendix \ref{cochain}).  The relation between the spacetime path integral
approach and the Hamiltonian approach is discussed in  Appendix \ref{pathHam}.

\subsection{A 3+1D model to realize a pure $Z_n$-1-SPT phase}

\subsubsection{The bulk theory and the boundary theory} \label{SecZn1SPT}

In this section, we will consider a 3+1D bosonic model on a spacetime complex
$\cM^4$, with $\Z_n$-valued dynamic field $ a^{\Z_n}_{ij}$ on the links $ij$ of
the complex $\cM^4$.  We also have a $\Z_n$-valued non-dynamical background
field $\hat B^{\Z_n}_{ijk}$  on the triangles $ijk$ of the complex $\cM^4$.
The path integral of our bosonic model is given by
\begin{align}
\label{Zn1SPT}
  Z &= \sum_{\{a^{\Z_n}\} } 
\ee^{2\pi \ii \int_{\cM^4} \frac{k}{n}
\gSq^2(\hat B^{\Z_n}+\dd a^{\Z_n})
},
\\
& (k,n)=(\text{integer}, \text{integer}) 
,
\nonumber 
\end{align}
where $\sum_{\{a^{\Z_n}\} }$ sums over $\Z_n$-valued 1-cochains $a^{\Z_n}$, and
$\hat B^{\Z_n}$ is a $\Z_n$-valued 2-cocycle
\begin{align}
\dd  \hat B^{\Z_n}\se{n}0.
\end{align}
Also $\gSq^2$ is the generalized Steenrod square defined by \eqn{Sqdef}.
We will show that the above model realizes a $Z_n$-1-SPT phase.

Since $\hat B^{\Z_n}$ and $a^{\Z_n}$ are $\Z_n$-valued, we require the action
amplitude $\ee^{2\pi \ii \int_{\cM^4} \frac{k}{n} (\hat B^{\Z_n}+\dd
a^{\Z_n})^2 }$ to be invariant under the transformation
\begin{align}
\label{gaugeZn}
 \hat B^{\Z_n} \to \hat B^{\Z_n} + n b^{\Z},\ \ \ \
 a^{\Z_n} \to a^{\Z_n} + n u^{\Z},
\end{align}
where $b^{\Z}$ and $u^{\Z}$ are any $\Z$ valued 2-cochain and 1-cochain.  (To
do the addition $a^{\Z_n} + n u^{\Z}$, we have lifted the $\Z_n$-value of
$a^{\Z_n}$ to $\Z$.)
From \eqn{Sqplus1}, we see that
\begin{align}
&\ \ \ \ 
\gSq^2(n b^{\Z} + \hat B^{\Z_n} ) 
=
 n^2 \gSq^2( b^{\Z}) 
+ \gSq^2(\hat B^{\Z_n})
+ 2 n b^{\Z} \hat B^{\Z_n} 
\nonumber\\
& 
+ n \dd \hat B^{\Z_n} \hcup{2} \dd b^{\Z}
-n\dd ( \hat B^{\Z_n} \hcup{1} b^{\Z})
-n\dd ( \dd \hat B^{\Z_n} \hcup{2} b^{\Z}
)
\nonumber\\
& \se{n}  \gSq^2(\hat B^{\Z_n}).
\end{align}
We see that the action amplitude
$\ee^{2\pi \ii \int_{\cM^4} \frac{k}{n}
(\hat B^{\Z_n}+\dd a^{\Z_n})^2
}$ is indeed invariant under \eqn{gaugeZn}
even when $\cM^4$ has a boundary.
The above result implies that the model has a $Z_n$-1-symmetry generated by
\begin{align}
\label{Zn1symm}
 a^{\Z_n} \to a^{\Z_n} +  \al^{\Z_n}, \ \ \ \dd  \al^{\Z_n} \se{n} 0,
\end{align}
even when $\cM^4$ has a boundary.  

In \eqn{Zn1SPT}, $\hat B^{\Z_n}$ is the $\Z_n$ background 2-connection to
describe the twist of the $Z_n$-1-symmetry.  The model has a $Z_n$ gauge
symmetry:
\begin{align}
\label{Zngauge}
 a^{\Z_n} \to a^{\Z_n} +  \hat a^{\Z_n},  \ \ \ \ \
 \hat B^{\Z_n} \to \hat B^{\Z_n} -  \dd\hat a^{\Z_n}. 
\end{align}
Using \eqn{Sqgauge} we find that
\begin{align}
&\ \ \ \ \gSq^2 (\hat B^{\Z_n}+\dd a^{\Z_n})
\\
&= \gSq^2 \hat B^{\Z_n} +2  \hat B^{\Z_n} \dd \hat a^{\Z_n}
+\dd [\gSq^2 \hat a^{\Z_n} - \dd \hat a^{\Z_n} \hcup{1} \hat B^{\Z_n} ]
\nonumber\\
&\se{2n} \gSq^2 \hat B^{\Z_n} 
+\dd [\hat a^{\Z_n} \dd \hat a^{\Z_n} - \dd \hat a^{\Z_n} \hcup{1} \hat B^{\Z_n} + \hat B^{\Z_n} \hat a^{\Z_n}]
\nonumber 
\end{align}
Therefore
\begin{align}
\ee^{2\pi \ii \int_{\cM^4} \frac{k}{n} \gSq^2 (\hat B^{\Z_n}+\dd a^{\Z_n}) }
=\ee^{2\pi \ii \int_{\cM^4} \frac{k}{n} \gSq^2 (\hat B^{\Z_n}) }
\end{align}
for closed spacetime $\cM^4$.  The model is exactly soluble and gapped for
closed spacetime $\cM^4$.

Eqn. \eq{Zn1SPT} has no topological order
since on closed spacetime and for $\hat B^{\Z_n}=0$
\begin{align}
 Z &= \sum_{\{a^{\Z_n}\} } \ee^{2\pi \ii \int_{\cM^4} \frac{k}{n}
(\dd a^{\Z_n})^2 } = \sum_{\{a^{\Z_n}\} } 1 =
n^{N_l},
\end{align}
where $N_l$ is the number of links in the spacetime complex $\cM^4$.
$n^{N_l}$ is the so called the volume term that is linear in the spacetime
volume.  The topological partition function $Z^\text{top}$ is given via
\begin{align}
 Z = \ee^{-\eps V} Z^\text{top}
\end{align}
where $V$ is the spacetime volume. (For a detailed discussion of the
non-universal volume term and the universal topological terms, see
\Ref{KW1458,WW180109938}.) After removing the volume term, the topological
partition function of the above model is $Z^\text{top}(\cM^4)=1$ for all closed
4-complex $\cM^4$.  Thus the above model has no topological order.  After we
turn on the flat $Z_n$ 2-connection $\hat B^{\Z_n}$, the topological partition
function of the model \eq{Zn1SPT} is
\begin{align}
Z^\text{top}(\cM^4, \hat B^{\Z_n})= \ee^{2\pi \ii \int_{\cM^4} \frac{k}{n}
(\hat B^{\Z_n})^2 },\ \ \ 
\dd \hat B^{\Z_n} \se{n}0.
\end{align}

The above 1-SPT invariant looks different for different $k$ mod $n$. But are 
they really different?  If we gauge the $Z_n$-1-symmetry, we turn the above
$Z_n$-1-SPT phase into a topological ordered phase described by a pure $Z_n$
2-gauge theory:\cite{ZW180809394} 
\begin{align}
  Z &= \sum_{\{\dd b^{\Z_n}\se{n}0\} } \ee^{2\pi \ii \int_{\cM^4} \frac{k}{n}
(\hat b^{\Z_n})^2 }. 
\end{align}
It turns out that the same topological order is also described by a
$Z_{\<2k,n\>}$ gauge theory.  The $Z_{\<2k,n\>}$ gauge theory has emergent
fermions iff $2kn/\<2k,n\>^2= \text{odd}$.  So the 1-SPT invariant is really
different at least when the pairs $[\<2k,n\>, \text{mod}(2kn/\<2k,n\>^2,2)]$
are different.

In \Ref{ZW180809394}, it was shown that $H^4(\cB(Z_n,2);\R/\Z)=\Z_n$ for
$n=\text{odd}$, and $H^4(\cB(Z_n,2);\R/\Z)=\Z_{2n}$ for $n=\text{even}$.  Thus
the above 1-SPT invariant is non-tirivial.  There are (at least) $n$ distinct
$Z_n$-1-SPT phases labeled by $k=0,\cdots,n-1$. 

To see the physical properties of the $Z_n$-1-SPT phase, we consider its 2+1D
boundary state described by 
\begin{align}
\label{Zn1ano}
 Z(\cB^3) & = 
\sum_{\{a^{\Z_n}\} } 
\ee^{2\pi \ii \int_{\cB^3} \frac{k}{n} a^{\Z_n} \dd a^{\Z_n} } 
\end{align}
where we have set the background 2-connection $\hat B^{\Z_n}=0$.
The boundary theory also has the $Z_n$-1-symmetry which is generated by
\eqn{Zn1symm}.  We like to point out that the action amplitude $\ee^{2\pi \ii
\int_{\cB^3} \frac{k}{n} a^{\Z_n} \dd a^{\Z_n} }$ for spacetime with boundary
$\prt\cB^3\neq 0$ is actually not invariant under the $Z_n$-1-symmetry
transformation.  Only the action amplitude for closed spacetime $\prt\cB^3= 0$
has the $Z_n$-1-symmetry.  This indecates that the $Z_n$-1-symmetry in the
2+1D model \eqn{Zn1ano} is anomalous.

The 2+1D model is not exactly soluble.
To have a soluble model, we restrict $a^{\Z_n}$ to be cocycles and obtain
\begin{align}
 Z(\cB^3) & = 
\hskip -1.3em
\sum_{\{\dd a^{\Z_n} \se{n} 0\} } 
\hskip -1.3em
\ee^{2\pi \ii \int_{\cB^3} \frac{k}{n} a^{\Z_n} \dd a^{\Z_n} } 
=
\sum_{\{\dd a^{\Z_n} \se{n} 0\} } 
1 .
\end{align}
The above model actually describes a
2+1D untwisted $Z_n$ gauge theory.

In the presence of the $Z_n$-flux described by 2-coboundary $\dd \hat a^{\Z_n}$
and the $Z_n$ charge described by worldline $C^1$, the above path integral is
modified. The new one is
obtained by adding the term $\ee^{ \frac{2\pi}{n} \ii \int_{C^1} a^{\Z_n}}$ and
then replacing $a^{\Z_n}$ by $a^{\Z_n}+\hat a^{\Z_n}$.  We find
\begin{align}
\label{Zn1SPTb}
 Z(\cB^3) & = 
\hskip -1.5em
\sum_{\{\dd a^{\Z_n} \se{n} 0\} } 
\hskip -1.3em
\ee^{2k\pi \ii \int_{\cB^3} a^{\Z_n} \Bs_n a^{\Z_n} 
+\frac{2k\pi}{n} \ii \int_{\cB^3} (2a^{\Z_n}\dd \hat a^{\Z_n} 
+\hat a^{\Z_n}\dd \hat a^{\Z_n})}
\nonumber \\
&\ \ \ \ \ \ \ \
\ee^{ \frac{2\pi}{n} \ii \int_{C^1} (a^{\Z_n}+\hat a^{\Z_n})}
.
\end{align}

Let $\t c_2$ be the Poincar\'e dual of the cycle $C^1$ which is a 2-cocycle on
the dual complex $\t\cB^3$. The above can be rewritten as
\begin{align}
 Z(\cB^3) & = 
\hskip -1.5em
\sum_{\{\dd a^{\Z_n} \se{n} 0\} } 
\hskip -1.3em
\ee^{ 
+2\pi \ii \int_{\cB^3} \frac{2k}{n} a^{\Z_n}\dd \hat a^{\Z_n} 
+\frac kn \hat a^{\Z_n}\dd \hat a^{\Z_n}
+\frac 1n \t c_2 (a^{\Z_n}+\hat a^{\Z_n})}
.
\end{align}
Now let us consider a bound state of $m_f$ $Z_n$-flux quanta
and $m_c$ unit of $Z_n$ charges.
Let 2-cocycle $\dd \hat a$ be the Poincar\'e dual of the worldline of such a bound state. The path integral
in presence of such a bound state is obtained
by setting $\dd \hat a^{\Z_n} = m_f \dd \hat a$ and
$\t c_2 = m_c \dd \hat {\t a}$. We get
\begin{align}
 Z(\cB^3) & = 
\hskip -1.5em
\sum_{\{\dd a^{\Z_n} \se{n} 0\} } 
\hskip -1.3em
\ee^{ 
2\pi \ii \int_{\cB^3} 
\frac{2k}{n} a^{\Z_n} \dd \hat a 
+\frac{m_c}{n} a^{\Z_n} \dd \hat {\t a} 
+\frac {km_f^2 } n \hat a\dd \hat a
+\frac {m_cm_f} n \hat a\dd \hat {\t a}
}
.
\end{align}
This suggests that the statistics of the bound state is given by $\th =2\pi(
\frac{km_f^2}{n} +\frac{m_c m_f}{n})=\pi \v m^\top K^{-1}\v m $, where
\begin{align}
 K = \bpm 
-2kn& n \\
n   & 0 \\
\epm
,
 K^{-1} = \bpm 
0  & \frac 1n \\
\frac 1n  &\frac {2k}n \\
\epm
, 
\v m =
\bpm
m_c\\
m_f\\
\epm .
\end{align}

The above statistics can be reproduced by a $U(1)\times U(1)$ Chern-Simons (CS)
theory. Thus the 2+1D bosonic model \eq{Zn1SPTb} can be described by
$U(1)\times U(1)$ CS theory 
\begin{align}
\label{KCS}
\cL= \pi K_{IJ} a_I\dd a_J  +\cdots 
,
\end{align}
where $a_I$'s are
1-forms and $a_I\dd a_J=a_I\wedge \dd a_J$ is a wedge product
of differential forms.  
The
$\cdots$ term makes $a_I$ to have a small curvature
\begin{align}
 \dd a_I \approx 0 
\end{align}

We can choose a new basis to rewrite \eqn{KCS} as
\begin{align}
\label{mCS}
\cL= \pi K'_{IJ} a'_I\dd a'_J  +\cdots 
,
\end{align}
where
\begin{align}
 K'&=\bpm 
0  & n \\
n  & 0 \\
\epm
= W^\top K W,\ \ \
W = \bpm 
1  & 0 \\
k  & 1 \\
\epm
\end{align}
Thus, the CS theory \eqn{KCS} always describes the same $Z_n$ gauge theory
\eqn{mCS}
regardless the value of $k$.  In the new bases, the excitations are labled by
\begin{align}
\v m' = (W^\top)^{-1} \v m\  \text{ or } \ 
\bpm
m_1'\\
m_2'\\
\epm
=
\bpm 
1  & -k \\
0  & 1 \\
\epm
\bpm
m_c\\
m_f\\
\epm .
\end{align}

The 2+1D bosonic model \eq{Zn1SPTb} without the $C^1$ term has a 
$Z_n$-1-symmetry \eq{Zn1symm}.  The model \eq{Zn1SPTb} can be described by a
$U(1)\times U(1)$ CS theory \eq{KCS} or \eq{mCS}.  The low energy allowed
excitations are described by $\v m_\text{a}^\top = (0,1)$, or $\v m_\text{a}'^\top = (-k,1)$
in the new basis.  Thus the $Z_n$-1-symmetry is generated by the excitation $\v
m'^\top_\text{t} =(k,1)$ which has a trivial mutual statistics with $\v m_\text{a}'$.  
However, such a $Z_n$-1-symmetry is anomalous (or non-on-site) when $k\neq 0 $
mod $n$, since the 2+1D theory is a boundary of the 3+1D hSPT phase.  We cannot
gauge it to obtain a $Z_n$ 2-gauge theory.  This is an example of emergent
anomalous $Z_n$-1-symmetry in a topologically ordered state.

We note that the the excitation $\v m'^\top_\text{t} =(k,1)$ has a statistics
$\th=2\pi \frac kn$, which is not bosonic when $k\neq 0 $ mod $n$.  Thus, the
$Z_n$-1-symmetry \eq{Zn1symm} is anomalous when the associated excitation is
not a boson.

\subsubsection{A conjecture to detect anomalous 1-symmetry}
\label{testanom}

In fact, the above discussions can be generalized to obtain emergent
(anomalous) 1-symmetry in a $U^\ka(1)$ CS theory \eq{KCS} described by a general
$K$-matrix.\cite{BW9045,FK9169,WZ9290} (For a more general discussion, see
\Ref{BH180309336}.) For a 2+1D bosonic topological order descried by even
$K$-matrix, an emergent higher symmetry is described by a set of low energy
allowd topological excitations $\cC_\text{a}=\{\v m_\text{a}\}$ which form a lattice.
All other non-trivial topological excitations not in the $\cC_\text{a}$ have very high
energies above $\La$.  Then the $K$ topological order plus the low energy
allowd topological excitations $\cC_\text{a}$ has an emergent higher symmetry below
the energy $\La$.  The  emergent higher symmetry is generated by the
topological excitations in $\cC_\text{t}=\{\v m_\text{t}\}$, which is formed by $\v
m_\text{t}$'s that satisfy
\begin{align}
 \v m_\text{a} K^{-1} \v m_\text{t} = \text{integer},
\ \ \ \forall \v m_\text{a} \in \cC_\text{a},
\end{align}
(\ie $\v m_\text{t}$ has a trivial mutual statistics with all low energy allowed
excitations $\v m_\text{a}$ in $\cC_\text{a}$).  Note that $\cC_\text{t}$ is also a lattice.  We see
that \frmbox{an emergent higher symmetry in a topological order can be fully
characterized by a subset $\cC_\text{a}$ of low energy allowed topological
excitations, which is closed under fusion and braiding.} 

If we include those allowed  low energy topological excitations, the action
amplitude will become
\begin{align}
\label{KCSa}
 \ee^{\ii \pi \int_{\cB^3} K_{IJ} a_I\dd a_J}
 \ee^{\ii 2\pi \int_{C^1} \v m_\text{a}^\top \v a}
\end{align}
where $C^1$ is the worldline of the $\v m_\text{a}$ excitation.  The above
action amplitude actually describes a boundary of a 3+1D hSPT phase described
by \eqn{K4DdadaC1} with a 1-symmetry (see \eqn{CSlatt1}).  In Section
\ref{Uk1}, we show that such an 1-symmetry happen to be the one described by
the lattice $\cC_\text{t}$ introduced above.  Also in Section \ref{Uk1}, we
will show that the 3+1D hSPT order is trivial iff $\pi \v m_\text{t}^\top
K^{-1} \v m_\text{t} = 2\pi \times \text{integer},\ \ \forall \v m_\text{t},\v
m_\text{t}' \in \cC_\text{t}$.  Therefore, \frmbox{for the higher symmetry
characterized by low energy allowed topological excitations $\cC_\text{a}$ in a
2+1D Abelian topological order, the higher symmetry is anormaly free iff
$\cC_\text{t}$ contains only bosons with tivial mutual statistics among them.
(Here $\cC_\text{t}$ is formed by the topological excitations, that have
trivial mutual statistics with all the topological excitations in
$\cC_\text{a}$).  } For a more general and detailed discussion, see
\Ref{BH180309336} and \Ref{KLTW}.

\subsubsection{Higher anomaly and phase transition}

In above, we see that the emergent (anomalous) higher symmetry is not a
property of a topologically ordered state.  It is a property of a pair: a
topologically ordered state plus its allowed low energy topological
excitations.

For example, a 2+1D untwisted $Z_n$-gauge theory has a $Z_n$-1-symmetry if we
only allow $Z_n$-flux and their fluctuations, and do not allow, for example,
any $Z_n$-charge and its fluctuations.  Such a $Z_n$-1-symmetry 
\begin{align}
\label{Zn1symm1}
 a^{\Z_n} \to a^{\Z_n} +  \al^{\Z_n}, \ \ \ \dd  \al^{\Z_n} \se{n} 0.
\end{align}
is anomaly free (\ie on-site and gaugable).  

Now let us start with the deconfined phase of the 2+1D untwisted $Z_n$-gauge
theory described by
\begin{align}
 Z(\cB^3) & = \sum_{\{\dd a^{\Z_n} \se{n} 0\} }  1 .
\end{align}
The deconfined phase has an anomaly-free $Z_n$-1-symmetry \eq{Zn1symm1}.  We
then increase the fluctuations of the $Z_n$-flux to drive a phase transition to
the confined phase.  The confined phase is described by
\begin{align}
\label{Zn1anofree}
 Z(\cB^3) & = \sum_{\{ a^{\Z_n} \} }  1 ,
\end{align}
which is a product state.  The product phase also has an anomaly-free $Z_n$
1-symmetry \eq{Zn1symm1}, which is the same as the deconfined phase.  Thus the
phase transition from the $Z_n$-gauge deconfined phase to the confined phase
(the product state) is an allowed phase transition.  Such a phase transition is
induced by the boson condensation of the  $Z_n$-flux quanta.

As a second example, let us consider the same 2+1D untwisted $Z_n$-gauge theory
described by the mutual CS theory \eq{mCS}, but with different allowed
topological excitations: the bound states of unit $Z_n$-flux and $-k$
$Z_n$-charges $\v m_\text{a}'^\top = (-k,1)$.  In this case, the system has a
different $Z_n$-1-symmetry.  When $k \neq 0$, the $Z_n$-1-symmetry
\eq{Zn1symm1} is \emph{anomalous} (\ie non-on-site and not gaugable).  This
means that if we want increase the fluctuations of the $\v m_\text{a}'^\top =
(-k,1)$ excitations to induce a phase transition, we get a phase described by
\eqn{Zn1ano}.  We cannot reach a product state described by \eqn{Zn1anofree}
which has a different \emph{anomaly-free} $Z_n$-1-symmetry.  This result is
expected.  When $k\neq 0$, the $\v m_\text{a}'^\top = (-k,1)$ excitation has a
statistics $\th=-\frac{2k}{n}\pi$.  The anyons cannot condense directly.
However, anyon-pairs or other proper clusters of anyons may condense to drive a
phase transition.  Previously, we believe that those condensations lead to
topologically ordered phases, but we are not totally sure.  

The result from this paper provides a proof for the above general belief, by
understanding it from a point of view of the anomaly matching of the $Z_n$
1-symmetry.  But why do we need to match the anomaly of the $Z_n$-1-symmetry?
This is because the theories with different anomalies of higher symmetry are
boundaries of different hSPT states in one higher dimension.  No matter how we
change the boundary interaction, we cannot change the hSPT order in one higher
dimension. Hence we cannot change the higher anomaly, unless we explicitly
break the higher symmetry on the boundary.  Thus for $k\neq 0 \text{ mod } n $
\frmbox{no matter how we condense the $\v m_\text{a}'^\top = (-k,1)$ excitation in a
2+1D $Z_n$-gauge theory (\ie the CS theory \eq{mCS}), we can never get the
trivial confined phase with no topological order.} On the other hand, if we
allow the fluctuations of the bound states of several different combinations of
$Z_n$-flux and $Z_n$-charges, then we may be able to induce the trivial
confined phase with no topological order.  In this case, the $Z_n$-1-symmetry
is explicitly broken and the anomaly matching of the $Z_n$-1-symmetry is
invalidated.

In general \frmbox{no matter how we condense the low energy allowed topological
excitation that form $\cC_\text{a}$, we can never get the trivial confined phase with
no topological order, if the higher symmetry characterize by $\cC_\text{a}$ is
anomalous (\ie if $\cC_\text{t}$ obtained from $\cC_\text{a}$ are not formed by boson
with trivial mutual statistics).}

\subsection{A $D$-dimensional model to realize a $Z_2$ $k$-SPT phase}

\subsubsection{The bulk theory and the boundary theory}

In the above, we have constructed models to realize $Z_n$-1-SPT phases in 3+1D.
Here we will construct models to realize pure $Z_2$ $k$-SPT phases in any
dimension:
\begin{align}
\label{Z2hSPT}
  Z(\cM^D) &= \sum_{\{a^{\Z_n}_k\} } \ee^{m\pi \ii \int_{\cM^D} 
\gSq^{D-k-1} (\hat B^{\Z_2}_{k+1} + \dd a^{\Z_2}_k) },
\end{align}
where $\dd \hat B^{\Z_2}_{k+1} \se{2} 0$, $m=0,1$.  The theory is well defined
even for $\cM^D$ with boundary,
since
\begin{align}
 \gSq^{D-k-1} (\hat B^{\Z_2}_{k+1} + 2 c_{k+1})
\se{2} \gSq^{D-k-1} \hat B^{\Z_2}_{k+1} 
\end{align}
where we have used \eqn{Sqplus1}.  
So the model has a $Z_2$
$k$-symmetry
\begin{align}
\label{Z2ksymm}
 a^{\Z_2}_k \to a^{\Z_2}_k +  \al^{\Z_2}_k,\ \ \ \dd  \al^{\Z_2}_k\se{2}0.
\end{align}
even when $\cM^D$ has a boundary.
We can also show that
\begin{align}
 \gSq^{D-k-1} (\hat B^{\Z_2}_{k+1} + \dd a^{\Z_2}_k) \se{2,\dd}
 \gSq^{D-k-1} \hat B^{\Z_2}_{k+1} ,
\end{align}
using \eqn{Sqgauge}.  
Thus, the hSPT phase is characterized by hSPT invariant
\begin{align}
 \label{Z2hSPTtop}
  Z^\text{top}(\cM^D, \hat B^{\Z_2}_{k+1}) &= \ee^{m\pi \ii \int_{\cM^D}
\gSq^{D-k-1} \hat B^{\Z_2}_{k+1} }.
\end{align}
for closed $\cM^D$.

One boundary of the above hSPT state is described by
(after setting $\hat B_{k+1}^{\Z_2}=0$)
\begin{align}
\label{Z2hSPTb}
  Z(\cB^{D_b}) &= \sum_{\{a^{\Z_2}_k\} } \ee^{m\pi \ii \int_{\cB^{D_b}} 
\gSq^{D_b-k} a^{\Z_2}_k },
\end{align}
where $D_b=D-1$ is the spacetime dimension of the boundary.  But such a
boundary theory is not exactly soluble.  An exactly soluble boundary is
described by
\begin{align}
\label{Z2hSPTb0}
  Z(\cB^{D_b}) &= 
\hskip -1em
\sum_{\{\dd a^{\Z_2}_k \se{2} 0\} } 
\hskip -1em
\ee^{m\pi \ii \int_{\cB^{D_b}} \gSq^{D_b-k} a^{\Z_2}_k  }
\end{align}
which describes a $Z_2$ $k$-gauge theory twisted by the topological term $
\ee^{m\pi \ii \int_{\cB^{D_b}} \gSq^{D_b-k} a^{\Z_2}_k  } $.  In the presence
of the higher $Z_2$-flux, the path integral becomes (after replacing
$a^{\Z_2}_k$ by $a^{\Z_2}_k+\hat a^{\Z_2}_k$)
\begin{align}
\label{Z2hSPTb1}
  Z(\cB^{D_b}) &= 
\hskip -1em
\sum_{\{\dd a^{\Z_2}_k \se{2} 0\} } 
\hskip -1em
\ee^{m\pi \ii \int_{\cB^{D_b}} \gSq^{D_b-k} (\hat a^{\Z_2}_k +a^{\Z_2}_k)  }
  \\
 &= 
\hskip -1em
\sum_{\{\dd a^{\Z_2}_k \se{2} 0\} } 
\hskip -1em
\ee^{m\pi \ii \int_{\cB^{D_b}} \gSq^{D_b-k} \hat a^{\Z_2}_k +\gSq^{D_b-k} a^{\Z_2}_k},
\nonumber 
\end{align}
where $\dd \hat a^{\Z_2}_k $ describes the higher $Z_2$-flux on the boundary.
The above model has a $Z_2$ $k$-symmetry \eq{Z2ksymm}.  The $Z_2$ $k$-symmetry
is anomalous for $m=1$ and anomaly-free for $m=0$.

\subsubsection{Higher anomaly and phase transition}

Now consider topologically ordered state in $D_b$ spacetime dimension
described by the deconfined phase of the $Z_2$ $k$-gauge theory \eq{Z2hSPTb0}.
We allow only the fluctuations of the higher $Z_2$-flux, and try to use them to
drive a phase transition.  Such a system has the $Z_2$ $k$-symmetry
\eq{Z2ksymm}.  Using the anomaly matching condition, we find that the phase
transition can nerve produce the confined phase with topological order, when
$m=1$.  On the other hand, when $m=0$, the tirivial confined phase can be reach
by the phase transition.

We like to stress that here we only ask can we obtain the product state from
the deconfined phase of the $Z_2$ $k$-gauge theory \eq{Z2hSPTb0} by the
fluctuations of the higher $Z_2$-flux only.  We find that we cannot obtain the
product state from the deconfined phase when $m=1$. However, if we include both
fluctuations of the higher $Z_2$-charge and the higher $Z_2$-flux, then we can
always obtain the product state from the deconfined phase regardless the value
of $m$.

As an application of the above result, let us consider the case with $D_b=4$
and $k=2$.  The deconfined phase of the 3+1D $Z_2$ $2$-gauge theory is
described by (with the $Z_2$ 2-flux)
\begin{align}
\label{Z2tgauge}
  Z(\cB^{4}) &= 
\hskip -1em
\sum_{\{\dd b^{\Z_2} \se{2} 0 \} } 
\hskip -1em
\ee^{m\pi \ii \int_{\cB^{4}} \gSq^{2} (\hat b^{\Z_2}+b^{\Z_2}) },
\end{align}
where $\dd \hat b^{\Z_2}$ is a fixed 3-coboundary describing the $Z_2$ 2-flux
and $b^{\Z_2}$ is a dynamical 2-cochain.

It is well known that a $Z_2$ 2-gauge theory in 3+1D is dual to a $Z_2$ gauge
theory (see for example \Ref{W161201418}).  The so called  $Z_2$ 2-flux in the
$Z_2$ 2-gauge theory correspond to the $Z_2$-charge in the $Z_2$ gauge theory.
In fact, the 3-coboundary $\dd \bar b^{\Z_2}$ is the Poincar\'e dual of the
worldline of the $Z_2$-charge in 3+1D spacetime.  When $m=0$, \eqn{Z2tgauge}
corresponds to a untwisted $Z_2$ gauge theory  where the $Z_2$ charge is a
boson and the $Z_2$ $2$-symmetry is anomaly-free.  When $m=1$, \eqn{Z2tgauge}
corresponds to a twisted $Z_2$ gauge theory where the $Z_2$ charge is a
fermion\cite{LW0316} and the $Z_2$ $2$-symmetry is anomalous. The result in
this section implies that \frmbox{ any $Z_2$ charge fluctuation and
condensations in the 3+1D bosonic topological order described by a twisted
$Z_2$ gauge theory \eq{Z2tgauge} cannot induce the trivial gapped phase with
no topological order.} In contrast, the $Z_2$-charge fluctuations and
condensations in the \emph{untwisted} $Z_2$ gauge theory can induce the trivial
product state.  Also, the $Z_2$-charge and $Z_2$-flux fluctuations and
condensations in the twisted $Z_2$ gauge theory can induce the trivial product
state.  The $Z_2$-flux fluctuations breaks the $Z_2$-2-symmetry and invalidate
the anomaly matching of the $Z_2$-2-symmetry.

\subsection{A $D$-dimensional model to realize a $Z_n$ $k$-SPT phase}

We can also construct models to realize more general pure $Z_n$ hSPT
phases.  For $D-k= \text{even}$, the following model
realizes a  $Z_n$ $k$-SPT phase.
\begin{align}
\label{ZnhSPT}
  Z &= \sum_{\{a^{\Z_n}_k\} } \ee^{2\pi \ii \int_{\cM^D} \frac{m}{n}
\gSq^{D-k-1} (\hat B^{\Z_n}_{k+1} + \dd a^{\Z_n}_k) },
\end{align}
The theory is well defined when $\dd \hat B^{\Z_n}_{k+1} \se{n} 0$, since
\begin{align}
 \gSq^{D-k-1} (\hat B^{\Z_n}_{k+1} + n c_{k+1})
\se{n} \gSq^{D-k-1} \hat B^{\Z_n}_{k+1} 
\end{align}
where we have used \eqn{Sqplus1}.  We can also show that
\begin{align}
 \gSq^{D-k-1} (\hat B^{\Z_n}_{k+1} + \dd a^{\Z_n}_k) \se{n,\dd}
 \gSq^{D-k-1} \hat B^{\Z_n}_{k+1} ,
\end{align}
using \eqn{Sqgauge} and $D-k-1 = \text{odd}$.  The model has a $Z_n$
$k$-symmetry
\begin{align}
\label{Znksymm}
 a^{\Z_n}_k \to a^{\Z_n}_k +  \al^{\Z_n}_k,\ \ \ \dd  \al^{\Z_n}_k\se{n}0.
\end{align}
and $Z_n$
$(k+1)$-gauge symmetry
\begin{align}
 \hat B_{k+1}^{\Z_n} \to \hat B_{k+1}^{\Z_n}+\dd a^{\Z_n}_k .
\end{align}
The $Z_n$
$k$-symmetry is anomaly free since it can be gauged.

Such a hSPT phase is characterized by hSPT invariant
\begin{align}
 \label{ZnhSPTtop}
  Z^\text{top}(\cM^D, \hat B^{\Z_n}_{k+1}) &= \ee^{2\pi \ii \int_{\cM^D} \frac{m}{n}
\gSq^{D-k-1} \hat B^{\Z_n}_{k+1} },
\end{align}
The  hSPT state can have a boundary described by
\begin{align}
\label{ZnhSPTb}
  Z &= \sum_{\{a^{\Z_n}_k\} } \ee^{2\pi \ii \int_{\cB^{D-1}} \frac{m}{n}
\gSq^{D-k-1} (a^{\Z_n}_k) },
\end{align}
after setting $\hat B^{\Z_n}_{k+1}=0$.
The boundary theory \eq{ZnhSPTb}
also has the $Z_n$ $k$-symmetry \eq{Znksymm} when $\cB^{D-1}$ has no boundary.
This can be shown by using \eqn{Sqgauge}.

We may choose $D=6$ and $k=2$
\begin{align}
\label{Zn2SPT6D}
  Z &= \sum_{\{a^{\Z_n}_2\} } \ee^{2\pi \ii \int_{\cM^4} \frac{m}{n}
\gSq^3 (\hat B^{\Z_n}_{3} + \dd a^{\Z_n}_2) }.
\end{align}
The model has a $Z_n$-2-symmetry
\begin{align}
 a^{\Z_n}_2 \to a^{\Z_n}_2 +  \al^{\Z_n}_2,\ \ \ \dd  \al^{\Z_n}_2\se{n} 0,
\end{align}
and realizes a $Z_n$-2-SPT phase.  A $Z_n$-2-symmetric boundary of such a 2-SPT
phase is described by (after setting $\hat B^{\Z_n}_{3}=0$):
\begin{align}
\label{Zn2SPT5D}
  Z(\cB^5) &
= 
\hskip -1.3em
\sum_{\{\dd a^{\Z_n}_2\se{n}0\} } 
\hskip -1.3em
\ee^{2\pi \ii \int_{\cB^5} \frac{m}{n} \gSq^3 (a^{\Z_n}_2) }
= 
\hskip -1.3em
\sum_{\{\dd a^{\Z_n}_2\se{n}0\} } 
\hskip -1.3em
\ee^{2\pi \ii \int_{\cB^5} \frac{m}{n} 
a^{\Z_n}_2\dd a^{\Z_n}_2 }.
\end{align}
The $Z_n$-2-symmetry on $\cB^5$ is anomalous when $m \neq 0$ mod $n$.  The
model can not reach to trivial gapped phase with no topological order even if
we allow fluctuations with $\dd a^{\Z_n}_2 \neq 0$, but do not allow the
fluctuations of the charges of $Z_n$ 2-gauge theory (which are closed strings).

\subsection{A 3+1D $U^\ka(1)$ bosonic model to realize a $Z_{k_1}\times
Z_{k_2}\times ... $-1-SPT phase}

\label{Uk1}

In this section, we will use a 3+1D $U^\ka(1)$ ``gauge theory'' in the confined
phase to realize some hSPT phase. Our model is a bosonic model defined on a
triangulated spacetime (with vertices labeled by $i,j,\cdots$.  On each link
$ij$, we have bosonic degrees of freedom described by $(a^\RZ_I)_{ij} \in
(-\frac12,\frac12]$, $I=1,\cdots,\ka$.  To write down the path integral of the
bosonic model, we start with 2+1D $U^\ka(1)$ Chern-Simons theory on spacetime
lattice $\cB^3$:\cite{DW}
\begin{align}
\label{CSlatt1}
Z=&\int D[a^\RZ_I]
\ee^{\ii 2\pi  \int_{\cB^3} \sum_{I<J} k_{IJ} \dd \big(a^{\R/\Z}_I(a^{\R/\Z}_J-\toZ{a^{\R/\Z}_J})\big) }
\nonumber \\
&
\ee^{\ii 2 \pi  \int_{\cB^3}  \sum_{I\leq J} k_{IJ}
a^\RZ_{I} (\dd a^\RZ_{J} -\toZ{\dd a^\RZ_J})-\toZ{\dd a^\RZ_I}a^\RZ_J
}
\nonumber\\
&
\ee^{- \int_{\cB^3} \sum_I \frac{|\dd a^{\R/\Z}_I - \toZ{\dd a^{\R/\Z}_I}|^2}{g_3}}
\end{align}
where $a^\RZ_I$ is a $\RZ$-valued 1-cochain, $\int D[a^\RZ_I]= \prod_{ij,I}
\int_{\-\frac12}^{\frac 12}\dd (a^\RZ_I)_{ij}$, and $k_{IJ}$ integers integers.
Since $a^\RZ_I$ is $\RZ$-valued, we require \eqn{CSlatt1} to have the following
gauge symmetry 
\begin{align} 
\label{aZgauge} 
a^\RZ_I \to a^\RZ_I + u_I^\Z 
\end{align}
for any $\Z$-valued 1-cochain $u_I^\Z$.  Eq. \eq{CSlatt1} satisfies this
condition even for $\cB^3$ with boundary, as shown in \Ref{DW}.

The path integral of the 3+1D bosonic model (for spacetime $\cM^4$ with or
without boundary) is obtained from \eqn{CSlatt1} by taking a derivative and
setting $g_3=\infty$:
\begin{align}
&\ \ \ \
 \ee^{\ii 2\pi  \int_{\cM^4} k_{IJ}\dd \big[ a^\RZ_{I} (\dd a^\RZ_{J} -\toZ{\dd a^\RZ_J})-\toZ{\dd a^\RZ_I}a^\RZ_J \big]}
\nonumber\\
&=
\ee^{\ii 2\pi  \int_{\cM^4} k_{IJ}(\dd a^\RZ_{I} -\toZ{\dd a^\RZ_I}) (\dd a^\RZ_{J} -\toZ{\dd a^\RZ_J}) }
\end{align}
We obtain a 3+1D bosonic model
on spacetime lattice
\begin{align}
\label{K4Ddada}
Z = &\int D[a^\RZ_I] 
\ee^{-  \int_{\cM^4} \sum_I\frac{|\dd a^\RZ_I-\toZ{\dd a^\RZ_I}|^2}{g} }
\\
&
 \ee^{\ii 2\pi \int_{\cM^4} \sum_{I\leq J}k_{IJ} (\dd a^\RZ_I-\toZ{\dd a^\RZ_I})(\dd a^\RZ_J-\toZ{\dd a^\RZ_J})}
.
\nonumber 
\end{align}
In the above, we have included an extra term $ \ee^{- \sum_I \int_{\cM^4}
\frac{|\dd a^\RZ_I-\toZ{\dd a^\RZ_I}|^2}{g} } $.  Without such a term,
\eqn{K4Ddada} reduces to \eqn{CSlatt1} when $\cM^4$ has a boundary $\cB^3=\prt
\cM^4$.

When $\cM^4$ has no boundary, by its construction from \eqn{CSlatt1},
\eqn{K4Ddada} can be simplified to
\begin{align}
\label{K4DdadaC}
Z(\cM^4) &= 
\int D[a^\RZ_I] 
\ee^{ -\int_{\cM^4}\frac{|\dd a^\RZ_I-\toZ{\dd a^\RZ_I}|^2}{g} },
\end{align}
We find that when $g\sim 0$, $\dd a^\RZ_I$ fluctuate weakly and the above model
describes the deconfined phase of the $U^\ka(1)$ gauge theory.  In this case,
the model is gapless.  In this limit, $\dd a^\RZ_I\sim 0$ or $\toZ{\dd
a^\RZ_I}=0$, and we can reduces \eqn{K4Ddada} to a familiar $U^\ka(1)$ gauge
theory with $2\pi$ quantized topological terms $ 2\pi \int_{\cM^4} \sum_{I\leq
J} k_{IJ} \dd a^\RZ_I\dd a^\RZ_J $ and the Maxwell terms
$\int_{\cM^4}\frac{|\dd a^\RZ_I|^2}{g}$:
\begin{align}
Z = \int D[a^\RZ_I] & \ee^{\ii 2 \pi \int_{\cM^4} \sum_{I\leq J} k_{IJ} 
\dd a^\RZ_I\dd a^\RZ_J
-\int_{\cM^4}\frac{|\dd a^\RZ_I|^2}{g} } 
.
\end{align}
In particular, when $\ka=1$, the above becomes
\begin{align}
Z = \int D[a] & \ee^{\ii 2\pi k \int_{\cM^4}  
\dd a\dd a
-\int_{\cM^4}\frac{|\dd a|^2}{g} } 
,
\end{align}
where $k=k_{11}$ is an integer.

When $g\sim \infty$, $\dd a^\RZ_I$ fluctuate strongly and the above model
describes the confined phase of the $U^\ka(1)$ gauge theory.  The model is
fully gapped.  For any closed $\cM^4$ and when $g=\infty$, the partition
function $Z(\cM^4) = \int D[a^\RZ_I]  =1$ since $\int_{-\frac12}^{\frac12} \dd (a^\RZ_I)_{ij}
=1$. Thus the topological partition function $Z^\text{top}(\cM^4)=1$ is trivial
for any closed $\cM^4$.  This implies that the $g=\infty$ confined phase is a
gapped phase with trivial topological order.

Regardless the value of $g$, let us include low energy allowed excitations
described by charges of the $U^\ka(1)$ gauge field.  The values of the charges
are encoded in integer vectors $\v m_\text{a}$.  In the $U(1)$ confined phase,
the so-called low energy allowed excitations becomes the particle-hole
fluctuations for the charges in  $\cC_\text{a}$.  Since the set of allowed
excitations is closed under the fusion, the allowed integer vectors $\v
m_\text{a}$ form a lattice $\cC_\text{a}$.  We like to point out that
$\cC_\text{a}$ includes the column vector of the $K$ matrix, which is given by
\begin{align}
 K_{II}=2k_{II}, \ \ \ \ \ \
K_{IJ}=K_{JI}=k_{IJ}, \ \  I < J.
\end{align}

To see this point, we note that for closed
$\cM^4$
\begin{align}
& \ \ \ \
\ee^{\ii 2\pi \int_{\cM^4} \sum_{I\leq J}k_{IJ} (\dd a^\RZ_I-\toZ{\dd a^\RZ_I})(\dd a^\RZ_J-\toZ{\dd a^\RZ_J})}
\nonumber\\
&=
\ee^{-\ii 2\pi \int_{\cM^4} \sum_{I,J}K_{IJ}  a^\RZ_I\dd \toZ{\dd a^\RZ_J}},
\end{align}
where $\dd \toZ{\dd a^\RZ_J}$ can be viewed as the Poincar\'e dual of the
worldline of the $U(1)$ monoples.  This implies that the effect of the
topological term $\ee^{\ii 2\pi \int_{\cM^4} \sum_{I\leq J}k_{IJ} (\dd
a^\RZ_I-\toZ{\dd a^\RZ_I})(\dd a^\RZ_J-\toZ{\dd a^\RZ_J})}$ is bind $U(1)$
monopoles with the $U(1)$ charges.  In particular, the monopole of the
$I^{\th}$ $U(1)$ field carries the $J^{\th}$ $U(1)$ charge $K_{IJ}$.  
For large $g$, the monopoles described by $\dd\toZ{\dd a^\RZ_I}$ are low energy
allowed excitations.  Those monopole excitations carry charges given by the
column vector of the $K$ matrix. So the column vectors of the $K$ matrix are
the charges for the allowed low energy excitations.

Let $\v m_\text{a}^\mu$, $\mu=1,2,\cdots,\ka$ be a basis of the $\cC_\text{a}$
lattice, and let $M_a$ is a square matrix whose columns are $\v m_\text{a}^\mu$
vectors.  If we do not have any extra low energy allowed charge excitation,
$M_a$ will be given by $K$.  In this case, our $U^\ka(1)$ model has maximal
1-symmetry.  Using Smith normal form, we can always choose a basis such are
that the square matrix $M_a$ is diagonal, \ie
\begin{align}
\label{madel}
 (\v m_\text{a}^\mu)_I = k_I \del_{\mu I}
.
\end{align}

The allowed charge excitations can be included in the path
integral via the Wilson-loop $C^1$
\begin{align}
\label{K4DdadaC1}
 Z &  = \int D[a^\RZ_I] 
\ee^{-  \int_{\cM^4} \sum_I\frac{|\dd a^\RZ_I-\toZ{\dd a^\RZ_I}|^2}{g} }
\ee^{\ii 2\pi \int_{C^1} \v m_\text{a}^\top \v a}
\nonumber \\
&
 \ee^{\ii 2\pi \int_{\cM^4} \sum_{I\leq J}k_{IJ} (\dd a^\RZ_I-\toZ{\dd a^\RZ_I})(\dd a^\RZ_J-\toZ{\dd a^\RZ_J})}
\end{align}
Such a model has 1-symmetries generated by
\begin{align}
\label{1symmZ}
 a^\RZ_I \to a^\RZ_I+ s_I \al^\Z
\end{align}
for $\cM^4$ with or without boundary. Here, $\al^\Z$ are arbitary $\Z$-valued
1-cocycles, and $\v s=(s_1,s_2,\cdots)^\top$ is an arbitrary rational vector
that satisfies
\begin{align}
 \v s^\top \v m_\text{a}\se{1} 0,\ \ \ \forall \v m_\text{a} \in \cC_\text{a}.
\end{align}
or
\begin{align}
 M_a^\top \v s\se{1} 0.
\end{align}
The above choices of $\v s$ ensure the invariance of $\ee^{\ii 2\pi \int_{C^1}
\v m_\text{a}^\top \v a}$.


The above implies that $K \v s \se{1} 0$ since $\cC_\text{a}$ contains the
columns of $K$.  
It is more convenient to introduce integer vectors
\begin{align}
 \v m_\text{t} = K\v s
\end{align}
to describe the 1-symmetries. $\v m_\text{t}$'s satisfy
\begin{align}
M_a^\top K^{-1} \v m_\text{t} \se{1} 0.
\end{align}
In fact, the integer vectors $\v m_\text{t}$ satisfying the above conditions
form a lattice $\cC_\text{t}$.  Let $\v m_\text{t}^\mu$ be a basis of the
lattice $\cC_\text{t}$.  The 1-symmetry is characterized by this
$\cC_\text{t}$ lattice.  For the special basis \eqn{madel}, $\v m_\text{t}^\mu$
and $\v s^\mu =K^{-1}\v m_\text{t}^\mu$ are given by
\begin{align}
 (\v m_\text{t}^\mu)_I = \frac{K_{I\mu}}{ k_\mu} \in \Z, \ \ \   
 (\v s^\mu)_I = \frac{\del_{I\mu}}{ k_\mu} ,  
\end{align}
where $\mu$ is not summed.

The above 1-symmetry is a $ Z_{k_1}\times Z_{k_2}\times \cdots$-1-symmetry,
with $k_I$ given by \eqn{madel}.  Such a 1-symmetry is defined on the spacetime
lattice with or without boundary, and are expected to be anomaly-free.  Thus
for large $g$, the gapped state described by \eqn{K4DdadaC1} is a state with
1-symmetry but no topological order.  In the following, we will try to
determine the hSPT order in such a gapped state.

To do so, let us gauge the 1-symmetries by
replacing $\dd a^\RZ_I$ with 
\begin{align}
b_I \equiv \dd a^\RZ_I -s_I^\mu \hat B^\Z_\mu, 
\end{align}
where $\v s^\mu = K^{-1} \v m_\text{t}^\mu$ and $\hat B^\Z_\mu$ are $\Z$-valued
2-cocycles:
\begin{align}
\label{K4DaBC1}
Z =  \int  D[a^\RZ_I] &
 \ee^{\ii 2\pi \int_{\cM^4} \sum_{I\leq J} k_{IJ} (b_I-\toZ{b_I})(b_J-\toZ{b_J}) }
\nonumber\\
& \ee^{\ii 2\pi \int_{C^1} \v m_\text{a}^\top \v a }
 \ee^{ -\int_{\cM^4}\frac{|b_I-\toZ{b_I}|^2}{g} }
.
\end{align}
In the above, we have replaced $\ee^{\ii 2\pi \int_{C^1} \v m_\text{a}^\top \v
a}$ by $ \ee^{\ii 2\pi \int_{D^2} (\v m_\text{a}^\top \dd \v a -\v
m_\text{a}^\top \v s^\mu \hat B^\Z_\mu )}= \ee^{\ii 2\pi (\int_{C^1} \v
m_\text{a}^\top \v a - \int_{D^2} \v m_\text{a}^\top \v s^\mu \hat B^\Z_\mu )}$
where $\prt D^2=C^1$.  Since $\v m_\text{a}^\top \v s^\mu$ are integers, $
\ee^{\ii 2\pi (\int_{C^1} \v m_\text{a}^\top \v a - \int_{D^2} \v
m_\text{a}^\top \v s^\mu \hat B^\Z_\mu )} =\ee^{\ii 2\pi \int_{C^1} \v
m_\text{a}^\top \v a } $, and $\ee^{\ii 2\pi \int_{C^1} \v m_\text{a}^\top \v
a}$ is unchanged under the gauging of the 1-symmetry.  
Note that the 1-symmetries are discrete
symmetries, and can be probed by flat 2-gauge connections.

The 2-gauged $U^\ka(1)$ theory \eq{K4DaBC1} still have the 1-symmetries
\eq{1symmZ} for $\cM^4$ with or without boundary.  In fact, it has the
following 2-gauge symmetries that include the 1-symmetries:
\begin{align}
\label{2gaugeZ}
 a^\RZ_I \to a^\RZ_I + s_I^\mu u^\Z_\mu,\ \ \ \
 \hat B^\Z_\mu \to \hat B^\Z_\mu +  \dd u^\Z_\mu 
\end{align}
for $\cM^4$ with or without boundary. Here $u^\Z_\mu$ are arbitrary $\Z$-valued
1-cochains.  This is because $b_I=\dd a^\RZ_I -  s_I^\mu \hat B^\Z_\mu$ is
invariant under the 2-gauge transformation \eq{2gaugeZ}.  
We also note that the 2-gauged theory \eq{K4DaBC1} has the
gauge symmetry \eqn{aZgauge} for $\cM^4$ with or without boundary.

When $\cM^4$ has no boundary and $g=\infty$, \eqn{K4DaBC1} can be 
rewritten as
\begin{align}
Z = \int D[a^\RZ_I] 
& \ee^{\ii \pi \int_{\cM^4} K_{IJ} 
(\dd a^\RZ_I-s_I^\mu \hat B^\Z_\mu)(\dd a^\RZ_J-s_J^\mu \hat B^\Z_\mu)
}
\nonumber\\
&
\ee^{\ii 2\pi \int_{C^1} \v m_\text{a}^\top \v a }
.
\end{align}
Without the charged excitations described by $\v m_\text{a}$, the partition function
becomes
\begin{align}
 Z(\cM^4, \hat B^\Z_\mu) &=
 Z^\text{top}(\cM^4, \hat B^\Z_\mu)
=\ee^{\ii \pi \int_{\cM^4} (\v s^\nu)^\top K \v s^\mu \hat B^\Z_\mu  \hat B^\Z_\nu }
\nonumber\\
&=\ee^{\ii \pi \int_{\cM^4} (\v m_\text{t}^\nu)^\top K^{-1} \v m_\text{t}^\mu \hat B^\Z_\mu  \hat B^\Z_\nu }
\end{align}
where $\v m_\text{t}^\mu \in \cC_\text{t}$.  We see that if $\cC_\text{t}$ satisfies
\begin{align}
\label{msKms}
 \v m_\text{t}^\top K^{-1} \v m_\text{t}' \se{2} 0,\ \ \ 
\forall \v m_\text{t}, \v m_\text{t}' \in \cC_\text{t}
\end{align}
then the hSPT invariant $\ee^{\ii \pi \int_{\cM^4} (\v m_\text{t}^\nu)^\top
K^{-1} \v m_\text{t}^\mu \hat B^\Z_\mu  \hat B^\Z_\nu }=1$ and is trivial.  The
confined phase  of our $U^\ka(1)$ model is a trivial hSPT phase protected by
the 1-symmetry characterized by $\cC_\text{t}$.  Otherwise, the confined phase
of our $U^\ka(1)$ model is a non-trivial hSPT phase.

To summarize, for our $U^\ka(1)$ model with low energy allowed charges in
$\cC_\text{a}$ \eq{K4DaBC1}, the 1-symmetry is characterized by a lattice
\begin{align}
 \cC_\text{t} = \{ \v m_\text{t}| \v m_\text{t}^\top K^{-1} \v m_\text{a} \se{1}0\ \forall \v m_\text{a} \in \cC_\text{a}\},
\end{align}
which is a $ Z_{k_1}\times Z_{k_2}\times \cdots$-1-symmetry.
The confined phase of our $U^\ka(1)$ model can be a non-trivial hSPT phase
protected by the 1-symmetry $\cC_\text{t}$. The confined phase is a trivial
hSPT phase iff $\cC_\text{t}$ satisfies \eqn{msKms}.
This supports our conjecture in Section \ref{testanom}.

We like to remark that for $\cM^4$ without boundary, our model reduces to
\eqn{K4DdadaC}.  Such a model have a $U^\ka(1)$-1-symmetry generated by
shifting $a^\RZ_I$ by $\R$-valued cocycles.  However, the  $U^\ka(1)$
1-symmetry is broken for the model with boundary and with the charge
excitations, \ie \eqn{K4DdadaC1} does not have the $U^\ka(1)$-1-symmetry.
However, the model \eq{K4DdadaC1} has an anomaly-free discrete 1-symmetry
generated by a subset of the  $U^\ka(1)$ 1-transformations, \ie \eqn{1symmZ}.
The model \eq{K4DdadaC1} realizes a hSPT phase for such an anomaly-free
discrete and finite 1-symmetry.  The finite 1-symmetry is a $ Z_{k_1}\times
Z_{k_2}\times \cdots$-1-symmetry where $k_I$ is given in \eqn{madel}.

\subsection{A model to realize a hSPT phase with a $U(1)$ $k$-symmetry}

In this section, we consider a model to realize a hSPT phase
with a continuous $U(1)$ $k$-symmetry:
\begin{align}
\label{ZU1hSPT}
& Z(\cM^D) = \sum_{\{ a_k^{\R/\Z}\}} 
\ee^{\pi \ii \int_{\cM^D}  \gSq^{D-k-2} \dd\toZ{\dd a_k^{\R/\Z}} 
}
,
\end{align}
where $ a^{\R/\Z}$ is a $\R/\Z$-valued $k$-cochain.  Since
$a_k^{\R/\Z}$ is $\R/\Z$-valued, the theory must also have the following gauge
symmetry, even for $\cM^D$ that has a boundary
\begin{align}
\label{au1Z}
 a_k^{\R/\Z} &\to a_k^{\R/\Z} + u_k^{\Z},
\end{align}
where $u_k^{\Z}$ is an arbitrary $\Z$-valued $k$-cochain.  We find that
\eqn{ZU1hSPT} indeed has such a gauge symmetry.  

The above theory has the following $U(1)$ $k$-symmetry,
even when $\cM^D$ has a boundary
\begin{align}
 a_k^{\R/\Z} &\to a_k^{\R/\Z} + \al^{\R/\Z}_k, \ \ \ \dd \al_k^{\R/\Z}\se{1}0
\end{align}
where $\al_k^{\R/\Z}$ is an arbitrary $\R/\Z$-valued $k$-cocycle.  
This implies that the model
\eq{ZU1hSPT} has an anomaly-free $U(1)$ $k$-symmetry.

Using \eqn{Sqd}, we can show that when $\cM^D$ is closed
\begin{align}
 \ee^{\pi \ii \int_{\cM^D}  \gSq^{D-k-2} \dd\toZ{\dd a_k^{\R/\Z}}} =1,\ \ \ 
\prt \cM^D=0.
\end{align}
Therefore, the corresponding topological partition function
$Z^\text{top}(\cM^D)=1$ for any closed $\cM^D$.  The model \eq{ZU1hSPT}
describes a phase with trivial topological order.

Here we would like to mention that when $D-k-2 = \text{odd}$ or when
$D-k-2\geq k+2$, we have (see \eq{Sqd1})
\begin{align}
 \gSq^{D-k-2} \dd\toZ{\dd a_k^{\R/\Z}}=\dd\gSq^{D-k-2} \toZ{\dd a_k^{\R/\Z}},
\end{align}
and
\begin{align}
 \ee^{\th \ii \int_{\cM^D}  \gSq^{D-k-2} \dd\toZ{\dd a_k^{\R/\Z}}} =1,\ \ \ 
\prt \cM^D=0,
\end{align}
for any $\th$. Thus in this case, we can tune $\pi$ in \eqn{ZU1hSPT}
continuously to $0$ without encounter phase transitions.  We see that when
$D-k-2 = \text{odd}$ or $D-k-2\geq k+2$, \eqn{ZU1hSPT} describes a trivial hSPT
phases.

When $D-k-2=0$
\begin{align}
 \gSq^{0} \dd\toZ{\dd a_k^{\R/\Z}}
\se{2} \dd\toZ{\dd a_k^{\R/\Z}}.
\end{align}
In this case
\begin{align}
 \ee^{\pi \ii \int_{\cM^D}  \gSq^{0} \dd\toZ{\dd a_k^{\R/\Z}}}
=\ee^{\pi \ii \int_{\cM^D}   \dd\toZ{\dd a_k^{\R/\Z}}} 
\end{align}
even when $\prt \cM^D \neq 0$.  When  $\prt \cM^D = 0$, $\ee^{\th \ii
\int_{\cM^D}   \dd\toZ{\dd a_k^{\R/\Z}}} =1$ for any $\th$. So we can tune
$\pi$ to $0$ without phase transitions.  We see that when $D-k-2 =0 $,
\eqn{ZU1hSPT} also describes a trivial hSPT phase.

When $D-k-2 = \text{even}$ and $0<D-k-2< k+2$,
\begin{align}
&\ \ \ \
 \gSq^{D-k-2} \dd\toZ{\dd a_k^{\R/\Z}}
\\
&=\dd\gSq^{D-k-2} \toZ{\dd a_k^{\R/\Z}} 
-2(-)^{k+1}\gSq^{D-k-1} \toZ{\dd a_k^{\R/\Z}} .
\nonumber 
\end{align}
Therefore
\begin{align}
 \ee^{\th \ii \int_{\cM^D}  \gSq^{D-k-2} \dd\toZ{\dd a_k^{\R/\Z}}} 
&=\ee^{-2\th \ii \int_{\cM^D}  (-)^{k+1}\gSq^{D-k-2} \toZ{\dd a_k^{\R/\Z}}} 
,
\nonumber\\
&\prt \cM^D=0.
\end{align}
Since $\gSq^{D-k-2} \toZ{\dd a_k^{\R/\Z}}$ is not a coboundary in general, the
action amplitude is $\ee^{\th \ii \int_{\cM^D}  \gSq^{D-k-2} \dd\toZ{\dd
a_k^{\R/\Z}}} =1$ only when $\th=0,\pi$. For other $\th$ the action amplitude
has a non-trivial phase, and the model may be gapless. In this case, $\th=0$
and $\th=\pi$ may correspond to two different hSPT phases.

To see if the  model \eq{ZU1hSPT} for $D-k-2 = \text{even}$ and
$0<D-k-2 < k+2$ describes a phase with a non-trivial hSPT
order or not, we gauge the $U(1)$ $k$-symmetry to obtain
\begin{align}
 Z = 
\hskip -0.5em
\sum_{\{ a_k^{\R/\Z}\}} 
\hskip -0.5em
\ee^{\pi \ii \int_{\cM^D}  \gSq^{D-k-2} \big(
\dd\toZ{\dd a_k^{\R/\Z}+\hat B_{k+1}^{\R/\Z}} 
\big)
}.
\end{align}
where the $\R/\Z$ valued 2-cochain $\hat B_{k+1}^{\R/\Z}$ is the background
2-connection for the twisted $U(1)$-1-symmetry.  Since $\hat B_{k+1}^{\R/\Z}$
is $\R/\Z$-valued, the action amplitude should have the following gauge
symmetry, even for $\cM^D$ that has a boundary,
\begin{align}
\label{Bu2Z}
  \hat B_{k+1}^{\R/\Z} \to  \hat B_{k+1}^{\R/\Z} + u_{k+1}^{\Z}
\end{align}
where $u_{k+1}^{\Z}$ is an arbitrary $\Z$-valued $(k+1)$-cochain.  But the
above the action amplitude does not have this gauge symmetry.
This problem can be fixed by including an additional term which
vanishes when $ \hat B_{k+1}^{\R/\Z}=0$:
\begin{align}
\label{ZU1hSPTG}
 Z = 
\hskip -0.5em
\sum_{\{ a_k^{\R/\Z}\}} 
\hskip -0.5em
\ee^{\pi \ii \int_{\cM^D}  \gSq^{D-k-2} \big(
\dd\toZ{\dd a_k^{\R/\Z}+\hat B_{k+1}^{\R/\Z}} 
-\toZ{\dd (\dd a_k^{\R/\Z}+\hat B_{k+1}^{\R/\Z})} \big)
}.
\end{align}
Such a theory has the
following 2-gauge symmetry, even when $\cM^D$ has a boundary
\begin{align}
 a_k^{\R/\Z} &\to a_k^{\R/\Z} + u_k^{\R/\Z}
\nonumber\\
 \hat B_{k+1}^{\R/\Z} &\to \hat B_{k+1}^{\R/\Z} - \dd u_k^{\R/\Z}
\end{align}
where $u_k^{\R/\Z}$ is an arbitrary $\R/\Z$-valued $k$-cochain.  

Using \eqn{Sqgauge}, we can show that, for closed $\cM^D$,
\begin{align}
&\ \ \ \ \ee^{\pi \ii \int_{\cM^D}  \gSq^{D-k-2} \big(
\dd\toZ{\dd a_k^{\R/\Z}+\hat B_{k+1}^{\R/\Z}} 
-\toZ{\dd (\dd a_k^{\R/\Z}+\hat B_{k+1}^{\R/\Z})} \big)
}
\nonumber \\
&=
\ee^{\pi \ii \int_{\cM^D}  \gSq^{D-k-2} 
\toZ{\dd \hat B_{k+1}^{\R/\Z}}
}.
\end{align}
Therefore, the corresponding topological partition function of the gauged model is given by
\begin{align}
Z^\text{top}(\cM^D,\hat B_{k+1}^{\R/\Z})= 
\ee^{\pi \ii \int_{\cM^D}  \gSq^{D-k-2} 
\toZ{\dd \hat B_{k+1}^{\R/\Z}} 
}
\end{align}
for any closed $\cM^D$.  This non-trivial hSPT invariant implies that the model
\eq{ZU1hSPT} or \eq{ZU1hSPTG} describes a phase with a non-trivial hSPT order,
when $D-k-2=\text{even}$ and $0< D-k-2 < k+2$ or when $D-k=\text{even}$ and
$k+3 \leq D \leq 2k+3$.  

When $k=1$, we have a
model to realize a non-trivial 4+1D $U(1)$-1-SPT phase
\begin{align}
\label{Z5D1SPTU1}
& Z(\cM^5) = \sum_{\{ a^{\R/\Z}\}} \ee^{\pi \ii \int_{\cM^5}  \gSq^{2} \dd\toZ{\dd a^{\R/\Z}} }.
\end{align}
When $k=2$, we have a  model to realize
a non-trivial 5+1D $U(1)$-2-SPT phase 
\begin{align}
\label{Z6D2SPTU1}
& Z(\cM^6) = \sum_{\{ b^{\R/\Z}\}} \ee^{\pi \ii \int_{\cM^6}  \gSq^{2} \dd\toZ{\dd b^{\R/\Z}} }.
\end{align}

It turns out that the 4+1D hSPT phase described by \eqn{Z5D1SPTU1} is very
important for condensed matter. This is because all the EM condensed matter
systems with dynamical EM fields must be a boundary of such a 4+1D hSPT phase
(see Section \ref{EMsys}).

\section{The topological robustness of emergent higher symmetry}
\label{robust}

\subsection{Translation invariant systems}

The lattice model \eq{H3dZ2} has an exact $Z_2$-1-symmetry
generate by the membrane operator \eq{memb},
since \emph{the $Z_2$ charges are not mobile}.
We can make the $Z_2$ charges mobile and break the
$Z_2$-1-symmetry by adding the term
\begin{align}
 \del H = -J \sum_{\<\v i\>} \si^x_{\v i} .
\end{align}
However when $U_2$ is very large, the $Z_2$ charges have a large energy gap of
order $|U_2|$.  The $Z_2$ charges do not even appear at low energies.  In this
case, we expect an emergence of $Z_2$-1-symmetry at low energies even
when $\del H\neq 0$.

Indeed, it was shown in \Ref{HW0541} that even though membrane operator
\eq{memb} does not commute with the perturbed Hamiltonian $H+\del H$, we can
define fattened  membrane operators
\begin{align} 
\label{fatmemb}
M_\text{fat-memb} = 
U_\text{LU}
\Big( \prod_{\v i \in \text{closed membrane}} \si^z_{\v i} \Big) 
U_\text{LU}^\dag ,
\end{align} 
where $U_\text{LU}$ is the local unitary operator defined in \Ref{CGW1038}.  We
can choose $U_\text{LU}$ such that the low energy eigenstates are also the
eigenstates of the fattened membrane operators.  This indicates an emergence of
$Z_2$-1-symmetry at low energies.

\Ref{HW0541} shows that such fattened  membrane operators can be found for any
local perturbation $\del H$ that can break any symmetries and higher
symmetries. Thus the  emergence of $Z_2$-1-symmetry at low energies is robust
against any local perturbation.  This represents a \emph{topological
robustness} of emergent of higher symmetry.  In general, we believe the
emergence of higher symmetry to be always topological, reflecting the
topological robustness of topological orders.

In fact $U_\text{LU}$ can be constructed using adiabatic
evolution:\cite{HW0541}
\begin{align}
 U_\text{LU}=T[\ee^{-\ii \int_0^1 \dd t H(t) }],\ \ \
H(t)\equiv H+t\del H.
\end{align}
The degenerate ground states $|\psi'_\al\>$ of $H+\del H$ can be obtained from
the degenerate ground states $|\psi_\al\>$ of $H$:
\begin{align}
 |\psi'_\al\> = U_\text{LU} |\psi_\al\>.
\end{align}
We see that fattened  membrane operators $M_\text{fat-memb}$ acts within the
ground state subspace of $H+\del H$, and generates the low energy emergent
$Z_2$-1-symmetry.

We like to remark that the generators of higher symmetry discussed in this
paper (regardless on-site or not) are always finite-depth local quantum
circuits.  The fattened  generators of higher symmetry are also finite-depth
local quantum circuits.  It is known that string operator that create a pair of
non-Abelian anyons are not finite-depth local quantum
circuits.\cite{BPh0110205,S181001986} The topological excitations associated
with the string operators that generate higher symmetry are always Abelian
anyons.  However, it is not proven that string operators that generate Abelian
anyons are always  finite-depth local quantum circuits.  We like to remark that
string operators (linear-depth local quantum circuits) that generate
non-Abelian anyons correspond to generalized higher symmetry, which is always
anomalous.\cite{KLTW}

\subsection{Emergent higher symmetry and many-body localization}

The lattice model \eq{H3dZ2} has an exact $Z_2$-1-symmetry for systems of
\emph{any size and at any energy}.  In the presence of a small perturbation
$\del H$, the model has an emergent  $Z_2$-1-symmetry for \emph{large systems
at low energies}.  Since the essence of $Z_2$-1-symmetry is that the pointlike
topological excitations are not mobile, we can use many-body localization to
realize a stronger  emergent $Z_2$-1-symmetry for \emph{large systems at any
energy}.\cite{HS13041158,BN13065753,CS13101096}  

We first consider the model
\begin{align}
\label{H3dZ2R}
H &=
-\sum_{\<\v i\v j\v k\v l\>}U_1(\<\v i\v j\v k\v l\>) \si^x_{\v i}\si^x_{\v j}\si^x_{\v k}\si^x_{\v l}
\\
&\ \ \ \
-\sum_{\<\v i\v j\v k\v l\v m\v n\>} U_2(\<\v i\v j\v k\v l\v m\v n\>)
\si^z_{\v i}\si^z_{\v j}\si^z_{\v k}\si^z_{\v l} \si^z_{\v m}\si^z_{\v n}
\nonumber 
\end{align}
where $U_1(\<\v i\v j\v k\v l\>)$ and $U_2(\<\v i\v j\v k\v l\v m\v n\>)$
strongly random positive numbers.  The random $U_1(\<\v i\v j\v k\v l\>)$ make
the $Z_2$-flux-loop $s$ to have a random tension.  The random $U_2(\<\v i\v j\v
K\v l\v m\v n\>)$ make the $Z_2$-charge $e$ to have a random energy.  In such a
model, there is no $Z_2$-flux-loop hopping term nor $Z_2$-charge hopping term.
The $Z_2$-flux-loop $s$ cannot change its shape and the $Z_2$-charge $e$ cannot
move around.  As a result, \eqn{H3dZ2R} has a $Z_2$-1-symmetry generated by
(see \eqn{WC2})
\begin{align}
  W(C^2)=\prod_{\v i \in C^{2}} \si^z_{\v i}
\end{align}
and a $Z_2$-2-symmetry generated by
\begin{align}
  W(\t C^1)=\prod_{\v i \in\t C^1} \si^x_{\v i}
\end{align}
where $\t C^1$ is a closed string formed by the links of the dual cubic lattice.

After we add a small perturbation $\del H$, due to the strong randomness of the
energies of the  $Z_2$-charge and the  $Z_2$-flux, many-body localization may
happen, and the $Z_2$-charge and the  $Z_2$-flux are still not mobile.  In this
case, there are emergent $Z_2$-1-symmetry and $Z_2$-2-symmetry for \emph{large
systems at any energy}.

\subsection{Continuous higher symmetry and gapless cases}

Next, we briefly discuss continuous higher symmetry and gapless
cases.  The emergence of 3+1D gapless $U(1)$ gauge theory is also accompanied
with an emergence of $U(1)$-1-symmetries, if the $U(1)$-charges and the
$U(1)$-monopoles have a large energy gap.  It was shown that the emergence of
such higher symmetries to be topological.\cite{HW0541} The  topological
robustness of the emergent $U(1)$-1-symmetries (which was called the local
$U(1)$ gauge symmetries in \Ref{HW0541}) is used to show the topological
robustness of the gapless $U(1)$ gauge theory: \frmbox{There are no local
perturbations that can open an energy gap for the gapless $U(1)$ gauge
bosons.\cite{HW0541}}

\section{Generic higher symmetry in spacetime lattice models}
\label{ST}

In this section, we will construct lattice model with a combined 0-symmetry and
1-symmetry.  The mixture of the 0-symmetry and 1-symmetry can be quite
non-trivial.  We also like to include background gauge field and higher gauge
field that describe the spacetime twist of  the 0-symmetry and 1-symmetry.  But
before describing the mixture of the 0-symmetry and 1-symmetry, we will first
review a particular construction of spacetime lattice models with global
on-site symmetry $G$ (\ie 0-symmetry).  This particular construction can be
generalized to obtain lattice model with a combined 0-symmetry and 1-symmetry.

\subsection{Models with 0-symmetry and 0-symmetry twist}

To describe a 0-symmetry
described by a finite group $G$, we consider a spacetime lattice model with a
field $g_i$ living on vertices.  The 0-symmetry lives on the closed
$D$-subcomplex of the dual spacetime complex $\t\cM^D$ (\ie the dual of the
vertices of $\cM^D$), which generate the following transformation 
\begin{align} 
g_i \to g g_i,\ \ \ g\in G.
\end{align} 
The 0-symmetry invariant lattice model \begin{align} Z
=\sum_{\{g_i\} } \ee^{-\int_{\cM^D} L(g_i)} \end{align} satisfies
\begin{align} 
L(g_i) = L(h g_i),\ \ \ h \in G.  
\end{align} 
The Lagrangian $L(g_i)$ (a $D$-cochain) can be ``gauged'' to obtain
$L(g_i, \hat A_{ij})$ with a non-dynamical flat gauge connection $\hat A_{ij}\in G$:
\begin{align} 
\hat A_{ij}\hat A_{jk}=\hat A_{ik}.  
\end{align}  
$\hat A_{ij}$ is also called the symmetry twist.  The ``gauged'' Lagrangian has
a 1-gauge symmetry 
\begin{align} 
L(g_i, \hat A_{ij}) = L(h_i g_i, h_i \hat A_{ij} h_j^{-1}), \ \ \ h_i \in G.  
\end{align} 
In the following, we will choose the value of the $g_i$ field to be the
symmetry group $G$. Using the above symmetry, we can rewrite 
\begin{align} 
L(g_i, \hat A_{ij}) = L(1, g_i^{-1}
\hat A_{ij} g_j)= L( A_{ij} ), 
\end{align} 
where $A_{ij}$ is the effective field 
\begin{align} 
A_{ij} = g_i^{-1} \hat A_{ij} g_j.  
\end{align} 
The partition function now can be written as 
\begin{align} 
\label{Za} Z
=\sum_{\{g_i\} } \ee^{-\int_{\cM^D} L(A_{ij})}, \ \ A_{ij} = g_i^{-1}
\hat A_{ij} g_j.  
\end{align} 
We remark that \eqn{Za} describes a $G$ symmetric system in a background of
twisted 0-symmetry. The twisted 0-symmetry is described by a connection
$\hat A_{ij}$.  We may also view the connection $\hat A_{ij}$ as a probe of
the $G$ 0-symmetry.

We also like to remark that the effective field $A_{ij}$ in \eqn{Za} describes a
flat connection
\begin{align}
\label{aaa}
 A_{ij}A_{jk}=A_{ik}.
\end{align}  
The summation $\sum_{\{g_i\} }$ sums over all gauge equivalent configuration
that correspond to the same flat $G$-bundle.  In fact, we can view $g_i$ as
the gauge transformation, and thus  $\sum_{\{g_i\} }$ sums over all gauge
transformations.

Last, we note that $L(A_{ij})$ can be viewed as a Lagrangian of a lattice gauge
theory (\ie 1-gauge theory).  Here we construct a lattice theory with a
0-symmetry twist by starting with a Lagrangian for lattice 1-gauge theory, and
doing the path integral by only summing over the 1-gauge configurations within
one gauge equivalent class.  We will use the similar approach to construct
lattice model with higher symmetry, with a higher symmetry twist.

\subsection{Models with a combined 0-symmetry and 1-symmetry and their twist}

To construct a model with a combined 0-symmetry and 1-symmetry, we include an
extra bosonic field $a_{ij}$ living on the links $ij$.  The value of $a_{ij}$
is taken from an Abelian group $\Pi_2$.  We start with the Lagrangian in terms
of the effective fields $A_{ij}$ and $B_{ijk}$.  Here $B_{ijk}$ is a
$\Pi_2$-valued 2-cochain field living on the triangles $ijk$.  The 1-cochain
field $A_{ij}$ is flat as before (see \eqn{aaa}).  The 2-cochain field
$B_{ijk}$ may not be flat
\begin{align}
 \dd B = n_3(A) ,
\end{align}
To understand $n_3(A)$, we note that, as explained in \Ref{ZW180809394}, the
field $A$ on the links satisfying \eq{aaa} define a  map $\cM^D
\xrightarrow{\phi} \cB G$ (or more precisely a homomorphism of simplicial
complexes).  Then $n_3(A)$ is given by $n_3(A)= \phi^* \bar n_3$, where $\bar
n_3 \in H^3(\cB G, \Pi_2)$.  Note that $\bar n_3$ is a cocycle on the
classifying space $\cB G$, while $n_3(A)$ lives on $\cM^D$.  Thus $n_3(A)$ is
the pullback of $\bar n_3$ on $\cB G$  by the homomorphism $\phi$.  We see that
the map $\phi$ must satisfy a property that the pullback of $\bar n_3$ is a
coboundary on $\cM^D$. (For details, see \Ref{ZW180809394,LW180901112}.) 

The higher gauge transformations are generated by
$g_i,a_{ij}$:
\begin{align}
\label{2gauge}
 A_{ij} &\to g_i^{-1} A_{ij} g_j,
\nonumber\\
B_{ijk} &\to B_{ijk} +a_{ij}+a_{jk}-a_{ik} + \xi_{ijk}(A_{ij},g_i),
\end{align}
where $\xi_{ijk}(A_{ij},g_i)$ is given by
\begin{align}
 \dd \xi(A_{ij},g_i) = n_3(g_i^{-1}A_{ij} g_j)-n_3(A_{ij}) .
\end{align}
Here \eqn{2gauge} is called a 2-gauge transformation.

Let $L(A_{ij}, B_{ijk})$ be a $D$-cochain that depends on $A$ and $B$. 
Then summing over all the  2-gauge transformations \eq{2gauge}
\begin{align}
\label{Zab}
  Z =\sum_{\{g_i,a_{ij}\} } \ee^{- \int_{\cM^D} L(A_{ij},B_{ijk})}
\end{align}
will give us a bosonic model with a non-trivially 
combined 0-symmetry and 1-symmetry. Here
\begin{align}
\label{AAbBBb}
 A_{ij} &= g_i^{-1}\hat A_{ij} g_j,
\nonumber\\
B_{ijk} &= \hat B_{ijk} +a_{ij}+a_{jk}-a_{ik} + \xi_{ijk}(\hat A_{ij},g_i),
\end{align}
$g_i,a_{ij}$ are dynamical fields,
and $\hat A_{ij}, \hat B_{ijk}$ are non-dynamical background 2-gauge connections
satisfying
\begin{align}
 \hat A_{ij}\hat A_{jk}=\hat A_{ik},\ \ \ \dd \hat B = n_3(\hat A).
\end{align}

Note that here $L(A_{ij}, B_{ijk})$ can be any function of $A_{ij}, B_{ijk}$.
In particular, it does not has to be invariant under the higher gauge
transformation \eq{AAbBBb}.  The model \eqn{Zab} has a combined global
0-symmetry and 1-symmetry when $\hat A_{ij}=1$ and $\hat B_{ijk}=0$.  The
combined 0-symmetry and 1-symmetry is generated by 
\begin{align}
 g_i \to g g_i,\ a_{ij} \to a_{ij}+ \al_{ij};\ \  
g\in G,\  \al_{ij}\in \Pi_2, \ \dd  \al =0.
\end{align}
(Note that $\xi_{ijk}(\hat A_{ij}=1,g_i)=\xi_{ijk}(\hat A_{ij}=1,gg_i)$.) In particular, the global
1-symmetry transformation changes the 1-cochain field $a$ by a cocycle.  (We
can view the 1-cochain field $a$ as a field on $(D-1)$-cells of the dual complex
$\t\cM^D$.  The global 1-symmetry transformation changes the 1-cochain field
$a$ by a constant on the $(D-1)$-cells of a closed the $(D-1)$-dimensional (or
codimension-1) complex in the dual complex $\t\cM^D$.

We point out that \eqn{Zab} describes a system with 0-symmetry and 1-symmetry
on a background of twisted 0-symmetry and 1-symmetry. The twisted 0-symmetry is
described by the 1-connection $\hat A_{ij} \in G$.  The twisted 1-symmetry is
described by the 2-connection $\hat B_{ijk} \in \Pi_2$, which is a
$\Pi_2$-valued 2-cochain satisfying $\dd \hat B = n_3(\hat A)$.

We like to remark that in our above construction of lattice models, we started
with a lattice 2-gauge theory.  However, in our construction, the 2-gauge
invariant field strength is a non-dynamical background field.  The pure 2-gauge
transformations are our dynamical fields.  Such a lattice model has a combined
global 0-symmetry and 1-symmetry.  We point out that the above construction can
also be used to construct lattice models with  a combined global 0-symmetry,
1-symmetry, and 2-symmetry, by starting with 3-gauge theories.  In general,
lattice models with higher symmetry can be constructed by starting from lattice
higher gauge theories,\cite{ZW180809394} where the higher field strength
corresponds to fixed higher symmetry twist, and the dynamical fields come from
the higher gauge transformations.

\section{Lattice models that realize higher SPT phases -- systematic
constructions}
\label{hSPT}

\subsection{Models realizing bosonic SPT phases}

After constructing models with on-site 0-symmetry \eqn{Za}, we can choose
$L(A)$ to be a $2\pi \ii \R/\Z$-valued cocycle
\begin{align}
\label{ZaSPT}
  Z(\hat A) =\sum_{\{g_i\} } \ee^{2\pi \ii \int_{\cM^D} \om_D(A_{ij})},\ \
A_{ij} = g_i^{-1} \hat A_{ij} g_j,
\end{align}
where $\om_D(A_{ij}) = \phi^* \bar \om_D$, $\bar \om_D$ is a cocycle $\bar
\om_D \in H^D(\cB G, \R/\Z)$  and $\cB G$ is the classifying space of $G$.
Note that $\bar \om_D$ lives on $\cB G$, while $\om_D(A_{ij})$ lives on
$\cM^D$.  Thus $\om_D(A_{ij})$ is the pullback of $\bar \om_D$ by the map
$\cM^D \xrightarrow{\phi} \cB G$ determined by the 1-cochain field $A_{ij}$:
$\om_D(A_{ij}) = \phi^* \bar \om_D$.  The above exactly soluble model realizes
a bosonic $G$-SPT state characterized by cocycle $\bar \om_D \in  H^D(\cB G,
\R/\Z)$.  For more details and a more precise description of the above model
and the notations, see, for examples, \Ref{W161201418} and \Ref{ZW180809394}.  

\subsection{Models realizing bosonic higher SPT phases}
\label{hSPTsec}

We have seen that using $\bar \om_D \in  H^D(\cB G, \R/\Z)$, we construct
exactly soluble bosonic models that realize SPT phases protected by symmetry
$G$.  Similarly, using $\bar \om_D \in  H^D[\cB(G,\Pi_2) ; \R/\Z]$, we
construct exactly soluble bosonic models that realize hSPT phases protected by
a combined 0-symmetry and 1-symmetry described by $\cB(G,\Pi_2)$.

Starting with the model \eq{Zab} with a combined 0-symmetry and 1-symmetry, we
can choose $L(A,B)$ to obtain an exactly soluble model
\begin{align}
\label{ZabSPT}
  Z &= \sum_{\{g_i,a_{ij}\} } \ee^{2\pi \ii \int_{\cM^D} \om_D(A,B)},
\nonumber\\
A_{ij} &= g_i^{-1} \hat A_{ij} g_j,
\nonumber\\
B_{ijk} &= \hat B_{ijk} +a_{ij}+a_{jk}-a_{ik} + \xi_{ijk}(\hat A,g),
\end{align}
where the dynamical field on vertices is $g_i \in G$ and the dynamical field on
links is $a_{\ij} \in \Pi_2$. Here $\om_D(A,B) = \phi^* \bar \om_D$, $\bar \om_D
\in H^D(\cB(G,\Pi_2), \R/\Z)$ and $\cB (G,\Pi_2)$ is the classifying space of a
2-group.\cite{ZW180809394}  We call $\bar \om_D$ a 2-group cocycle. Also 
$\phi$ is the map $\cM^D \xrightarrow{\phi} \cB(G,\Pi_2)$ as determined by the
fields $A_{ij}, B_{ijk}$.  In fact, the homomorphism $\phi$ and the fields $A,
B$ on $\cM^D$ are directly related in the following way
\begin{align}
 A = \phi^* \bar A,\ \ \ \
 B = \phi^* \bar B
\end{align}
where $\bar A$ is the $G$-valued cannonical 1-cochain on  $\cB(G,\Pi_2)$ and
$\bar B$ is the $\Pi_2$-valued cannonical 2-cochain on  $\cB(G,\Pi_2)$.  For
more details, see, for example, \Ref{ZW180809394}. 

\frmbox{The model \eq{Zab} realizes a hSPT phase with a higher
symmetry described by 2-group $\cB(G,\Pi_2)$.\cite{TK151102929}  The hSPT
phases are systematically constructed via $\R/\Z$-valued $D$-cocycles $\om_D$
on the classifying space $\cB(G,\Pi_2)$.}

For more general higher group $\cB(G,\Pi_2,\Pi_3,\cdots)$, we note that higher
group admits a special one-vertex triangulation. The resulting complex is a
simplicial set (see for example \Ref{ZW180809394}).  We will use the same
symbol $\cB(G,\Pi_2,\Pi_3,\cdots)$ to denote such a simplicial set.  Using the
simplicial set, an exactly soluble local bosonic model that realize a higher
gauge theory with gauge group $\cB(G,\Pi_2,\Pi_3,\cdots)$ can be
constructed\cite{ZW180809394}
\begin{align}
\label{Zhgauge}
  Z(\cM^D) &= \sum_{\phi} \ee^{2\pi \ii \int_{\cM^D} \phi^* \bar \om_D},
\end{align}
where $\sum_{\phi}$ sums over all the simplicial-complex homomorphisms $\cM^D \xrightarrow{\phi}
\cB(G,\Pi_2,\Pi_3,\cdots)$.  Here $\bar \om_D$ is a $\R/\Z$ valued $D$-cocycle
on $\cB(G,\Pi_2,\Pi_3,\cdots)$:
\begin{align}
 \bar \om_D \in H^D[\cB(G,\Pi_2,\Pi_3,\cdots);\R/\Z],
\end{align}
and $\phi^* \bar \om_D$ is the pullback of the cocycle on
$\cB(G,\Pi_2,\Pi_3,\cdots)$ to $\cM^D$.  We note that the model \eq{Zhgauge}
realizes a topologically ordered phase described by a higher gauge theory
(which is not a hSPT phase).

To obtain a model that realizes a hSPT phase with trivial topological order, we
note that the  simplicial-complex homomorphisms $\cM^D \xrightarrow{\phi}
\cB(G,\Pi_2,\Pi_3,\cdots)$ can be divied into many different homopoty classes.
Each class correspond to gauge equivalent configurations.  So we can label the
homopoty classes as $[\phi]$, which are formed by all the configurations that
are homotopic to $\phi$.  We may also label the homopoty classes as $[A,
B,\cdots]$ where $A, B,\cdots$ are the higher gauge connections, and $[A,
B,\cdots]$ are formed by all the configurations that are gauge equivalent (\ie
homotopic) to $A, B,\cdots$.  $[\phi]$ and $[A, B,\cdots]$ are just two
notations for the same thing.  

Now we generate the gauge equivalent configurations in the class $[A,
B,\cdots]$ by gauge transformations $g_, a_{ij},\cdots$
\begin{align}
 A^g,B^{a,g},\cdots  \in [A, B,\cdots] .
\end{align}
We may also rewrite the above as
\begin{align}
 \phi^{g,a,\cdots} \in [\phi],
\end{align}
where $ \phi^{g,a}$ is the homomorphism obtained from $\phi$ by 2-gauge
transformation $g,a$ (see \eqn{2gauge}).  We note that the number of gauge
transformations $g,a,\cdots$ and the number of configurations in $[\phi]$ may
not be the same, since some different gauge transformations may give rise to
the same homomorphism $ \phi^{g,a,\cdots} = \phi^{g',a',\cdots}$.

With the above notation, we can write down the local bosonic model that
realize a hSPT phase
\begin{align}
\label{ZhSPT}
  Z(\cM^D,\hat \phi) &= \sum_{g,a,\cdots} 
\ee^{2\pi \ii \int_{\cM^D} \hat \phi^{*;g,a,\cdots} \bar \om_D}.
\end{align}
Comparing to the higher gauge theory \eqn{Zhgauge} here we just change the
dynamics of the field $\phi$ by restricting it to a homotopy class $[\hat
\phi]$.  We note that $\phi$ in each homotopy class $[\phi]$ give rise to the
same $\ee^{2\pi \ii \int_{\cM^D} \phi^* \bar \om_D}$ for closed $\cM^D$.  Thus
\begin{align}
\label{ZhSPT1}
  Z(\cM^D,\hat \phi) &= 
\big(|G|^{N_v} |\Pi_2|^{N_l} |\Pi_3|^{N_t} \cdots \big)
\ee^{2\pi \ii \int_{\cM^D} \hat \phi^* \bar \om_D},
\end{align}
where $N_v$, $N_l$, $N_t$, $\cdots$ are the numbers of
vertices, links, triangles, $\cdots$, in the spacetime complex $\cM^D$.
We see that the topological partition function is given by
\begin{align}
\label{ZhSPTtop}
  Z^\text{top}(\cM^D,\hat \phi) &= 
\ee^{2\pi \ii \int_{\cM^D} \hat \phi^* \bar \om_D},
\end{align}
which is the hSPT invariant charactering the hSPT phase.

\subsection{More general bosonic hSPT phases}
\label{SPTSO}

The bosonic model eqn. \eq{ZhSPT} does not realize all possible  bosonic hSPT
phases.  To obtain more general bosonic hSPT phases protected by higher
symmetry described by higher group $\cB(G,\Pi_2,\cdots)$, we can replace the
symmetry group $G$ by $G^{SO}$, as proposed in \Ref{W1477} and
\Ref{LW180901112}:
\begin{align}
 G_{SO} = G \gext SO_\infty .
\end{align}
We arrive at the following local bosonic model
\begin{align}
\label{ZhSPTSO}
  Z(\cM^D,\hat \phi_{GSO}) &= 
\sum_{g^{G_{SO}},a^{\Pi_2}} 
\ee^{2\pi \ii \int_{\cM^D} \hat \phi_{GSO}^{*;g^{G_{SO}},a^{\Pi_2}} \bar \om_D^{GSO}},
\end{align}
where $\bar \om_D^{GSO} \in H^D[\cB(G_{SO},\Pi_2,\cdots);\R/\Z)$, $\hat
\phi_{GSO}$ is a simplicial-complex homomorphism $\cM^D \xrightarrow{\hat
\phi_{GSO}}\cB(G_{SO},\Pi_2,\cdots)$, and $ \sum_{g^{G_{SO}},a^{\Pi_2}} $ sums
over all the gauge transformations described by $g^{G_{SO}},a^{\Pi_2},\cdots$.
Also, $\hat \phi_{GSO}^{g^{G_{SO}},a^{\Pi_2}}$ is the homomorphism obtained from
the $\hat \phi_{GSO}$ by gauge transformation $g^{G_{SO}},a^{\Pi_2},\cdots$.  We
stress that here $\hat \phi_{GSO}$ is not an arbitrary homomorpism from $\cM^D$
to $\cB(G_{SO},\Pi_2,\cdots)$.  We note that a  homomorpism $\hat \phi_{GSO}$
give rise to a $G_{SO}$ gauge configuration $\hat A^{GSO} = \hat \phi^*_{GSO}
\bar A^{GSO}$ on $\cM^D$, where $\bar A^{GSO}$ is the canonical 1-cochain on
$\cB(G_{SO},\Pi_2,\cdots)$.  Since $\hat A^{GSO}_{ij} \in G_{SO}$, we can use
the natural projection $G_{SO} \xrightarrow{\pi} SO_\infty$ to obtain $\hat
A^{SO}_{ij} =\pi(\hat A^{GSO}_{ij}) \in SO_\infty$.  We require $\hat
A^{SO}_{ij}$ to be the connection of the tangent bundle of $\cM^D$.  The
resulting model \eq{ZhSPTSO} realizes more general bosonic hSPT phases.

In the presence of time-reversal symmetry, the symmetry group is given by
$G=G_0\gext Z_2^T$, where $Z_2^T$ is the time-reversal symmetry group.  In this
case we replace $G$ by $G_O$, as proposed in \Ref{W1477} and \Ref{LW180901112}:
\begin{align}
 G_O = G\gext SO_\infty =G_0\gext Z_2^T \gext SO_\infty = G_0\gext O_\infty
\end{align}
since $O_\infty =Z_2^T\gext SO_\infty$.
We obtain the following local bosonic model
\begin{align}
\label{ZhSPTO}
  Z(\cM^D,\hat \phi_{GO}) &= 
\sum_{g^{G_O},a^{\Pi_2}} 
\ee^{2\pi \ii \int_{\cM^D} \hat \phi_{GO}^{*;g^{G_O},a^{\Pi_2}} \bar \om_D^{GO}},
\end{align}
where $\bar \om_D^{GO} \in H^D[\cB(G_{O},\Pi_2,\cdots);\R/\Z)$, $\hat \phi_{GO}$
is a simplicial-complex homomorphism $\cM^D \xrightarrow{\hat
f_{GO}}\cB(G_{O},\Pi_2,\cdots)$, and $ \sum_{g^{G_O},a^{\Pi_2}} $ sums over all the
higher gauge transformations of the higher group $\cB(G_{O},\Pi_2,\cdots)$.
Again, $\hat \phi_{GSO}$ is not an arbitrary homomorphism from $\cM^D$ to
$\cB(G_{SO},\Pi_2,\cdots)$.  We require $\hat A^{O}_{ij} =\pi(\hat A^{GO}_{ij})
\in O_\infty$ to be the connection of the tangent bundle of $\cM^D$,  where
$\hat A^{GO} = \hat \phi^*_{GO} \bar A^{GO}$.  The resulting model \eq{ZhSPTO}
realizes more general bosonic hSPT phases.

\subsection{Fermionic hSPT phases}
\label{SPTSOF}

With the above general construction of bosonic models to realize bosonic hSPT
phases, we can use higher dimensional bosonization\cite{W161201418,KT170108264}
to obtain  fermionic models to realize fermionic hSPT phases.  Such a
construction is closely related to the fermion worldline
decoration\cite{LW180901112}, and does not produce all possible fermionic hSPT
phases.

Without time reversal symmetry, we consider the following bosonized
local fermion model
\begin{align}
\label{ZhSPTSOF}
  Z(\cM^D,\hat \phi_{GfSO}) &= 
\hskip -1em
\sum_{g^{G_{fSO}},a^{\Pi_2}} 
\hskip -1em
\ee^{2\pi \ii \int_{\cM^D} \hat \phi_{GfSO}^{*;g^{G_{fSO}},a^{\Pi_2}} \bar \nu_D^{GfSO}}
\times
\nonumber\\
&
\ee^{2\pi \ii \int_{\cN^{D+1}} \hat \phi_{SO}^{*;g^{G_{fSO}},a^{\Pi_2}} \bar \om_{D+1}^{SO}}.
\end{align}
The model is build using the following data
\begin{enumerate}
\item
A fermion higher symmetry described by higher group
$\cB(G_{fSO},\Pi_2,\cdots)$, where 
\begin{align} 
G_{fSO} = G_f \gext SO_{\infty} 
\end{align} 
and $G_f=Z_2^f\gext G_b$ is the fermion 0-symmetry
group.  The  higher group $\cB(G_{fSO},\Pi_2,\cdots)$ has the canonical
$G_{fSO}$-valued 1-cochain $\bar A^{GfSO}$, the canonical $\Pi_2$-valued
2-cochain $\bar B^{GfSO}$ that satisfy $\dd \bar B^{GfSO} =\bar n_3(\bar
A^{GfSO})$, \etc\ (see \Ref{ZW180809394} and \Ref{LW180901112}).

\item
A higher group $\cB_f(G_{fSO},1;Z_2,D-1)$ with the canonical
$SO_\infty$-valued 1-cochain $\bar A^{SO}$, the canonical $\Z_2$-valued
$(D-1)$-cochain $\bar f_{D-1}$ that satisfy $\dd \bar f_{D-1} \se{2}0$ (see
\Ref{ZW180809394} and \Ref{LW180901112}).

\item
A $\RZ$-valued $(D+1)$-cocycle 
\begin{align}
\label{bomd2}
\bar \om_{D+1}^{SO}\se{1} \frac12 \Sq^2 \bar f_{D-1} +
\frac12 \bar f_{D-1} \bar \w_2(\bar A^{SO})
\end{align}
on the higher group $\cB_f(SO_\infty,1;Z_2,D-1)$.
\item
A trivialization homomorphisms $ \vphi: \cB (G_{fSO},\Pi_2.\cdots) \to
\cB_f(SO_\infty,1;Z_2,D-1)$.
\item
A choice of trivialization, \ie a $\RZ$-valued $D$-cochain on $\cB
(G_{fSO},\Pi_2.\cdots)$ that satisfies 
\begin{align}
-\dd \bar\nu_{D}^{GfSO} \se{1} \vphi^* \bar \om_{D+1}^{SO}.  
\end{align}
\end{enumerate}
The above data, in fact, gives us a partial classification of fermionic hSPT
phases without time reversal symmetry.

Now let us explain the compact notation \eqn{ZhSPTSOF} that describes the model
\begin{enumerate}
\item
The model is defined on a spacetime complex $\cM^D$.
\item
The model has a higher symmetry described by a higher group
$\cB(G_{fSO},\Pi_2,\cdots)$.  However, there is a twist of the higher symmetry
described by the background higher connection on $\cM^D$.  Such a background
higher connection is encoded in $\hat \phi^{GfSO}$, which is a
simplicial-complex homomorphism $\cM^D \xrightarrow{\hat
\phi^{GfSO}}\cB(G_{fSO},\Pi_2,\cdots)$.  $\hat \phi_{GfSO}$ is not an arbitrary
homomorphism.  We require $\hat A^{SO}_{ij} =\pi(\hat A^{GfSO}_{ij}) \in
SO_\infty$ to be the connection of the tangent bundle of $\cM^D$,  where $\hat
A^{GfSO} = \hat \phi^*_{GfSO} \bar A^{GfSO}$ and $\pi$ is the natural projection
$G_{fSO} \xrightarrow{\pi} SO_\infty$.  

\item
$\sum_{g^{G_{fSO}},a^{\Pi_2},\cdots}$ is a summation of all the higher gauge
transformations described by $g^{G_{fSO}},a^{\Pi_2},\cdots$ (see \eqn{2gauge}).
Here $g^{G_{fSO}}$ lives on the vertices of $\cM^D$: $g^{G_{fSO}}_i \in
G_{fSO}$, $a^{\Pi_2}$ lives on the links of $\cM^D$: $a^{G_{fSO}}_{ij} \in
\Pi_2$, \etc. $g^{G_{fSO}},a^{\Pi_2},\cdots$ are the dynamical fields in our
model.

\item
$\hat \phi_{GfSO}^{g^{G_{fSO}},a^{\Pi_2}}$ is the higher connection obtained
from the background higher connection $\hat \phi_{GfSO}$ via the higher gauge
transformation $g^{G_{fSO}},a^{\Pi_2},\cdots$.

\item
$\cN^{D+1}$ is a $(D+1)$-dimensional complex whose boundary is $\cM^D$:
$\prt \cN^{D+1} = \cM^D$.

\item
$\hat \phi_{SO}^{g^{G_{fSO}},a^{\Pi_2}}$
is a simplicial-complex homomorphism
$\cN^{D+1} \xrightarrow{\hat \phi_{SO}^{g^{G_{fSO}},a^{\Pi_2}}}
\cB_f(SO_\infty,1;\Z_2,D-1)$.
When restricted to the boundary $\cM^D=\prt \cN^{D+1}$,
it satisfies
$\hat \phi_{SO}^{g^{G_{fSO}},a^{\Pi_2}}= \vphi
\hat \phi_{GfSO}^{g^{G_{fSO}},a^{\Pi_2}}$:
\[
\xymatrix{
 & & \cB(G_{fSO},\Pi_2,\cdots) \ar[d]^{\vphi} \\
\prt \cN^{D+1} \ar[urr]^{\hat \phi_{GfSO}^{g^{G_{fSO}},a^{\Pi_2}}} \ar[rr]^{\hat \phi_{SO}^{g^{G_{fSO}},a^{\Pi_2}}} & & \cB_f(SO_\infty,1;\Z_2,D-1)
}
\]

\end{enumerate}
The resulting model \eq{ZhSPTSOF} realizes fermionic hSPT phases without time
reversal symmetry.

When we have only the usual global symmetry, \ie when
$\cB(G_{fSO},\Pi_2,\cdots)=\cB G_{fSO}$, the above model \eq{ZhSPTSOF} reduces
to the one descussed in \Ref{LW180901112}, which realizes fermionic SPT phases.
When $\bar \om_{D+1}^{SO}=0$, the model \eq{ZhSPTSOF} reduces to
\eqn{ZhSPTSO} which realizes bosonic hSPT phases.

To include time reversal symmetry, we simply replace $SO_\infty$ by $O_\infty$
in the above construction.  However, now $\bar \om_{D+1}^{O}$ has two choices:
If the fermions are Kramers singlet, we have 
\begin{align}
\bar \om_{D+1}^{SO}\se{1} \frac12 \Sq^2 \bar f_{D-1} +
\frac12 \bar f_{D-1} \bar \w_2(\bar A^{O})
\end{align}
If the fermions are Kramers doublets, we have 
\begin{align}
\bar \om_{D+1}^{SO}\se{1} \frac12 \Sq^2 \bar f_{D-1} +
\frac12 \bar f_{D-1} [\bar \w_2(\bar A^{O}) + \bar \w_1^2(\bar A^{O})]
\end{align}

\subsection{The usefulness of hSPT phases}

But hSPT phases do not exist in  natural condensed matter systems. This is
because, similar to the usual SPT phase, a hSPT phase requires a higher
symmetry.  Without higher symmetry, we simply do not have distinct hSPT phases.
But in the usual condensed matter systems, we do not have higher symmetry.  In
the EM condensed matter systems with dynamical electromagnetic field, we either
have gapless photon modes or have non-trivial topological orders (where the
higher symmetry is spontaneously broken due to the superconductivity in the EM
condensed matter systems).  This is why we do not have hSPT phases in natural
condensed matter systems.  (But we may construct fine-tuned toy models
experimentally to realize higher symmetries and hSPT phases.)

However, understanding the hSPT phases is still important in condensed matter.
This is because the emergent higher symmetries in topological orders may be
anomalous, which has physical consequences.  We need to understand hSPT phases
in order to understand the anomalous higher symmetry via the boundary of hSPT
states.\cite{W1313,TK151102929}  In the following, we will study a few simple
hSPT phases. In particular, we will give examples of topological orders with
emergent (anomalous) higher symmetries.

\section{Lattice models that realize topological orders with higher symmetry}

\subsection{The first type of constructions}

We have constructed models to realize a hSPT phase with a combined 0-symmetry
and 1-symmetry, where we have summed over all 1-cochains $a_{ij}$ in \eq{Zab}.
To realize a topologically ordered phase with a combined 0-symmetry and
1-symmetry, we can change the dynamics of the $a_{ij}$ field, by instead
summing over only all 1-cocycles $a_{ij}$ that satisfy $\dd a=0$:
\begin{align}
\label{ZabTO}
  Z &=\sum_{\{g_i,\dd a = 0\} } \ee^{2\pi \ii \int_{\cM^D} \om_D(A,B)} ,
\nonumber\\
A_{ij} &= g_i^{-1} \hat A_{ij} g_j,\ \ \
B = \hat B + \dd a + \xi(\hat A,g),
\end{align}
Since $a_{ij}\in \Pi_2$ (\ie $a$ is a $\Pi_2$-valued 1-cocycle), the above
model realizes a topologically ordered state described by Abelian gauge theory
with $\Pi_2$ gauge group.  The model also has a combined 0-symmetry and
1-symmetry. In particular, the 1-symmetry is generated by shifting $a$ by
$\Pi_2$-valued 1-cocycles.

The flux of the $\Pi_2$ gauge theory is described by a $D-2$-cycle $\t F_{D-2}$
in the dual spacetime complex $\t\cM^D$.  The charge of the $\Pi_2$ gauge
theory is described by a $1$-cycle $C^1$ (the worldline of the charge) in the
spacetime complex $\cM^D$.
In the presence of the flux and charge, the path integral becomes
\begin{align}
\label{ZabTOs}
  Z &=\sum_{\{g_i,\dd a = *\t F_{D-2}\}} \ee^{2\pi \ii \int_{\cM^D} \om_D(A,B)+2\pi\ii \int_{C^1} a} ,
\nonumber\\
A_{ij} &= g_i^{-1} \hat A_{ij} g_j,\ \ \
B = \hat B + \dd a + \xi(\hat A,g),
\end{align}
where $*\t F_{D-2}$ is a two cocycle -- the Poincar\'e dual of $D-2$ cycle $\t
F_{D-2}$.  

We note that the gauge charges are not mobile in the exactly soluble model
\eqn{ZabTO}. To make the gauge charges mobile, we need to add terms like
$\ee^{-\la \int_{I_1} a}$ where $I_1$ is a 1-chain in the spacetime complex
$\cM^D$.  Shifting $a$ by a cocycle will change $\ee^{-\la \int_{I_1} a}$.  So
the term $\ee^{-\la \int_{I_1} a}$ will break the 1-symmetry.  The term
$\ee^{2\pi \ii \int_{C^1} a}$ will also break the 1-symmetry if $C^1$ is not a
1-boundary.  Thus if we only allow the gauge flux, the $\Pi_2$-gauge theory
\eq{ZabTOs} (without the $C^1$ term) will have a combined 0-symmetry and
1-symmetry.

We like to stress that the above 1-symmetry can be emergent in the weak
coupling phase of the gauge theory.  As long as the gauge flux to be the only
low energy excitations, we will have the emergent 1-symmetry in the in the weak
coupling gauge theory, and we will have a combined 0-symmetry and 1-symmetry at
low energies.

\subsection{The second type of constructions}

In the second type of constructions, we start with a model that realize a hSPT
state with no topological order (see \eqn{ZabSPT})
\begin{align}
\label{ZabSPT1}
  Z &= \sum_{\{g_i,a_{ij}\} } \ee^{2\pi \ii \int_{\cM^D} \om_D(A,B)},
\nonumber\\
A_{ij} &= g_i^{-1} \hat A_{ij} g_j,
\nonumber\\
B_{ijk} &= \hat B_{ijk} +a_{ij}+a_{jk}-a_{ik} + \xi_{ijk}(\hat A,g),
\end{align}
where $\hat A_{ij}, g_i $ belong to a group $G$ and $\hat B_{ijk}, a_{ij}$
belong to an Abelian group $ \Pi_2$. Also $\om_D$ is a $\R/\Z$-valued 2-group
cocycle.  We note that the above model is construct by starting with a fixed
background 2-connection described by $\hat A_{ij}$ and $\hat B_{ijk}$, and then
include the ``pure'' 2-gauge transformations described by $g_i$ and $a_{ij}$ as
dynamical fields.

To obtain a model that realizes a topological order, we can partially gauge the
2-group. This way, we obtain a combined  1-gauge and a 2-gauge theory with a
combined 0-symmetry and 1-symmetry.  To do so, we assume $G=G^g\gext G^s$ and
$\Pi_2=\Pi_2^g\gext \Pi_2^s$.  $G^s$ will be our 0-symmetry group and $G^g$ the
gauge group of the 1-gauge theory.  We label the group elements of $G$ by a
pair 
\begin{align}
 g = (g^g, g^s), \ \ \
g^g \in G^g,\ g^s \in G^s,\ g_i \in G.  
\end{align}
So we can denote the effective field $A_{ij}$ as a pair $A_{ij} =
(A_{ij}^g,A_{ij}^s)$. We have
\begin{align}
\label{Atop}
 A_{ij} &= (g^g_i, g^s_i)^{-1}(a_{ij}^g, \hat A_{ij}^s) (g^g_j, g^s_j)
 = g_i^{-1}(a_{ij}^g, \hat A_{ij}^s) g_j,
\nonumber\\
& g_i^g, a_{ij}^g \in G^g,\ g_i^s, \hat A_{ij}^s \in G^s,\ g_i \in G.  
\end{align}
Here $\hat A_{ij}^s$ is the flat connection
describing the $G^s$ symmetry twist.
$a_{ij}^g$ is the dynamical gauge field
satisfying
\begin{align}
\label{acond}
 a_{ij}^g a_{jk}^g=a_{ik}^g,
\end{align}
so that $a_{ij}^g$ describes the deconfined phase of the $G^g$ 1-gauge theory.
$g_i \in G$ is a dynamical scalar field carrying both symmetry charge and gauge
charge.

Similarly, we denote the elements in
$\Pi_2$ with a pair
\begin{align}
 h = (h^g, h^s), \ \ \
h^g \in \Pi_2^g,\ h^s \in \Pi_2^s,\ h \in \Pi_2.  
\end{align}
So we can write the effective dynamical field $B$ as
\begin{align}
\label{Btop}
B_{ijk} &= 
(b_{ijk}^g, \hat B_{ijk}^s) 
+a_{ij}+a_{jk}-a_{ik} + \xi_{ijk}[(a_{ij}^g, \hat A_{ij}^s),g_i],
\end{align}
Here $\hat B_{ijk}^s\in \Pi_2^s$ is the background 2-connection describing the
1-symmetry twist of $\Pi_2^s$.  $b_{ijk}^g\in \Pi_2^g$ is the dynamical 2-gauge
field of $\Pi_2^g$, that satisfy
\begin{align}
\label{bcond}
 \dd (b^g, \hat B^s) = n_3[(a^g, \hat A^s)] .
\end{align}
$a_{ij} \in \Pi_2$ is a dynamical 1-cochain field.

Now, we can write down our partially gauged hSPT model:
\begin{align}
  Z &= \sum_{\{g_i,a_{ij},a_{ij}^g,b_{ijk}^g\} } \ee^{2\pi \ii \int_{\cM^D} \om_D(A,B)},
\end{align}
where the 1-cochain effective field $A$ and the 2-cochain effective field $B$
are given by \eqn{Atop} and \eqn{Btop}.  The dynamical fields $g_i \in G$ and
$a_{ij} \in \Pi_2$ can fluctuate arbitrarily.  The dynamical fields $a^g \in
G^g$ and $b^g \in \Pi_2^g$ cannot fluctuate arbitrarily and should satisfy the
conditions \eqn{acond} and \eqn{bcond}, so that they describe deconfined phase
of a 2-gauge theory.  

We see that  partially gauging is simply
making part of the background connections dynamical
\begin{align}
 \hat A_{ij} &= (\hat A^g_{ij}, \hat A^s_{ij}) \to (a^g_{ij}, \hat A^s_{ij}),
\nonumber\\
 \hat B_{ijk} &= (\hat B^g_{ijk}, \hat B^s_{ijk}) \to (b^g_{ijk}, \hat B^s_{ijk}).
\end{align}
plus ``pure'' 2-gauge fluctuations described by dynamical fields $g_i$ and
$a_{ij}$.  We note that, in our construction, the combined 1-gauge and 2-gauge
theory plus its combined 0-symmetry and 1-symmetry together is described a
2-group $\cB(G,\Pi_2)$ and a 2-group cocycle $\om_D$.

More generally, we may consider a simplicial-complex homomorphis $\vphi$ between two higher
groups
\begin{align}
 \cB(G,\Pi_2,\Pi_3,\cdots) \xrightarrow{\vphi}
 \cB(G^s,\Pi_2^s,\Pi_3^s,\cdots) .
\end{align}
We assume $\vphi$ to be surjective.
Then using a $\R/\Z$-valued cocycle $\bar \om_D$ on $\cB(G,\Pi_2,\Pi_3,\cdots)$,
we can construct the following local bosonic model (see Section \ref{hSPTsec})
\begin{align}
Z(\cM^D,\hat \phi^s)&= 
\sum_{\{g,a,\cdots\}} 
\sum_{\{\phi|{\vphi \phi=\hat \phi^s}\}} 
\ee^{2\pi \ii \int_{\cM^D} \phi^{*;g,a,\cdots}\bar \om_D}.
\end{align}
The above model realize a topological order with a higher symmetry described by
the higher group $\cB(G^s,\Pi_2^s,\Pi_3^s,\cdots)$.  Here $\phi$ is a complex
homomorphism from $\cM^D$ to $\cB(G,\Pi_2,\Pi_3,\cdots)$ and $\hat \phi^s$ is a
fixed simplicial-complex homomorphism from $\cM^D$ to $\cB(G^s,\Pi_2^s,\Pi_3^s,\cdots)$:
\[
\xymatrix{
 & \cB(G,\Pi_2,\cdots) \ar[d]^{\vphi} \\
\cM^D \ar[ur]^\phi \ar[r]^{ \hat \phi^s} &  \cB(G^s,\Pi_2^s,\cdots)
}
\]
$\hat \phi^s$ is the background connection that describes the higher symmetry
twist on $\cM^D$.  The summation $\sum_{\{\phi|{\vphi \phi=\hat \phi^s}\}}$
sums over all the homomorphisms $\phi$ such that $\vphi \phi=\hat \phi^s$.  The
summation $\sum_{\{g,a,\cdots\}}$ sums over all the higher gauge
transformations of the higher group $\cB(G,\Pi_2,\Pi_3,\cdots)$.

\section{EM condensed matter systems and their higher symmetry}
\label{EMsys}

In our theoretical descriptions of condensed matter systems, we usually ignore
the dynamical electromagnetic (EM) field. Those usual condensed matter theories
in general do not have higher symmetries.  However, more accurate theoretical
descriptions of condensed matter systems should contain the dynamical EM field.
In this section, we point out that those EM condensed matter theories with
dynamical EM field actually have an \emph{anomalous} higher symmetry if we
ignore the magnetic monopoles.  

The reason is very simple.  A dynamical $U(1)$ gauge theory in 3+1D does not
have higher symmetry if the mobile $U(1)$ charges and $U(1)$ monopoles appear
in the interested energy scales (which are about 1eV for condensed matter
physics).  However, for the dynamical $U(1)$ EM gauge theory in condensed
matter, although the mobile $U(1)$ charges appear at energy scales of 1eV (the
energy gap of an insulator), the $U(1)$ monopoles can appear only beyond
100GeV.  So the dynamical $U(1)$ EM gauge theory in condensed matter can be
viewed as a dynamical $U(1)$ gauge theory without mobile $U(1)$ monopoles. Such
a dynamical U(1) gauge theory has a higher symmetry.\cite{GW14125148} 

Using the general picture of higher symmetry and its relation to topological
excitations developed in Section \ref{SecZn1SPT}, we see that the emergent
higher symmetry in the EM condensed matter systems is characterized by low
energy allowed EM charge excitations.  It is generated by the topological
excitations with trivial mutual statistics with the EM charge excitations.  In
other words, the higher symmetry is generated by the electric charged charge
excitations.  
This appearance of higher symmetry in condensed matter systems has been noticed
in some recent papers.\cite{GI161007392,GS181104879}

To see the higher symmetry in EM condensed matter systems more explicitly, let
us consider a dynamical $U(1)$ gauge lattice gauge theory described by a lattice
rotor model where the rotor angles correspond to the gauge connection of the
\emph{dual} $\t U(1)$ gauge field:\cite{MS0204,W0313,LW0622}
\begin{align}
 H &= 
U \sum_{\t{\v i}} \Big(\sum_{\t{\v j} \text{ next to }\t{\v i}} \t L_{\t{\v i}\t{\v j}}\Big)^2
+g \sum_{\<\t{\v i}\t{\v j}\>} \t L_{\t{\v i}\t{\v j}}
\nonumber\\
&\ \ \ \ 
-J \sum_{\<\t{\v i}\t{\v j}\t{\v k}\t{\v l}\>} 
\Big( 
\t L_{\t{\v i}\t{\v j}}^+
\t L_{\t{\v j}\t{\v k}}^+
\t L_{\t{\v k}\t{\v l}}^+
\t L_{\t{\v l}\t{\v i}}^+ + h.c. \Big)
\end{align}
where $\t{\v i},\t{\v j},\t{\v k},\t{\v l}$ label the sites of a dual cubic
lattice $\t \cM^3$, and $\t L_{\v i\t{\v j}}=-\ii \prt_{\t\th_{\t{\v i}\t{\v
j}}}$ is the angular momentum of the rotor $\t\th_{\t{\v i}\t{\v
j}}=-\t\th_{\t{\v j}\t{\v i}}$ living on the link $\t{\v i}\t{\v j}$.  Also $\t
L^\pm_{\t{\v i}\t{\v j}}= \ee^{\pm \ii \t\th_{\t{\v i}\t{\v j}}}$.  The
summation $\sum_{\<\v i\t{\v j}\t{\v k}\t{\v l}\>}$ sums over all the square
faces $\<\t{\v i}\t{\v j}\t{\v k}\t{\v l}\>$ of the cubic lattice.  The charge
of the dual $\t{U}(1)$ corresponds to the monopole of the EM $U(1)$.  The
monopole of the dual $\t{U}(1)$ corresponds to the charge of the EM
$U(1)$.

The higher symmetry is given by
\begin{align}
 W_{C^2} = \ee^{\ii \vphi \sum_{\<\t{\v i}\t{\v j}\> \in C^2 } \t L_{\t{\v i}\t{\v j}}}.
\end{align}
Where $C^2$ is a closed surface formed by the square faces the cubic lattice
$\cM^3$.  Since $\cM^3$ and $\t\cM^3$ are dual to each other, the links
$\<\t{\v i}\t{\v j}\>$ in the dual lattice $\t\cM^3$ correspond to the square
faces of the lattice $\cM^3$.  
Since $C^2$ has a codimension-1 in the 3-dimensional space,
the higher symmetry is a $\t U(1)$-1-symmetry.
The above $\t U(1)$ 1-higher symmetry is generated by
\begin{align}
 W_{\t{\v i}} = \ee^{\ii \vphi \sum_{\t{\v j} \text{ next to }\t{\v i}} \t L_{\t{\v i}\t{\v j}}}.
\end{align}
which leaves the above Hamiltonian invariant.  $W_{\t{\v i}}$ is also called
the local gauge symmetry.  Thus the $U(1)$-1-symmetry is simply the dual $\t
U(1)$ gauge symmetry.  Such an $U(1)$-1-symmetry forbids the term
$\sum_{\<\t{\v i}\t{\v j}\>} L^+_{\v i\t{\v j}}$ in the Hamiltonian.  So the
charge of the dual $\t{U}(1)$ (\ie the monopole of the EM $U(1)$) is not
mobile.

We can also describe the above dual $\t U(1)$ gauge theory using path integral
of cochain fields on spacetime complex $\cM^4$:
\begin{align}
\label{Ztab}
 Z = \sum_{\{ \t a^{\R/\Z}, \cdots\}} \ee^{-\int_{\cM^4} L( \dd \t a^{\R/\Z}, \cdots)}
\end{align}
where $\sum_{\{ \t a^{\R/\Z}, \cdots\}}$ sums over $\R/\Z$-valued 1-cochains $\t a^{\R/\Z}$ on
spacetime dual complex $\t \cM^4$, and possibly some other EM neutral bosonic
fields represented by $\cdots$.  Since $\t a^{\R/\Z}$ is $\R/\Z$-valued,  $L( \dd \t a^{\R/\Z},
\cdots)$ is invariant under
\begin{align}
\label{aaZ}
 \t a^{\R/\Z} \to \t a^{\R/\Z}+ \t \al^\Z,
\end{align}
where $\t \al^\Z$ is any $\Z$-valued 1-cochain.  The model has a $\t U(1)$
1-symmetry, which  is generated by shifting $\t a^{\R/\Z}$ by $\R/\Z$-valued
1-cocycles $\t \al$: 
\begin{align}
\label{tU11symm}
\t a^{\R/\Z} \to \t a^{\R/\Z}+ \t \al^{\R/\Z},\ \ \ \ \dd \t \al^{\R/\Z}=0.
\end{align}

When $L( \dd \t a^{\R/\Z}, \cdots)$ restricts the fluctuations to be $\dd \t
a^{\R/\Z} \approx \text{integer cochain}$, then $\t a^{\R/\Z}$ will describe a
$\t U(1)$ gauge field (the EM field) in semiclassical limit.  

We note that $\dd \toZ{\dd \t a^{\R/\Z}}$ is a $\Z$-valued 3-coboundary that
corresponds to the Poincar\'e dual of the wordlines of monopoles of the $\t
U(1)$ gauge field.  The monopoles of the $\t U(1)$ gauge field is the electric
charges.  So the Poincar\'e dual of $\dd \toZ{\dd \t a^{\R/\Z}}$ is the world
line of the electric charges.  We also note that $\dd \toZ{\dd \t a^{\R/\Z}}$
is invariant under \eqn{aaZ}, and is thus physical.

Now we see that if $L( \dd \t a^{\R/\Z}, \cdots)$ further restricts the
fluctuations to be $\dd \t a^{\R/\Z} \approx 0$, in this case $\t
a^{\R/\Z}$ will describes a deconfined phase of $\t U(1)$ gauge field $\t
a^{\R/\Z}$, which corresponds to an EM insulator.  On the order hand, if  $L(
\dd \t a^{\R/\Z}, \cdots)$ restricts $\dd \t a^{\R/\Z} \approx \text{integer
cochain}$, and allow strong fluctuations of $\dd \toZ{\dd \t a^{\R/\Z}}$, the
system will be in a metallic or a superconducting phase.

However, in the model \eq{Ztab} worldlines of electric charges, described by
$*\dd \toZ{\dd \t a^{\R/\Z}}$, are bosons.  So the model \eq{Ztab} does not
describe the EM condensed matter systems, where the odd charges of EM $U(1)$
are always fermions.  To make the electric charge worldline $*\dd \toZ{\dd \t
a^{\R/\Z}}$ to describe the worldlines of fermions, we can use the high
dimensional bosonization.\cite{W161201418,KT170108264} Thus the correct models
that describe EM condensed matter systems are given by
\begin{align}
\label{ZtabF}
 Z = \sum_{\{ \t a^{\R/\Z}, \cdots\}} &\ee^{-\int_{\cM^4} L(\dd \t a^{\R/\Z}, \cdots)
+\pi\ii \int_{\cM^4} \hat A^{\Z_2} \dd\toZ{\dd \t a^{\R/\Z}}}
\nonumber\\
&\ee^{\pi \ii \int_{\cN^5}  \gSq^2 \dd\toZ{\dd \t a^{\R/\Z}} +\w_2 \dd\toZ{\dd \t a^{\R/\Z}}}
,
\end{align}
where $\prt \cN^5 =\cM^4$, $\w_k$ is the $k^\text{th}$ Stiefel-Whitney class of
the tangent bundle of $\cN^5$, and $\hat A^{\Z_2}$ is a spin structure
\begin{align}
\dd  \hat A^{\Z_2} \se{2} \w_2.
\end{align}
The term $ \ee^{\pi \ii \int_{\cN^5} \gSq^2 \dd \toZ{\dd \t a^{\R/\Z}} }$ makes
$*\dd \toZ{\dd \t a^{\R/\Z}}$ to describe the worldlines of fermions.  
For details, see \Ref{LW180901112}.  

We like to remark that even though 
\begin{align}
\ee^{\pi \ii \int_{\cN^5}  \gSq^2 \dd\toZ{\dd \t a^{\R/\Z}} }
=\ee^{\pi \ii \int_{\cM^4}  \gSq^2 \toZ{\dd \t a^{\R/\Z}} },
\end{align}
the term $\ee^{\pi \ii \int_{\cN^5}  \gSq^2 \dd\toZ{\dd \t a^{\R/\Z}} }$ cannot
come from a local Lagragian on $\cM^4$ such as $\ee^{\pi \ii \int_{\cM^4}
\gSq^2 \toZ{\dd \t a^{\R/\Z}} }$. This is because $\gSq^2 \toZ{\dd \t
a^{\R/\Z}}$ mod 2 is not invariant under under the gauge transformation
\eq{aaZ}, and is not allowed in a 3+1D Lagrangian.  Thus $\ee^{\pi \ii
\int_{\cN^5}  \gSq^2 \dd\toZ{\dd \t a^{\R/\Z}} }$ is an intrinsic $\cN^5$ term
like the Wess-Zumino-Witten term.\cite{WZ7195,W8322}

The model \eq{ZtabF} still has the $\t U(1)$-1-symmetry \eq{tU11symm}.  To find
out if such a $\t U(1)$-1-symmetry is anomalous or not, we note that $\ee^{\pi
\ii \int_{\cN^5}  \gSq^2 \dd\toZ{\dd \t a^{\R/\Z}}}$ is the action amplitude
\eq{Z5D1SPTU1}, describing a 4+1D local bosonic model with the $\t U(1)$
1-symmetry \eq{tU11symm}.  A EM condensed matter system \eqn{ZtabF} is always a
boundary of such a 4+1D $\t U(1)$-1-symmetric model.  It was shown that the
action amplitude \eq{Z5D1SPTU1} describes a non-trivial hSPT phase of $\t U(1)$
1-symmetry.  Thus the $\t U(1)$-1-symmetry in the EM condensed matter systems
is always anomalous.

Eqn. \eq{ZtabF} describe a generic EM condensed matter system without time
reversal symmetry. In the presence of time reversal symmetry, the EM $U(1)$
odd-charged fermions must also be a Kramers doublet.  In this case  EM
condensed matter systems are described by\cite{LW180901112}
\begin{align}
\label{ZtabFT}
 Z = \sum_{\{ \t a^{\R/\Z}, \cdots\}} &\ee^{-\int_{\cM^4} L(\dd \t a^{\R/\Z}, \cdots)+\pi\ii \int_{\cM^4} \hat A^{\Z_2} \dd\toZ{\dd \t a^{\R/\Z}}}
\nonumber\\
&\ee^{\pi \ii \int_{\cN^5}  \gSq^2 \dd\toZ{\dd \t a^{\R/\Z}} +(\w_2+\w_1^2) \dd \toZ{\dd \t a^{\R/\Z}}}
,
\end{align}
where the spin structure $\hat A^{\Z_2}$ now satisfies  
\begin{align}
\dd  \hat A^{\Z_2} \se{2} \w_2 +\w_1^2  .
\end{align}
Similarly, the  $\t U(1)$-1-symmetry in \eqn{ZtabFT} is also anomalous.

We note that the  EM condensed matter systems described by \eqn{ZtabF} or
\eqn{ZtabFT} are actually a bosonic theory, since all the charge neutral
excitations are bosons.  In fact  \eqn{ZtabF} and \eqn{ZtabFT} themselves are
bosonic theories, which are boundaries of a bosonic theory with trivial
topological order.  So \eqn{ZtabF} and \eqn{ZtabFT} are really 3+1D local
bosonic theories.

The  anomalous $\t U(1)$-1-symmetry in \eqn{ZtabF} or \eqn{ZtabFT} implies that
\frmbox{all the gapped liquid phases\cite{ZW1490,SM1403} of all the EM
condensed matter systems must have a bosonic topological orders.} Here
``bosonic topological orders'' means topological orders in local bosonic
systems.  The 3+1D bosonic topological orders are classified in
\Ref{LW170404221}, \Ref{LW180108530}, and \Ref{ZW180809394}.

We know that any condensation of even-number of electrons can only break the
$U(1)$ gauge symmetry down to $Z_n$ gauge symmetry ($n=\text{even}\neq 0$).  So
the induced gapped phases (the supercoducting phases) have non-trivial
topological orders described by $Z_n$ gauge theory.  However, in addition to
the boson-condensation of the electron clusters,  the electrons may also form a
non-trivial topological state.  One may wonder whether this extra topological
order can cancel the  topological order of the $Z_n$ gauge theory, and give
rise to a trivial product state. Using the higher anomaly, we find that no
matter how electric charges fluctuate and condense, their induced gapped liquid
phase of EM condensed matter systems must have a non-trivial bosonic
topological order.

This work is motivated by the presentations on higher form symmetry and higher
anomaly in field theories, in the workshop ``Developments in Quantum Field
Theory and Condensed Matter Physics'' at the Simons Center for Geometry and
Physics, Stony Brook University.  I would lile to thank Jason Alicea, Liang
Kong, Gil Refael, Tian Lan for many helpful discussions.  This research is
partially supported by NSF Grant No.  DMR-1506475 and DMS-1664412.

\appendix
\allowdisplaybreaks

\section{Space-time complex, cochains, and cocycles} 

\label{cochain}

In this paper, we consider models defined on a spacetime lattice.  A
spacetime lattice is a triangulation of the $D$-dimensional spacetime $M^D$,
which is denoted by $\cM^D$.  We will also call the triangulation $\cM^D$ as a
spacetime complex, which is formed by simplices -- the vertices, links,
triangles, \etc.  We will use $i,j,\cdots$ to label vertices of the spacetime
complex.  The links of the complex (the 1-simplices) will be labeled by
$(i,j),(j,k),\cdots$.  Similarly, the triangles of the complex  (the
2-simplices)  will be labeled by $(i,j,k),(j,k,l),\cdots$.

\begin{figure}[t]
\begin{center}
\includegraphics[scale=0.5]{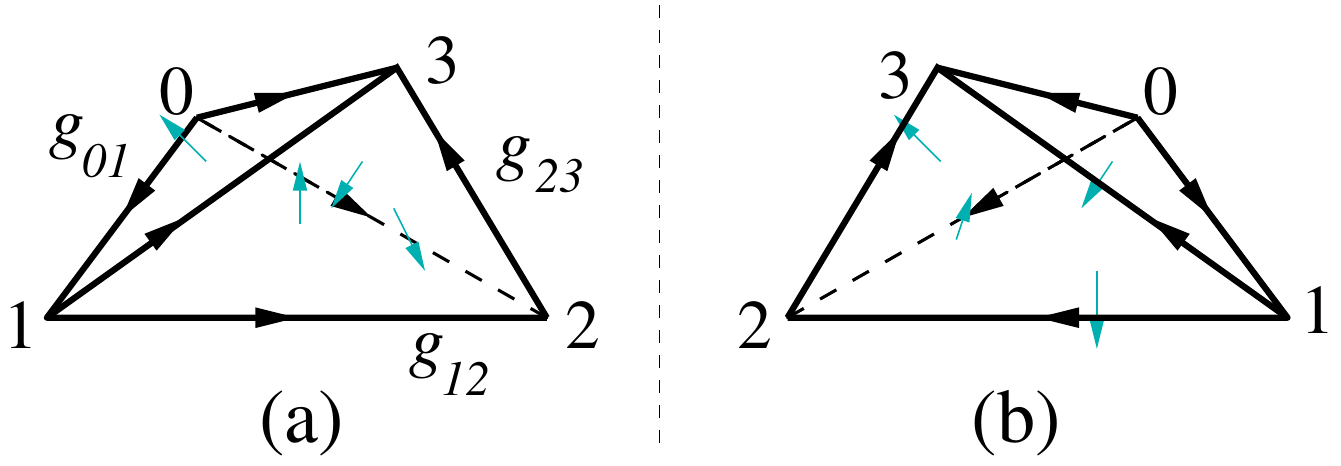} 
\end{center}
\caption{ (Color online) Two branched simplices with opposite orientations.
(a) A branched simplex with positive orientation and (b) a branched simplex
with negative orientation.  }
\label{mir}
\end{figure}

In order to define a generic lattice theory on the spacetime complex
$\cM^D$ using local Lagrangian term on each simplex, it is important
to give the vertices of each simplex a local order.  A nice local scheme to
order  the vertices is given by a branching
structure.\cite{C0527,CGL1314,CGL1204} A branching structure is a choice of
orientation of each link in the $d$-dimensional complex so that there is no
oriented loop on any triangle (see Fig. \ref{mir}).

The branching structure induces a \emph{local order} of the vertices on each
simplex.  The first vertex of a simplex is the vertex with no incoming links,
and the second vertex is the vertex with only one incoming link, \etc.  So the
simplex in  Fig. \ref{mir}a has the following vertex ordering: $0,1,2,3$.

The branching structure also gives the simplex (and its sub-simplices) a
canonical orientation.  Fig. \ref{mir} illustrates two $3$-simplices with
opposite canonical orientations compared with the 3-dimension space in which
they are embedded.  The blue arrows indicate the canonical orientations of the
$2$-simplices.  The black arrows indicate the canonical orientations of the
$1$-simplices.

Given an Abelian group $(\M, +)$, an $n$-cochain $f_n$ is an assignment of
values in $\M$ to each $n$-simplex, for example a value $f_{n;i,j,\cdots,k}\in
\M$ is assigned to $n$-simplex $(i,j,\cdots,k)$.  So \emph{a cochain $f_n$ can
be viewed as a bosonic field on the spacetime lattice}. 

$\M$ can also be viewed a $\Z$-module (\ie a vector space with
integer coefficient) that also allows scaling by an integer:  
\begin{align}
	 x+y &= z,\ \ \ \ x*y=z, \ \ \ \ mx=y,
	\nonumber\\
	x,y,z & \in \M,\ \ \ m \in \Z.
\end{align}
The direct sum of two modules
$\M_1\oplus \M_2$ (as vector spaces) is equal to the direct product of the two
modules (as sets):
\begin{align}
 \M_1\oplus \M_2 \stackrel{\text{as set}}{=} \M_1\times \M_2
\end{align}

\begin{figure}[tb]
\begin{center}
\includegraphics[scale=0.5]{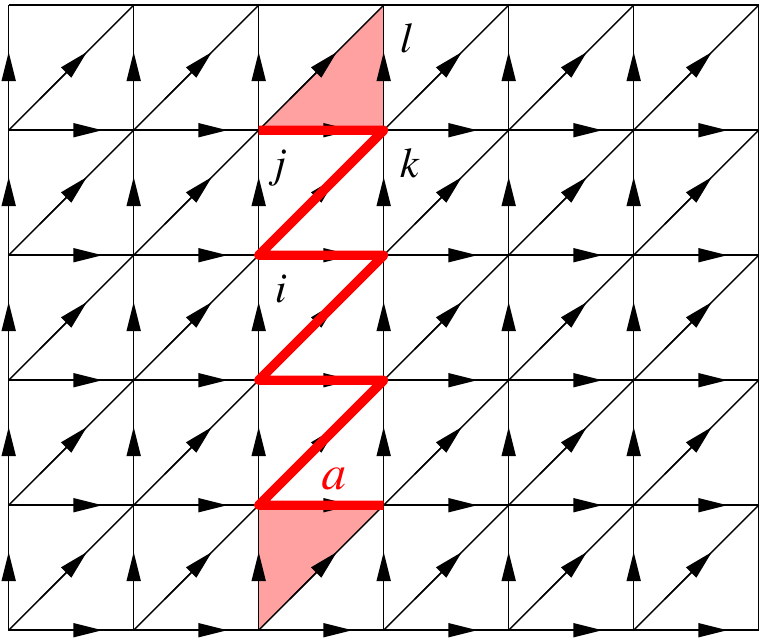} \end{center}
\caption{ (Color online)
A 1-cochain $a$ has a value $1$ on the red links: $ a_{ik}=a_{jk}= 1$ and a
value $0$ on other links: $ a_{ij}=a_{kl}=0 $.  $\dd a$ is non-zero on the
shaded triangles: $(\dd a)_{jkl} = a_{jk} + a_{kl} - a_{jl}$.  For such
1-cohain, we also have $a\smile a=0$.  So when viewed as a $\Z_2$-valued cochain,
$\Bs_2 a \neq a\smile a$ mod 2.
}
\label{dcochain}
\end{figure}

We like to remark that a simplex $(i,j,\cdots,k)$ can have two different
orientations. We can use $(i,j,\cdots,k)$ and $(j,i,\cdots,k)=-(i,j,\cdots,k)$
to denote the same simplex with opposite orientations.  The value
$f_{n;i,j,\cdots,k}$ assigned to the simplex with opposite  orientations should
differ by a sign: $f_{n;i,j,\cdots,k}=-f_{n;j,i,\cdots,k}$.  So to be more
precise $f_n$ is a linear map $f_n: n\text{-simplex} \to \M$. We can denote the
linear map as $\<f_n, n\text{-simplex}\>$, or
\begin{align}
 \<f_n, (i,j,\cdots,k)\> = f_{n;i,j,\cdots,k} \in \M.
\end{align}
More generally, a \emph{cochain} $f_n$ is a linear map
of $n$-chains:
\begin{align}
	f_n:  n\text{-chains} \to \M,
\end{align}
or (see Fig. \ref{dcochain})
\begin{align}
 \<f_n, n\text{-chain}\> \in \M,
\end{align}
where a \emph{chain} is a composition of simplices. For example, a 2-chain can
be a 2-simplex: $(i,j,k)$, a sum of two 2-simplices: $(i,j,k)+(j,k,l)$, a more
general composition of 2-simplices: $(i,j,k)-2(j,k,l)$, \etc.  The map $f_n$ is
linear respect to such a composition.  For example, if a chain is $m$ copies of
a simplex, then its assigned value will be $m$ times that of the simplex.
$m=-1$ correspond to an opposite orientation.  

We will use $C^n(\cM^D;\M)$ to denote the set of all
$n$-cochains on $\cM^D$.  $C^n(\cM^D;\M)$ can also be
viewed as a set all $\M$-valued fields (or paths) on  $\cM^D$.  Note
that $C^n(\cM^D;\M)$ is an Abelian group under the $+$-operation.

The total spacetime lattice $\cM^D$ correspond to a $D$-chain.  We
will use the same $\cM^D$ to denote it.  Viewing $f_D$ as a linear
map of $D$-chains, we can define an ``integral'' over $\cM^D$:
\begin{align}
 \int_{\cM^D} f_D &\equiv \<f_D,\cM^D\>
\\
&=\sum_{(i_0,i_1,\cdots,i_D)}
s_{i_0i_1\cdots i_D} (f_D)_{i_0,i_1,\cdots,i_D}
.
\nonumber 
\end{align}
Here $s_{i_0i_1\cdots i_D}=\pm 1$, such that a $D$-simplex in the $D$-chain
$\cM^D$ is given by $s_{i_0i_1\cdots i_D} (i_0,i_1,\cdots,i_D)$.

\begin{figure}[tb]
\begin{center}
\includegraphics[scale=0.5]{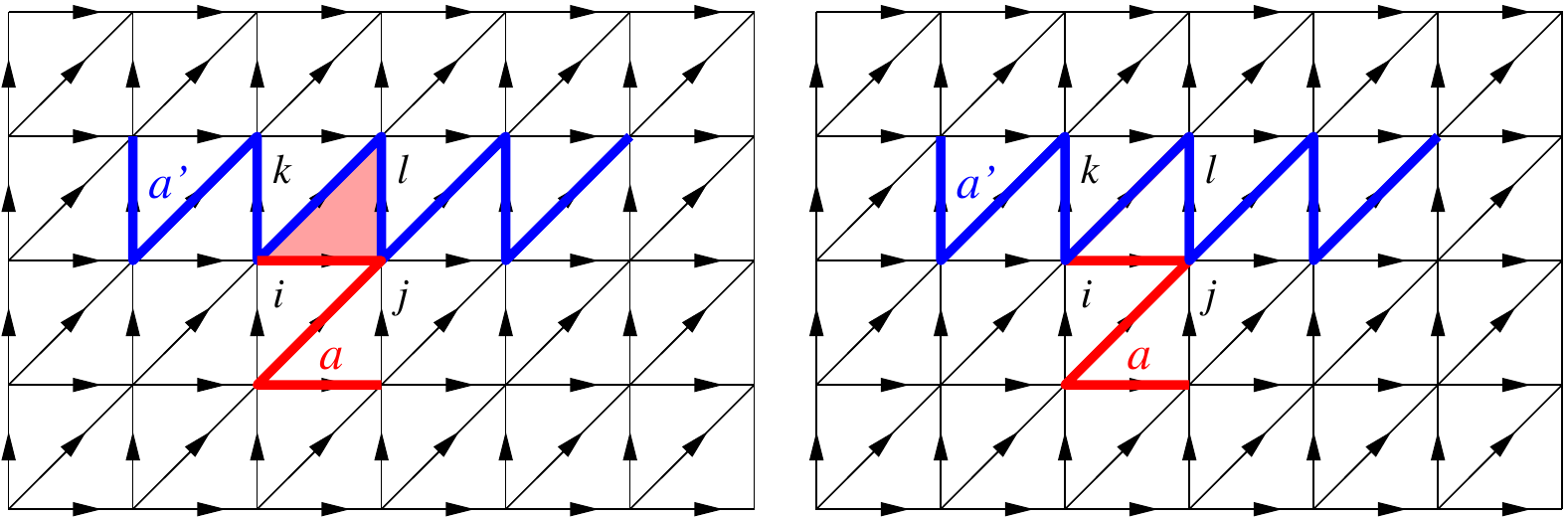} \end{center}
\caption{ (Color online)
A 1-cochain $a$ has a value $1$ on the red links, Another
1-cochain $a'$ has a value $1$ on the blue links.
On the left, $a\smile a'$ is non-zero on the shade triangles:
$(a\smile a')_{ijl}=a_{ij}a'_{jl}=1$.
On the right, $a'\smile a$ is zero on every triangle.
Thus $a\smile a'+a'\smile a$ is not a coboundary.
}
\label{cupcom}
\end{figure}

We can define a derivative operator $\dd$ acting on an $n$-cochain $f_n$, which
give us an $(n+1)$-cochain (see Fig. \ref{dcochain}):
\begin{align} 
\label{eq:differential}
&\ \ \ \ \<\dd f_n, (i_0i_1i_2\cdots i_{n+1})\>
\nonumber\\
&=\sum_{m=0}^{n+1} (-)^m 
\<f_n, (i_0i_1i_2\cdots\hat i_m\cdots i_{n+1})\>
\end{align}
where $i_0i_1i_2\cdots \hat i_m \cdots i_{n+1}$ is the sequence
$i_0 i_1 i_2 \cdots i_{n+1}$ with $i_m$ removed, and
$i_0, i_1,i_2 \cdots i_{n+1}$ are the ordered vertices of the $(n+1)$-simplex
$(i_0 i_1 i_2 \cdots i_{n+1})$.

A cochain $f_n \in C^n(\cM^D;\M)$ is called a \emph{cocycle} if $\dd
f_n=0$.   The set of cocycles is denoted by $Z^n(\cM^D;\M)$.  A
cochain $f_n$ is called a \emph{coboundary} if there exist a cochain $f_{n-1}$
such that $\dd f_{n-1}=f_n$.  The set of coboundaries is denoted by
$B^n(\cM^D;\M)$.  Both $Z^n(\cM^D;\M)$ and
$B^n(\cM^D;\M)$ are Abelian groups as well.  Since $\dd^2=0$, a
coboundary is always a cocycle: $B^n(\cM^D;\M) \subset
Z^n(\cM^D;\M)$.  We may view two  cocycles differ by a coboundary as
equivalent.  The equivalence classes of cocycles, $[f_n]$, form the so called
cohomology group denoted by \begin{align} H^n(\cM^D;\M)=
Z^n(\cM^D;\M)/ B^n(\cM^D;\M), \end{align}
$H^n(\cM^D;\M)$, as a group quotient of $Z^n(\cM^D;\M)$ by
$B^n(\cM^D;\M)$, is also an Abelian group.

For the $\Z_N$-valued cocycle $x_n$, $\dd x_n \se{N} 0$. Thus 
\begin{align}
\label{BsDef}
 \Bs_N x_n \equiv \frac1N \dd x_n 
\end{align}
is a $\Z$-valued cocycle. Here $\Bs_N$ is Bockstrin homomorphism.

We notice the above definition for cochains still makes sense if we
have a non-Abelian group $(G, \cdot)$ instead of an Abelian group $(\M,
+)$, however the differential $\dd$ defined by \eqn{eq:differential}
will not satisfy $\dd \circ \dd = 1$, except for the first two
$\dd$'s. That is, one may still make sense of 0-cocycle and 1-cocycle,
but no more further naively by formula
\eqn{eq:differential}. For us, we only use non-Abelian 1-cocycle in
this article. Thus it is ok.  Non-Abelian cohomology is then thoroughly
studied in mathematics motivating concepts such as gerbes to enter.

From two cochains $f_m$  and $h_n$, we can construct a third cochain
$p_{m+n}$ via the cup product (see Fig. \ref{cupcom}):
\begin{align}
p_{m+n} &= f_m \smile h_n ,
\nonumber\\
\<p_{m+n}, (0 \to {m+n})\> 
&= 
\<f_m, (0 \to m)\> \times
\nonumber\\
&\ \ \ \ 
\<h_n,(m \to {m+n}) \>,
\end{align}
where $i\to j$ is the consecutive sequence from $i$ to $j$: 
\begin{align}
i\to j\equiv i,i+1,\cdots,j-1,j. 
\end{align}
Note that the above definition applies to cochains with global.

The cup product has the following property 
\begin{align}
\label{cupprop}
 \dd(h_n \smile f_m) &= (\dd h_n) \smile f_m + (-)^n h_n \smile (\dd f_m) 
\end{align}
for  cochains with global or local values.  
We see that $h_n \smile f_m $ is a
cocycle if both $f_m$ and $h_n$ are cocycles.  If both $f_m$ and $h_n$ are
cocycles, then $f_m \smile h_n$ is a coboundary if one of $f_m$ and $h_n$ is a
coboundary.  So the cup product is also an operation on cohomology groups
$\hcup{} : H^m(M^D;\M)\times H^n(M^D;\M) \to H^{m+n}(M^D;\M)$.  The cup product
of two \emph{cocycles} has the following property (see Fig. \ref{cupcom}) 
\begin{align}
 f_m \smile h_n &= (-)^{mn} h_n \smile f_m + \text{coboundary}
\end{align}

We can also define higher cup product $f_m \hcup{k} h_n$ which gives rise to a
$(m+n-k)$-cochain \cite{S4790}:
\begin{align}
\label{hcupdef}
&\ \ \ \
 \<f_m \hcup{k} h_n, (0,1,\cdots,m+n-k)\> 
\nonumber\\
&
 = 
\hskip -1em 
\sum_{0\leq i_0<\cdots< i_k \leq n+m-k} 
\hskip -3em  
(-)^p
\<f_m,(0 \to i_0, i_1\to i_2, \cdots)\>\times
\nonumber\\
&
\ \ \ \ \ \ \ \ \ \
\ \ \ \ \ \ \ \ \ \
\<h_n,(i_0\to i_1, i_2\to i_3, \cdots)\>,
\end{align} 
and $f_m \hcup{k} h_n =0$ for  $k<0$ or for $k>m \text{ or } n$.  Here $i\to j$
is the sequence $i,i+1,\cdots,j-1,j$, and $p$ is the number of permutations to
bring the sequence
\begin{align}
 0 \to i_0, i_1\to i_2, \cdots; i_0+1\to i_1-1, i_2+1\to i_3-1,\cdots
\end{align}
to the sequence
\begin{align}
 0 \to m+n-k.
\end{align}
For example
\begin{align}
&
 \<f_m \hcup1 h_n, (0\to m+n-1)\> 
 = \sum_{i=0}^{m-1} (-)^{(m-i)(n+1)}\times
\nonumber\\
&
\<f_m,(0 \to i, i+n\to m+n-1)\>
\<h_n,(i\to i+n)\>.
\end{align} 
We can see that $\hcup0 =\smile$.  
Unlike cup product at $k=0$, the higher cup product of two
cocycles may not be a cocycle. For cochains $f_m, h_n$, we have
\begin{align}
\label{cupkrel}
& \dd( f_m \hcup{k} h_n)=
\dd f_m \hcup{k} h_n +
(-)^{m} f_m \hcup{k} \dd h_n+
\\
& \ \ \
(-)^{m+n-k} f_m \hcup{k-1} h_n +
(-)^{mn+m+n} h_n \hcup{k-1} f_m 
\nonumber 
\end{align}

Let $f_m$ and $h_n$ be cocycles and $c_l$ be a chain, from \eqn{cupkrel} we
can obtain
\begin{align}
\label{cupkrel1}
 & \dd (f_m \hcup{k} h_n) = (-)^{m+n-k} f_m \hcup{k-1} h_n 
\nonumber\\
&
\ \ \ \ \ \ \ \ \ \
 \ \ \ \ \ \ \
+ (-)^{mn+m+n}  h_n \hcup{k-1} f_m,
\nonumber\\
 & \dd (f_m \hcup{k} f_m) = [(-)^k+(-)^m] f_m \hcup{k-1} f_m,
\nonumber\\
& \dd (c_l\hcup{k-1} c_l + c_l\hcup{k} \dd c_l)
= \dd c_l\hcup{k} \dd c_l 
\nonumber\\
&\ \ \ -[(-)^k-(-)^l]
(c_l\hcup{k-2} c_l + c_l\hcup{k-1} \dd c_l) .
\end{align}

From \eqn{cupkrel1}, we see that, for $\Z_2$-valued cocycles $z_n$,
\begin{align}
 \Sq^{n-k}(z_n) \equiv z_n\hcup{k} z_n
\end{align}
is always a cocycle.  Here $\Sq$ is called the Steenrod square.  More generally
$h_n \hcup{k} h_n$ is a cocycle if $n+k =$ odd and $h_n$ is a cocycle.
Usually, the Steenrod square is defined only for $\Z_2$-valued cocycles or
cohomology classes.  Here, we like to define a generalized
Steenrod square for $\M$-valued
cochains $c_n$:
\begin{align}
\label{Sqdef}
 \gSq^{n-k} c_n \equiv c_n\hcup{k} c_n +  c_n\hcup{k+1} \dd c_n .
\end{align}
From \eqn{cupkrel1}, we see that
\begin{align}
\label{Sqd1}
 \dd \gSq^{k} c_n &= \dd(
c_n\hcup{n-k} c_n +  c_n\hcup{n-k+1} \dd c_n )
\\
&= \gSq^k \dd c_n +(-)^{n}
\begin{cases}
0, & k=\text{odd} \\ 
2  \gSq^{k+1} c_n  & k=\text{even} \\ 
\end{cases}
.
\nonumber 
\end{align}
In particular, when $c_n$ is a $\Z_2$-valued cochain, we have
\begin{align}
\label{Sqd}
  \dd \gSq^{k} c_n \se{2} \gSq^k \dd c_n.
\end{align}

Next, let us consider the action of $\gSq^k$ on the sum of two
 $\M$-valued cochains $c_n$ and $c_n'$:
\begin{align}
\label{Sqplus1}
& \gSq^{k} (c_n+c_n')
 = \gSq^{k} c_n + \gSq^k c_n' +
\nonumber\\
&\ \ \
 c_n \hcup{n-k} c_n' + c_n' \hcup{n-k} c_n 
+ c_n \hcup{n-k+1} \dd c_n' + c_n' \hcup{n-k+1} \dd c_n 
\nonumber\\
&=\gSq^{k} c_n + \gSq^k c_n' 
+[1 + (-)^k]c_n \hcup{n-k} c_n'
\nonumber\\
&\ \ \
-(-)^{n-k} [ - (-)^{n-k} c_n' \hcup{n-k} c_n + (-)^n c_n \hcup{n-k} c_n']
\nonumber\\
&\ \ \
+ c_n \hcup{n-k+1} \dd c_n' + c_n' \hcup{n-k+1} \dd c_n
\nonumber\\
& = 
\gSq^{k} c_n + \gSq^k c_n' 
+[1 + (-)^k]c_n \hcup{n-k} c_n'
\nonumber\\
&
+(-)^{n-k} [ \dd c_n' \hcup{n-k+1} c_n +(-)^n c_n' \hcup{n-k+1} \dd c_n
]
\nonumber\\
&
-(-)^{n-k} 
\dd (c_n'\hcup{n-k+1}c_n) 
+c_n \hcup{n-k+1} \dd c_n'+ c_n' \hcup{n-k+1} \dd c_n
\nonumber\\
&=
\gSq^{k} c_n + \gSq^k c_n'  
+[1 + (-)^k]c_n \hcup{n-k} c_n'
\nonumber \\
&\  \ \
+[1+(-)^{k}]c_n' \hcup{n-k+1} \dd c_n 
-(-)^{n-k} \dd (c_n'\hcup{n-k+1}c_n)
\nonumber\\
&\ \ \
-[(-)^{n-k+1}\dd c_n' \hcup{n-k+1} c_n
- c_n \hcup{n-k+1} \dd c_n']
\nonumber\\
&=
\gSq^{k} c_n + \gSq^k c_n'  
+[1 + (-)^k]c_n \hcup{n-k} c_n'
\nonumber \\
&\  \ \
+[1+(-)^{k}]c_n' \hcup{n-k+1} \dd c_n 
-(-)^{n-k} \dd (c_n'\hcup{n-k+1}c_n)
\nonumber\\
&\ \ \
-\dd (\dd c_n'\hcup{n-k+2} c_n )
+ \dd c_n'\hcup{n-k+2} \dd c_n 
\nonumber\\
&=
\gSq^{k} c_n + \gSq^k c_n'  
+ \dd c_n'\hcup{n-k+2} \dd c_n 
\nonumber \\
&\ \ \
+[1+(-)^{k}][c_n \hcup{n-k} c_n'+ c_n' \hcup{n-k+1} \dd c_n] 
\nonumber\\
&\ \ \
-(-)^{n-k} \dd (c_n'\hcup{n-k+1}c_n)
-\dd (\dd c_n'\hcup{n-k+2} c_n )
.
\end{align}
We see that, if one of the $c_n$ and $c_n'$ is a cocycle,
\begin{align}
\label{Sqplus}
  \gSq^{k} (c_n+c_n') \se{2,\dd} \gSq^{k} c_n + \gSq^k c_n' .
\end{align}
We also see that
\begin{align}
\label{Sqgauge}
&\ \ \ \
 \gSq^{k} (c_n+\dd f_{n-1})
\\
& = \gSq^{k} c_n + \gSq^k \dd f_{n-1} +
[1+(-)^k] \dd f_{n-1}\hcup{n-k} c_n
\nonumber\\
&\ \ \
-(-)^{n-k} \dd (c_n\hcup{n-k+1}\dd f_{n-1})
-\dd (\dd c_n\hcup{n-k+2} \dd f_{n-1} )
\nonumber\\
& = \gSq^{k} c_n 
+ [1+(-)^k] [\dd f_{n-1}\hcup{n-k} c_n +(-)^n \gSq^{k+1}f_{n-1}]
\nonumber\\
&
+\dd [\gSq^k  f_{n-1}
-(-)^{n-k} c_n \hskip -0.5em \hcup{n-k+1} \hskip -0.5em \dd f_{n-1}
-\dd c_n \hskip -0.5em \hcup{n-k+2}  \hskip -0.5em \dd f_{n-1} ]
\nonumber\\
& = \gSq^{k} c_n 
+ [1+(-)^k] [c_n\hcup{n-k}  \dd f_{n-1} +(-)^n \gSq^{k+1}f_{n-1}]
\nonumber\\
&
+\dd [\gSq^k  f_{n-1} -(-)^{n-k} \dd f_{n-1} \hcup{n-k+1} c_n ]
.
\nonumber 
\end{align}
Using \eqn{Sqplus2}, we can also obtain the following result
if $\dd c_n =  \text{even}$
\begin{align}
\label{Sqplus2}
& \ \ \ \
 \gSq^k (c_n+2c_n')
\nonumber\\
& \se{4} \gSq^k c_n+2 \dd (c_n\hcup{n-k+1} c_n') +2 \dd c_n\hcup{n-k+1} c_n'
\nonumber\\
& \se{4} \gSq^k c_n+2 \dd (c_n\hcup{n-k+1} c_n') 
\end{align}

As another application, we note that, for a $\M$-valued cochain $m_d$ and using
\eqn{cupkrel},
\begin{align}
\label{Sq1Bs}
& \gSq^1(m_{d}) = m_{d}\hcup{d-1} m_{d} + m_{d}\hcup{d} \dd m_{d}
\nonumber\\
&=\frac12 (-)^{d} 
[\dd (m_{d}\hcup{d} m_{d}) 
-\dd m_{d} \hcup{d} m_{d}] 
+\frac12  m_{d} \hcup{d} \dd m_{d} 
\nonumber\\
&=
(-)^{d} \Bs_2 (m_{d}\hcup{d} m_{d}) -(-)^d \Bs_2 m_{d} \hcup{d} m_{d}
+  m_{d} \hcup{d} \Bs_2 m_{d}
\nonumber\\
&=
(-)^{d} \Bs_2  \gSq^0 m_{d} 
-2 (-)^d \Bs_2 m_{d} \hcup{d+1} \Bs_2 m_{d}
\nonumber\\
&=
(-)^{d} \Bs_2 \gSq^0 m_{d} 
-2 (-)^d \gSq^0 \Bs_2 m_{d} 
\end{align}
This way, we obtain a relation between Steenrod square and Bockstein
homomorphism, when $m_d$ is a $\Z_2$-valued cochain
\begin{align}
\label{Sq1Bs2}
  \gSq^1(m_{d}) \se{2} \Bs_2 m_{d} ,
\end{align}
where we have used $\gSq^0 m_{d}= m_d$ for $\Z_2$-valued cochain.

For a $k$-cochain $a_k$, $k=\text{odd}$, we find that
\begin{align}
&\ \ \ \
 \gSq^k a_k = a_ka_k+a_k\hcup{1}\dd a_k
\\
&=
\frac12 [\dd a_k \hcup{1} a_k -a_k \hcup{1}\dd a_k -\dd(a_k\hcup{1}a_k) ]
+a_k\hcup{1}\dd a_k
\nonumber\\
&= \frac12 [\dd a_k \hcup{2}\dd a_k-\dd (\dd a_k\hcup{2} a_k)] 
-\frac12 \dd(a_k\hcup{1}a_k)
\nonumber\\
&= \frac14 \dd (\dd a_k \hcup{3}\dd a_k) 
-\frac12 \dd(a_k\hcup{1}a_k+\dd a_k\hcup{2} a_k)
\nonumber 
\end{align}
Thus $\gSq^k a_k$ is always a $\Q$-valued conboundary, when $k$ is odd.

\section{Path integral and Hamiltonian}
\label{pathHam}

\begin{figure}[tb]
\begin{center}
\includegraphics[scale=0.5]{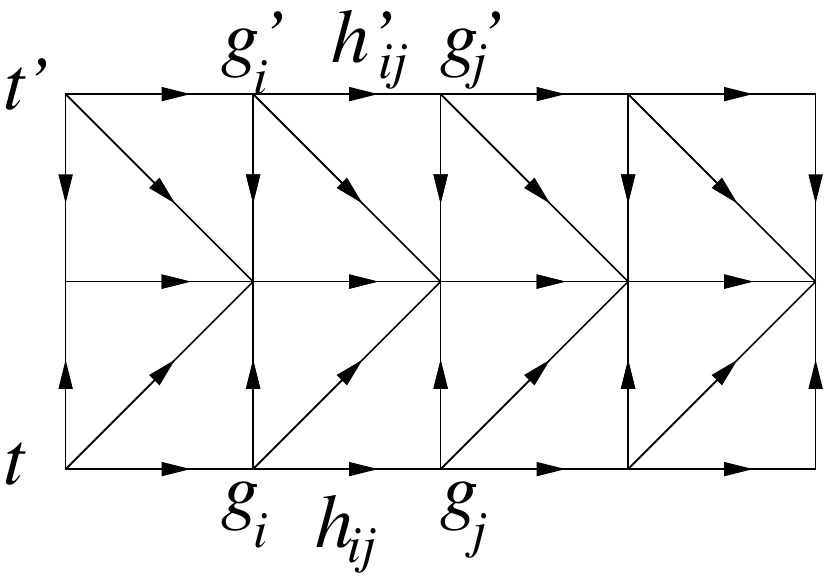} \end{center}
\caption{
Each time-step of evolution is given by the path integral on a particular form
of branched graph.  Here is an example in 1+1D. 
}
\label{tStep}
\end{figure}

Consider a spacetime complex of topology $M_\text{space}\times I$ where
$I=[t,t']$ represents the time dimension and $M_\text{space}$ is a closed space
complex (see Fig. \ref{tStep}).  The spacetime complex $M_\text{space}\times
I$ has two boundaries: one at time $t$ and another at time $t'$.  A path
integral on the spacetime complex $M_\text{space}\times I$ give us an
amplitude $Z[\{ g_i', h_{ij}',\cdots  \}, \{ g_i, h_{ij},\cdots \}]$ from a
configuration $\{ g_i, h_{ij},\cdots \}$ at $t$ to another configuration $\{
g_i', h_{ij}',\cdots  \}$ at $t'$.  Here, $\{ g_i, h_{ij},\cdots \}$ and $\{
g_i', h_{ij}',\cdots  \}$ are the degrees of freedom on the boundaries (see
Fig. \ref{tStep}).  We like to interpret $Z[\{ g_i', h_{ij}',\cdots  \}, \{
g_i, h_{ij},\cdots  \}]$ as the amplitude of an evolution in imaginary time by
a Hamiltonian:
\begin{align}
& \ \ \ \
 Z[\{ g_i', h_{ij}',\cdots  \}, \{ g_i, h_{ij},\cdots \}] 
\nonumber\\
&=\< g_i', h_{ij}',\cdots  | \ee^{-(t'-t)H} |
g_i, h_{ij},\cdots  \> .
\end{align}
However, such an interpretation may not be valid since $ Z[\{ g_i',
h_{ij}',\cdots  \}, \{ g_i, h_{ij},\cdots \}]$ may not give raise to a
hermitian matrix.  

To have a path integral that give rise to a hermitian matrix $H$, we require
the path integral defined on the branched graphs to have a ``reflection''
property. 
The imaginary-time path integral (or partition function) has a form
\begin{align}
\label{Zpath}
 Z=\sum_{ \{g_i\},\{h_{ij}\},\cdots } \ee^{-S(\{g_i\},\{h_{ij}\},\cdots)}
\end{align}
 where the total action amplitude $\ee^{-S}$ for
a configuration $\{g_i\},\{h_{ij}\},\cdots$ (or a path) is given by
\begin{align}
\label{eS}
\ee^{-S}=
\prod_{(ij \cdots k)} [V_{ij \cdots k}(\{g_i\},\{h_{ij}\},\cdots)]^{s_{ij \cdots k}}.
\end{align}
Here $\prod_{(ij \cdots k)}$ is the product over all the $n$-cells $(ij \cdots
k)$.  Note that the contribution from an $n$-cell $(ij \cdots k)$ is
$V_{ij \cdots k}(\{g_i\},\{h_{ij}\},\cdots)$ or $V^*_{ij \cdots k}(\{g_i\},\{h_{ij}\},\cdots)$
depending on the orientation $s_{ij \cdots k}$ of the cell (see Fig. \ref{mir}).

Such a path integral give rise to the hermitian Hamiltonian evolution.  The key
is to require that each time-step of evolution is given by  branched graphs of
the form in Fig. \ref{tStep}.  One can show that $Z[\{ g_i', h_{ij}',\cdots
\}, \{ g_i, h_{ij},\cdots \}]$ obtained by summing over all in the internal
indices in the  branched graphs Fig. \ref{tStep} has a form
\begin{align}
&\ \ \ \
 Z[\{ g_i', h_{ij}',\cdots  \}, \{ g_i, h_{ij},\cdots \}]
\\
&=\sum_{\{ g_i'', h_{ij}'',\cdots  \}}
 U^*[\{ g_i'', h_{ij}'',\cdots  \}, \{ g_i', h_{ij}',\cdots \}]
\nonumber\\
&\ \ \ \ \ \ \ \ \ \ \ \ \ \ \ \ \
 U[\{ g_i'', h_{ij}'',\cdots  \}, \{ g_i, h_{ij},\cdots \}]
\nonumber
\end{align}
and represents a positive-definite hermitian matrix.  Thus the path integral of
the form \eq{eS} always correspond to a Hamiltonian evolution in imaginary
time.  In fact, the above $Z[\{ g_i', h_{ij}',\cdots  \}, \{ g_i, h_{ij},\cdots
\}]$ can be viewed as an imaginary-time evolution $T=\ee^{-\Del \tau H}$ for a
single time step.

\bibliography{../../bib/all,../../bib/publst,./local} 

\begin{thebibliography}{97}%
\makeatletter
\providecommand \@ifxundefined [1]{%
 \@ifx{#1\undefined}
}%
\providecommand \@ifnum [1]{%
 \ifnum #1\expandafter \@firstoftwo
 \else \expandafter \@secondoftwo
 \fi
}%
\providecommand \@ifx [1]{%
 \ifx #1\expandafter \@firstoftwo
 \else \expandafter \@secondoftwo
 \fi
}%
\providecommand \natexlab [1]{#1}%
\providecommand \enquote  [1]{``#1''}%
\providecommand \bibnamefont  [1]{#1}%
\providecommand \bibfnamefont [1]{#1}%
\providecommand \citenamefont [1]{#1}%
\providecommand \href@noop [0]{\@secondoftwo}%
\providecommand \href [0]{\begingroup \@sanitize@url \@href}%
\providecommand \@href[1]{\@@startlink{#1}\@@href}%
\providecommand \@@href[1]{\endgroup#1\@@endlink}%
\providecommand \@sanitize@url [0]{\catcode `\\12\catcode `\$12\catcode
  `\&12\catcode `\#12\catcode `\^12\catcode `\_12\catcode `\%12\relax}%
\providecommand \@@startlink[1]{}%
\providecommand \@@endlink[0]{}%
\providecommand \url  [0]{\begingroup\@sanitize@url \@url }%
\providecommand \@url [1]{\endgroup\@href {#1}{\urlprefix }}%
\providecommand \urlprefix  [0]{URL }%
\providecommand \Eprint [0]{\href }%
\providecommand \doibase [0]{http://dx.doi.org/}%
\providecommand \selectlanguage [0]{\@gobble}%
\providecommand \bibinfo  [0]{\@secondoftwo}%
\providecommand \bibfield  [0]{\@secondoftwo}%
\providecommand \translation [1]{[#1]}%
\providecommand \BibitemOpen [0]{}%
\providecommand \bibitemStop [0]{}%
\providecommand \bibitemNoStop [0]{.\EOS\space}%
\providecommand \EOS [0]{\spacefactor3000\relax}%
\providecommand \BibitemShut  [1]{\csname bibitem#1\endcsname}%
\let\auto@bib@innerbib\@empty
\bibitem [{\citenamefont {Kitaev}(2003)}]{K032}%
  \BibitemOpen
  \bibfield  {author} {\bibinfo {author} {\bibfnamefont {A.~Y.}\ \bibnamefont
  {Kitaev}},\ }\href@noop {} {\bibfield  {journal} {\bibinfo  {journal} {Ann.
  Phys. (N.Y.)}\ }\textbf {\bibinfo {volume} {303}},\ \bibinfo {pages} {2}
  (\bibinfo {year} {2003})}\BibitemShut {NoStop}%
\bibitem [{\citenamefont {Wen}(2003{\natexlab{a}})}]{W0303}%
  \BibitemOpen
  \bibfield  {author} {\bibinfo {author} {\bibfnamefont {X.-G.}\ \bibnamefont
  {Wen}},\ }\href {\doibase 10.1103/physrevlett.90.016803} {\bibfield
  {journal} {\bibinfo  {journal} {Phys. Rev. Lett.}\ }\textbf {\bibinfo
  {volume} {90}},\ \bibinfo {pages} {016803} (\bibinfo {year}
  {2003}{\natexlab{a}})},\ \Eprint {http://arxiv.org/abs/quant-ph/0205004}
  {quant-ph/0205004} \BibitemShut {NoStop}%
\bibitem [{\citenamefont {Levin}\ and\ \citenamefont {Wen}(2003)}]{LW0316}%
  \BibitemOpen
  \bibfield  {author} {\bibinfo {author} {\bibfnamefont {M.}~\bibnamefont
  {Levin}}\ and\ \bibinfo {author} {\bibfnamefont {X.-G.}\ \bibnamefont
  {Wen}},\ }\href {\doibase 10.1103/physrevb.67.245316} {\bibfield  {journal}
  {\bibinfo  {journal} {Phys. Rev. B}\ }\textbf {\bibinfo {volume} {67}},\
  \bibinfo {pages} {245316} (\bibinfo {year} {2003})},\ \Eprint
  {http://arxiv.org/abs/cond-mat/0302460} {cond-mat/0302460} \BibitemShut
  {NoStop}%
\bibitem [{\citenamefont {{Yoshida}}(2011)}]{Y10074601}%
  \BibitemOpen
  \bibfield  {author} {\bibinfo {author} {\bibfnamefont {B.}~\bibnamefont
  {{Yoshida}}},\ }\href {\doibase 10.1016/j.aop.2010.10.009} {\bibfield
  {journal} {\bibinfo  {journal} {Annals of Physics}\ }\textbf {\bibinfo
  {volume} {326}},\ \bibinfo {pages} {15} (\bibinfo {year} {2011})},\ \Eprint
  {http://arxiv.org/abs/1007.4601} {arXiv:1007.4601} \BibitemShut {NoStop}%
\bibitem [{\citenamefont {{Bomb{\'{\i}}n}}(2014)}]{B11072707}%
  \BibitemOpen
  \bibfield  {author} {\bibinfo {author} {\bibfnamefont {H.}~\bibnamefont
  {{Bomb{\'{\i}}n}}},\ }\href {\doibase 10.1007/s00220-014-1893-4} {\bibfield
  {journal} {\bibinfo  {journal} {Communications in Mathematical Physics}\
  }\textbf {\bibinfo {volume} {327}},\ \bibinfo {pages} {387} (\bibinfo {year}
  {2014})},\ \Eprint {http://arxiv.org/abs/1107.2707} {arXiv:1107.2707}
  \BibitemShut {NoStop}%
\bibitem [{\citenamefont {Wen}(1989)}]{W8987}%
  \BibitemOpen
  \bibfield  {author} {\bibinfo {author} {\bibfnamefont {X.~G.}\ \bibnamefont
  {Wen}},\ }\href {\doibase 10.1103/physrevb.40.7387} {\bibfield  {journal}
  {\bibinfo  {journal} {Phys. Rev. B}\ }\textbf {\bibinfo {volume} {40}},\
  \bibinfo {pages} {7387} (\bibinfo {year} {1989})}\BibitemShut {NoStop}%
\bibitem [{\citenamefont {Wen}\ and\ \citenamefont {Niu}(1990)}]{WN9077}%
  \BibitemOpen
  \bibfield  {author} {\bibinfo {author} {\bibfnamefont {X.~G.}\ \bibnamefont
  {Wen}}\ and\ \bibinfo {author} {\bibfnamefont {Q.}~\bibnamefont {Niu}},\
  }\href {\doibase 10.1103/physrevb.41.9377} {\bibfield  {journal} {\bibinfo
  {journal} {Phys. Rev. B}\ }\textbf {\bibinfo {volume} {41}},\ \bibinfo
  {pages} {9377} (\bibinfo {year} {1990})}\BibitemShut {NoStop}%
\bibitem [{\citenamefont {Wen}(1990)}]{W9039}%
  \BibitemOpen
  \bibfield  {author} {\bibinfo {author} {\bibfnamefont {X.~G.}\ \bibnamefont
  {Wen}},\ }\href {\doibase 10.1142/s0217979290000139} {\bibfield  {journal}
  {\bibinfo  {journal} {Int. J. Mod. Phys. B}\ }\textbf {\bibinfo {volume}
  {04}},\ \bibinfo {pages} {239} (\bibinfo {year} {1990})}\BibitemShut
  {NoStop}%
\bibitem [{\citenamefont {{Nussinov}}\ and\ \citenamefont
  {{Ortiz}}(2009{\natexlab{a}})}]{NOc0605316}%
  \BibitemOpen
  \bibfield  {author} {\bibinfo {author} {\bibfnamefont {Z.}~\bibnamefont
  {{Nussinov}}}\ and\ \bibinfo {author} {\bibfnamefont {G.}~\bibnamefont
  {{Ortiz}}},\ }\href {\doibase 10.1073/pnas.0803726105} {\bibfield  {journal}
  {\bibinfo  {journal} {Proceedings of the National Academy of Science}\
  }\textbf {\bibinfo {volume} {106}},\ \bibinfo {pages} {16944} (\bibinfo
  {year} {2009}{\natexlab{a}})},\ \Eprint
  {http://arxiv.org/abs/cond-mat/0605316} {arXiv:cond-mat/0605316} \BibitemShut
  {NoStop}%
\bibitem [{\citenamefont {{Nussinov}}\ and\ \citenamefont
  {{Ortiz}}(2009{\natexlab{b}})}]{NOc0702377}%
  \BibitemOpen
  \bibfield  {author} {\bibinfo {author} {\bibfnamefont {Z.}~\bibnamefont
  {{Nussinov}}}\ and\ \bibinfo {author} {\bibfnamefont {G.}~\bibnamefont
  {{Ortiz}}},\ }\href {\doibase 10.1016/j.aop.2008.11.002} {\bibfield
  {journal} {\bibinfo  {journal} {Annals of Physics}\ }\textbf {\bibinfo
  {volume} {324}},\ \bibinfo {pages} {977} (\bibinfo {year}
  {2009}{\natexlab{b}})},\ \Eprint {http://arxiv.org/abs/cond-mat/0702377}
  {arXiv:cond-mat/0702377} \BibitemShut {NoStop}%
\bibitem [{\citenamefont {{Gaiotto}}\ \emph {et~al.}(2015)\citenamefont
  {{Gaiotto}}, \citenamefont {{Kapustin}}, \citenamefont {{Seiberg}},\ and\
  \citenamefont {{Willett}}}]{GW14125148}%
  \BibitemOpen
  \bibfield  {author} {\bibinfo {author} {\bibfnamefont {D.}~\bibnamefont
  {{Gaiotto}}}, \bibinfo {author} {\bibfnamefont {A.}~\bibnamefont
  {{Kapustin}}}, \bibinfo {author} {\bibfnamefont {N.}~\bibnamefont
  {{Seiberg}}}, \ and\ \bibinfo {author} {\bibfnamefont {B.}~\bibnamefont
  {{Willett}}},\ }\href {\doibase 10.1007/JHEP02(2015)172} {\bibfield
  {journal} {\bibinfo  {journal} {Journal of High Energy Physics}\ }\textbf
  {\bibinfo {volume} {2}},\ \bibinfo {pages} {172} (\bibinfo {year} {2015})},\
  \Eprint {http://arxiv.org/abs/arXiv:1412.5148} {arXiv:1412.5148} \BibitemShut
  {NoStop}%
\bibitem [{\citenamefont {Hastings}\ and\ \citenamefont {Wen}(2005)}]{HW0541}%
  \BibitemOpen
  \bibfield  {author} {\bibinfo {author} {\bibfnamefont {M.~B.}\ \bibnamefont
  {Hastings}}\ and\ \bibinfo {author} {\bibfnamefont {X.-G.}\ \bibnamefont
  {Wen}},\ }\href {\doibase 10.1103/physrevb.72.045141} {\bibfield  {journal}
  {\bibinfo  {journal} {Phys. Rev. B}\ }\textbf {\bibinfo {volume} {72}},\
  \bibinfo {pages} {045141} (\bibinfo {year} {2005})},\ \Eprint
  {http://arxiv.org/abs/cond-mat/0503554} {cond-mat/0503554} \BibitemShut
  {NoStop}%
\bibitem [{\citenamefont {{Baez}}\ and\ \citenamefont
  {{Schreiber}}(2007)}]{BSm0511710}%
  \BibitemOpen
  \bibfield  {author} {\bibinfo {author} {\bibfnamefont {J.~C.}\ \bibnamefont
  {{Baez}}}\ and\ \bibinfo {author} {\bibfnamefont {U.}~\bibnamefont
  {{Schreiber}}},\ }\href@noop {} {\bibfield  {journal} {\bibinfo  {journal}
  {Contemp. Math. 431, AMS, Providence, Rhode Island}\ ,\ \bibinfo {pages} {7}}
  (\bibinfo {year} {2007})},\ \Eprint {http://arxiv.org/abs/math/0511710}
  {math/0511710} \BibitemShut {NoStop}%
\bibitem [{\citenamefont {{Girelli}}\ \emph {et~al.}(2008)\citenamefont
  {{Girelli}}, \citenamefont {{Pfeiffer}},\ and\ \citenamefont
  {{Popescu}}}]{GP07083051}%
  \BibitemOpen
  \bibfield  {author} {\bibinfo {author} {\bibfnamefont {F.}~\bibnamefont
  {{Girelli}}}, \bibinfo {author} {\bibfnamefont {H.}~\bibnamefont
  {{Pfeiffer}}}, \ and\ \bibinfo {author} {\bibfnamefont {E.~M.}\ \bibnamefont
  {{Popescu}}},\ }\href {\doibase 10.1063/1.2888764} {\bibfield  {journal}
  {\bibinfo  {journal} {Journal of Mathematical Physics}\ }\textbf {\bibinfo
  {volume} {49}},\ \bibinfo {pages} {032503} (\bibinfo {year} {2008})},\
  \Eprint {http://arxiv.org/abs/0708.3051} {arXiv:0708.3051} \BibitemShut
  {NoStop}%
\bibitem [{\citenamefont {{Baez}}\ and\ \citenamefont
  {{Huerta}}(2011)}]{BH10034485}%
  \BibitemOpen
  \bibfield  {author} {\bibinfo {author} {\bibfnamefont {J.~C.}\ \bibnamefont
  {{Baez}}}\ and\ \bibinfo {author} {\bibfnamefont {J.}~\bibnamefont
  {{Huerta}}},\ }\href {\doibase 10.1007/s10714-010-1070-9} {\bibfield
  {journal} {\bibinfo  {journal} {General Relativity and Gravitation}\ }\textbf
  {\bibinfo {volume} {43}},\ \bibinfo {pages} {2335} (\bibinfo {year}
  {2011})},\ \Eprint {http://arxiv.org/abs/1003.4485} {arXiv:1003.4485}
  \BibitemShut {NoStop}%
\bibitem [{\citenamefont {{Kapustin}}\ and\ \citenamefont
  {{Thorngren}}(2013)}]{KT13094721}%
  \BibitemOpen
  \bibfield  {author} {\bibinfo {author} {\bibfnamefont {A.}~\bibnamefont
  {{Kapustin}}}\ and\ \bibinfo {author} {\bibfnamefont {R.}~\bibnamefont
  {{Thorngren}}},\ }\href@noop {} {\  (\bibinfo {year} {2013})},\ \Eprint
  {http://arxiv.org/abs/1309.4721} {arXiv:1309.4721} \BibitemShut {NoStop}%
\bibitem [{\citenamefont {{Sharpe}}(2015)}]{S150804770}%
  \BibitemOpen
  \bibfield  {author} {\bibinfo {author} {\bibfnamefont {E.}~\bibnamefont
  {{Sharpe}}},\ }\href {\doibase 10.1002/prop.201500048} {\bibfield  {journal}
  {\bibinfo  {journal} {Fortschritte der Physik}\ }\textbf {\bibinfo {volume}
  {63}},\ \bibinfo {pages} {659} (\bibinfo {year} {2015})},\ \Eprint
  {http://arxiv.org/abs/1508.04770} {arXiv:1508.04770} \BibitemShut {NoStop}%
\bibitem [{\citenamefont {{Thorngren}}\ and\ \citenamefont {{von
  Keyserlingk}}(2015)}]{TK151102929}%
  \BibitemOpen
  \bibfield  {author} {\bibinfo {author} {\bibfnamefont {R.}~\bibnamefont
  {{Thorngren}}}\ and\ \bibinfo {author} {\bibfnamefont {C.}~\bibnamefont {{von
  Keyserlingk}}},\ }\href@noop {} {\  (\bibinfo {year} {2015})},\ \Eprint
  {http://arxiv.org/abs/1511.02929} {arXiv:1511.02929} \BibitemShut {NoStop}%
\bibitem [{\citenamefont {Bullivant}\ \emph
  {et~al.}(2017{\natexlab{a}})\citenamefont {Bullivant}, \citenamefont
  {Calcada}, \citenamefont {K\'ad\'ar}, \citenamefont {Martin},\ and\
  \citenamefont {Martins}}]{BM160606639}%
  \BibitemOpen
  \bibfield  {author} {\bibinfo {author} {\bibfnamefont {A.}~\bibnamefont
  {Bullivant}}, \bibinfo {author} {\bibfnamefont {M.}~\bibnamefont {Calcada}},
  \bibinfo {author} {\bibfnamefont {Z.}~\bibnamefont {K\'ad\'ar}}, \bibinfo
  {author} {\bibfnamefont {P.}~\bibnamefont {Martin}}, \ and\ \bibinfo {author}
  {\bibfnamefont {J.~F.}\ \bibnamefont {Martins}},\ }\href {\doibase
  10.1103/PhysRevB.95.155118} {\bibfield  {journal} {\bibinfo  {journal} {Phys.
  Rev.}\ }\textbf {\bibinfo {volume} {B95}},\ \bibinfo {pages} {155118}
  (\bibinfo {year} {2017}{\natexlab{a}})},\ \Eprint
  {http://arxiv.org/abs/1606.06639} {arXiv:1606.06639} \BibitemShut {NoStop}%
\bibitem [{\citenamefont {{Costa de Almeida}}\ \emph
  {et~al.}(2017)\citenamefont {{Costa de Almeida}}, \citenamefont
  {{Ibieta-Jimenez}}, \citenamefont {{Lorca Espiro}},\ and\ \citenamefont
  {{Teotonio-Sobrinho}}}]{CT171104186}%
  \BibitemOpen
  \bibfield  {author} {\bibinfo {author} {\bibfnamefont {R.}~\bibnamefont
  {{Costa de Almeida}}}, \bibinfo {author} {\bibfnamefont {J.~P.}\ \bibnamefont
  {{Ibieta-Jimenez}}}, \bibinfo {author} {\bibfnamefont {J.}~\bibnamefont
  {{Lorca Espiro}}}, \ and\ \bibinfo {author} {\bibfnamefont {P.}~\bibnamefont
  {{Teotonio-Sobrinho}}},\ }\href@noop {} {\  (\bibinfo {year} {2017})},\
  \Eprint {http://arxiv.org/abs/1711.04186} {arXiv:1711.04186} \BibitemShut
  {NoStop}%
\bibitem [{\citenamefont {{Tachikawa}}(2017)}]{T171209542}%
  \BibitemOpen
  \bibfield  {author} {\bibinfo {author} {\bibfnamefont {Y.}~\bibnamefont
  {{Tachikawa}}},\ }\href@noop {} {\  (\bibinfo {year} {2017})},\ \Eprint
  {http://arxiv.org/abs/1712.09542} {arXiv:1712.09542} \BibitemShut {NoStop}%
\bibitem [{\citenamefont {Parzygnat}(2018)}]{P180201139}%
  \BibitemOpen
  \bibfield  {author} {\bibinfo {author} {\bibfnamefont {A.~J.}\ \bibnamefont
  {Parzygnat}},\ }\emph {\bibinfo {title} {{Two-dimensional algebra in lattice
  gauge theory}}},\ \href
  {https://inspirehep.net/record/1653107/files/arXiv:1802.01139.pdf} {Ph.D.
  thesis},\ \bibinfo  {school} {Connecticut U.} (\bibinfo {year} {2018}),\
  \Eprint {http://arxiv.org/abs/1802.01139} {arXiv:1802.01139} \BibitemShut
  {NoStop}%
\bibitem [{\citenamefont {{Delcamp}}\ and\ \citenamefont
  {{Tiwari}}(2018)}]{DT180210104}%
  \BibitemOpen
  \bibfield  {author} {\bibinfo {author} {\bibfnamefont {C.}~\bibnamefont
  {{Delcamp}}}\ and\ \bibinfo {author} {\bibfnamefont {A.}~\bibnamefont
  {{Tiwari}}},\ }\href {\doibase 10.1007/JHEP10(2018)049} {\bibfield  {journal}
  {\bibinfo  {journal} {Journal of High Energy Physics}\ }\textbf {\bibinfo
  {volume} {2018}},\ \bibinfo {pages} {49} (\bibinfo {year} {2018})},\ \Eprint
  {http://arxiv.org/abs/1802.10104} {arXiv:1802.10104} \BibitemShut {NoStop}%
\bibitem [{\citenamefont {{Lake}}(2018)}]{L180207747}%
  \BibitemOpen
  \bibfield  {author} {\bibinfo {author} {\bibfnamefont {E.}~\bibnamefont
  {{Lake}}},\ }\href@noop {} {\  (\bibinfo {year} {2018})},\ \Eprint
  {http://arxiv.org/abs/1802.07747} {arXiv:1802.07747} \BibitemShut {NoStop}%
\bibitem [{\citenamefont {{Cordova}}\ \emph {et~al.}(2018)\citenamefont
  {{Cordova}}, \citenamefont {{Dumitrescu}},\ and\ \citenamefont
  {{Intriligator}}}]{CI180204790}%
  \BibitemOpen
  \bibfield  {author} {\bibinfo {author} {\bibfnamefont {C.}~\bibnamefont
  {{Cordova}}}, \bibinfo {author} {\bibfnamefont {T.~T.}\ \bibnamefont
  {{Dumitrescu}}}, \ and\ \bibinfo {author} {\bibfnamefont {K.}~\bibnamefont
  {{Intriligator}}},\ }\href@noop {} {\  (\bibinfo {year} {2018})},\ \Eprint
  {http://arxiv.org/abs/1802.04790} {arXiv:1802.04790} \BibitemShut {NoStop}%
\bibitem [{\citenamefont {{Hofman}}\ and\ \citenamefont
  {{Iqbal}}(2019)}]{HI180209512}%
  \BibitemOpen
  \bibfield  {author} {\bibinfo {author} {\bibfnamefont {D.~M.}\ \bibnamefont
  {{Hofman}}}\ and\ \bibinfo {author} {\bibfnamefont {N.}~\bibnamefont
  {{Iqbal}}},\ }\href {\doibase 10.21468/SciPostPhys.6.1.006} {\bibfield
  {journal} {\bibinfo  {journal} {SciPost Phys.}\ }\textbf {\bibinfo {volume}
  {6}},\ \bibinfo {pages} {006} (\bibinfo {year} {2019})},\ \Eprint
  {http://arxiv.org/abs/1802.09512} {arXiv:1802.09512} \BibitemShut {NoStop}%
\bibitem [{\citenamefont {{Bouzid}}\ and\ \citenamefont
  {{Tahiri}}(2018)}]{BT180300529}%
  \BibitemOpen
  \bibfield  {author} {\bibinfo {author} {\bibfnamefont {B.}~\bibnamefont
  {{Bouzid}}}\ and\ \bibinfo {author} {\bibfnamefont {M.}~\bibnamefont
  {{Tahiri}}},\ }\href {\doibase 10.5506/APhysPolB.49.1885} {\bibfield
  {journal} {\bibinfo  {journal} {Acta Physica Polonica B}\ }\textbf {\bibinfo
  {volume} {49}},\ \bibinfo {pages} {1885} (\bibinfo {year} {2018})},\ \Eprint
  {http://arxiv.org/abs/1803.00529} {arXiv:1803.00529} \BibitemShut {NoStop}%
\bibitem [{\citenamefont {{Benini}}\ \emph {et~al.}(2018)\citenamefont
  {{Benini}}, \citenamefont {{Cordova}},\ and\ \citenamefont
  {{Hsin}}}]{BH180309336}%
  \BibitemOpen
  \bibfield  {author} {\bibinfo {author} {\bibfnamefont {F.}~\bibnamefont
  {{Benini}}}, \bibinfo {author} {\bibfnamefont {C.}~\bibnamefont {{Cordova}}},
  \ and\ \bibinfo {author} {\bibfnamefont {P.-S.}\ \bibnamefont {{Hsin}}},\
  }\href@noop {} {\  (\bibinfo {year} {2018})},\ \Eprint
  {http://arxiv.org/abs/1803.09336} {arXiv:1803.09336} \BibitemShut {NoStop}%
\bibitem [{\citenamefont {Nikolaus}\ and\ \citenamefont
  {Waldorf}(2018)}]{NW180400677}%
  \BibitemOpen
  \bibfield  {author} {\bibinfo {author} {\bibfnamefont {T.}~\bibnamefont
  {Nikolaus}}\ and\ \bibinfo {author} {\bibfnamefont {K.}~\bibnamefont
  {Waldorf}},\ }\href@noop {} {\  (\bibinfo {year} {2018})},\ \Eprint
  {http://arxiv.org/abs/1804.00677} {arXiv:1804.00677} \BibitemShut {NoStop}%
\bibitem [{\citenamefont {{Zhu}}\ \emph {et~al.}(2018)\citenamefont {{Zhu}},
  \citenamefont {{Lan}},\ and\ \citenamefont {{Wen}}}]{ZW180809394}%
  \BibitemOpen
  \bibfield  {author} {\bibinfo {author} {\bibfnamefont {C.}~\bibnamefont
  {{Zhu}}}, \bibinfo {author} {\bibfnamefont {T.}~\bibnamefont {{Lan}}}, \ and\
  \bibinfo {author} {\bibfnamefont {X.-G.}\ \bibnamefont {{Wen}}},\ }\href@noop
  {} {\  (\bibinfo {year} {2018})},\ \Eprint {http://arxiv.org/abs/1808.09394}
  {arXiv:1808.09394} \BibitemShut {NoStop}%
\bibitem [{\citenamefont {{Harlow}}\ and\ \citenamefont
  {{Ooguri}}(2018)}]{HO181005338}%
  \BibitemOpen
  \bibfield  {author} {\bibinfo {author} {\bibfnamefont {D.}~\bibnamefont
  {{Harlow}}}\ and\ \bibinfo {author} {\bibfnamefont {H.}~\bibnamefont
  {{Ooguri}}},\ }\href@noop {} {\  (\bibinfo {year} {2018})},\ \Eprint
  {http://arxiv.org/abs/1810.05338} {arXiv:1810.05338} \BibitemShut {NoStop}%
\bibitem [{\citenamefont {{Wan}}\ and\ \citenamefont
  {{Wang}}(2018{\natexlab{a}})}]{WW181211968}%
  \BibitemOpen
  \bibfield  {author} {\bibinfo {author} {\bibfnamefont {Z.}~\bibnamefont
  {{Wan}}}\ and\ \bibinfo {author} {\bibfnamefont {J.}~\bibnamefont {{Wang}}},\
  }\href@noop {} {\  (\bibinfo {year} {2018}{\natexlab{a}})},\ \Eprint
  {http://arxiv.org/abs/1812.11968} {arXiv:1812.11968} \BibitemShut {NoStop}%
\bibitem [{\citenamefont {{Wan}}\ and\ \citenamefont
  {{Wang}}(2018{\natexlab{b}})}]{WW181211967}%
  \BibitemOpen
  \bibfield  {author} {\bibinfo {author} {\bibfnamefont {Z.}~\bibnamefont
  {{Wan}}}\ and\ \bibinfo {author} {\bibfnamefont {J.}~\bibnamefont {{Wang}}},\
  }\href@noop {} {\  (\bibinfo {year} {2018}{\natexlab{b}})},\ \Eprint
  {http://arxiv.org/abs/1812.11967} {arXiv:1812.11967} \BibitemShut {NoStop}%
\bibitem [{\citenamefont {{Guo}}\ \emph {et~al.}(2018)\citenamefont {{Guo}},
  \citenamefont {{Ohmori}}, \citenamefont {{Putrov}}, \citenamefont {{Wan}},\
  and\ \citenamefont {{Wang}}}]{GW181211959}%
  \BibitemOpen
  \bibfield  {author} {\bibinfo {author} {\bibfnamefont {M.}~\bibnamefont
  {{Guo}}}, \bibinfo {author} {\bibfnamefont {K.}~\bibnamefont {{Ohmori}}},
  \bibinfo {author} {\bibfnamefont {P.}~\bibnamefont {{Putrov}}}, \bibinfo
  {author} {\bibfnamefont {Z.}~\bibnamefont {{Wan}}}, \ and\ \bibinfo {author}
  {\bibfnamefont {J.}~\bibnamefont {{Wang}}},\ }\href@noop {} {\  (\bibinfo
  {year} {2018})},\ \Eprint {http://arxiv.org/abs/1812.11959}
  {arXiv:1812.11959} \BibitemShut {NoStop}%
\bibitem [{\citenamefont {{Wan}}\ and\ \citenamefont
  {{Wang}}(2018{\natexlab{c}})}]{WW181211955}%
  \BibitemOpen
  \bibfield  {author} {\bibinfo {author} {\bibfnamefont {Z.}~\bibnamefont
  {{Wan}}}\ and\ \bibinfo {author} {\bibfnamefont {J.}~\bibnamefont {{Wang}}},\
  }\href@noop {} {\  (\bibinfo {year} {2018}{\natexlab{c}})},\ \Eprint
  {http://arxiv.org/abs/1812.11955} {arXiv:1812.11955} \BibitemShut {NoStop}%
\bibitem [{\citenamefont {Bullivant}\ \emph
  {et~al.}(2017{\natexlab{b}})\citenamefont {Bullivant}, \citenamefont
  {Calcada}, \citenamefont {K\'ad\'ar}, \citenamefont {Martins},\ and\
  \citenamefont {Martin}}]{BM170200868}%
  \BibitemOpen
  \bibfield  {author} {\bibinfo {author} {\bibfnamefont {A.}~\bibnamefont
  {Bullivant}}, \bibinfo {author} {\bibfnamefont {M.}~\bibnamefont {Calcada}},
  \bibinfo {author} {\bibfnamefont {Z.}~\bibnamefont {K\'ad\'ar}}, \bibinfo
  {author} {\bibfnamefont {J.~F.}\ \bibnamefont {Martins}}, \ and\ \bibinfo
  {author} {\bibfnamefont {P.}~\bibnamefont {Martin}},\ }\href@noop {} {\
  (\bibinfo {year} {2017}{\natexlab{b}})},\ \Eprint
  {http://arxiv.org/abs/1702.00868} {arXiv:1702.00868} \BibitemShut {NoStop}%
\bibitem [{\citenamefont {{Kobayashi}}\ \emph {et~al.}(2018)\citenamefont
  {{Kobayashi}}, \citenamefont {{Shiozaki}}, \citenamefont {{Kikuchi}},\ and\
  \citenamefont {{Ryu}}}]{KR180505367}%
  \BibitemOpen
  \bibfield  {author} {\bibinfo {author} {\bibfnamefont {R.}~\bibnamefont
  {{Kobayashi}}}, \bibinfo {author} {\bibfnamefont {K.}~\bibnamefont
  {{Shiozaki}}}, \bibinfo {author} {\bibfnamefont {Y.}~\bibnamefont
  {{Kikuchi}}}, \ and\ \bibinfo {author} {\bibfnamefont {S.}~\bibnamefont
  {{Ryu}}},\ }\href@noop {} {\  (\bibinfo {year} {2018})},\ \Eprint
  {http://arxiv.org/abs/1805.05367} {arXiv:1805.05367} \BibitemShut {NoStop}%
\bibitem [{\citenamefont {Zeng}\ and\ \citenamefont {Wen}(2015)}]{ZW1490}%
  \BibitemOpen
  \bibfield  {author} {\bibinfo {author} {\bibfnamefont {B.}~\bibnamefont
  {Zeng}}\ and\ \bibinfo {author} {\bibfnamefont {X.-G.}\ \bibnamefont {Wen}},\
  }\href {\doibase 10.1103/physrevb.91.125121} {\bibfield  {journal} {\bibinfo
  {journal} {Phys. Rev. B}\ }\textbf {\bibinfo {volume} {91}},\ \bibinfo
  {pages} {125121} (\bibinfo {year} {2015})},\ \Eprint
  {http://arxiv.org/abs/arXiv:1406.5090} {arXiv:1406.5090} \BibitemShut
  {NoStop}%
\bibitem [{\citenamefont {{Swingle}}\ and\ \citenamefont
  {{McGreevy}}(2016)}]{SM1403}%
  \BibitemOpen
  \bibfield  {author} {\bibinfo {author} {\bibfnamefont {B.}~\bibnamefont
  {{Swingle}}}\ and\ \bibinfo {author} {\bibfnamefont {J.}~\bibnamefont
  {{McGreevy}}},\ }\href {\doibase 10.1103/PhysRevB.93.045127} {\bibfield
  {journal} {\bibinfo  {journal} {Phys. Rev. B}\ }\textbf {\bibinfo {volume}
  {93}},\ \bibinfo {pages} {045127} (\bibinfo {year} {2016})},\ \Eprint
  {http://arxiv.org/abs/arXiv:1407.8203} {arXiv:1407.8203} \BibitemShut
  {NoStop}%
\bibitem [{\citenamefont {Morandi}(1997)}]{Mor97}%
  \BibitemOpen
  \bibfield  {author} {\bibinfo {author} {\bibfnamefont {P.~J.}\ \bibnamefont
  {Morandi}},\ }\href
  {https://web.nmsu.edu/~pamorand/notes/GroupExtensions.pdf} {\bibfield
  {journal} {\bibinfo  {journal} {https://web.nmsu.edu/$\sim$pamorand/notes/}\
  } (\bibinfo {year} {1997})}\BibitemShut {NoStop}%
\bibitem [{\citenamefont {Fannes}\ \emph {et~al.}(1992)\citenamefont {Fannes},
  \citenamefont {Nachtergaele},\ and\ \citenamefont {Werner}}]{FNW9243}%
  \BibitemOpen
  \bibfield  {author} {\bibinfo {author} {\bibfnamefont {M.}~\bibnamefont
  {Fannes}}, \bibinfo {author} {\bibfnamefont {B.}~\bibnamefont
  {Nachtergaele}}, \ and\ \bibinfo {author} {\bibfnamefont {R.~F.}\
  \bibnamefont {Werner}},\ }\href@noop {} {\bibfield  {journal} {\bibinfo
  {journal} {Commun. Math. Phys.}\ }\textbf {\bibinfo {volume} {144}},\
  \bibinfo {pages} {443} (\bibinfo {year} {1992})}\BibitemShut {NoStop}%
\bibitem [{\citenamefont {Verstraete}\ \emph {et~al.}(2005)\citenamefont
  {Verstraete}, \citenamefont {Cirac}, \citenamefont {Latorre}, \citenamefont
  {Rico},\ and\ \citenamefont {Wolf}}]{VCL0501}%
  \BibitemOpen
  \bibfield  {author} {\bibinfo {author} {\bibfnamefont {F.}~\bibnamefont
  {Verstraete}}, \bibinfo {author} {\bibfnamefont {J.~I.}\ \bibnamefont
  {Cirac}}, \bibinfo {author} {\bibfnamefont {J.~I.}\ \bibnamefont {Latorre}},
  \bibinfo {author} {\bibfnamefont {E.}~\bibnamefont {Rico}}, \ and\ \bibinfo
  {author} {\bibfnamefont {M.~M.}\ \bibnamefont {Wolf}},\ }\href@noop {}
  {\bibfield  {journal} {\bibinfo  {journal} {Phys. Rev. Lett.}\ }\textbf
  {\bibinfo {volume} {94}},\ \bibinfo {pages} {140601} (\bibinfo {year}
  {2005})},\ \Eprint {http://arxiv.org/abs/quant-ph/0410227} {quant-ph/0410227}
  \BibitemShut {NoStop}%
\bibitem [{\citenamefont {Kitaev}(2006)}]{K062}%
  \BibitemOpen
  \bibfield  {author} {\bibinfo {author} {\bibfnamefont {A.}~\bibnamefont
  {Kitaev}},\ }\href@noop {} {\bibfield  {journal} {\bibinfo  {journal} {Annals
  of Physics}\ }\textbf {\bibinfo {volume} {321}},\ \bibinfo {pages} {2}
  (\bibinfo {year} {2006})},\ \Eprint {http://arxiv.org/abs/cond-mat/0506438}
  {cond-mat/0506438} \BibitemShut {NoStop}%
\bibitem [{\citenamefont {{Rowell}}\ \emph {et~al.}(2009)\citenamefont
  {{Rowell}}, \citenamefont {{Stong}},\ and\ \citenamefont {{Wang}}}]{RSW0777}%
  \BibitemOpen
  \bibfield  {author} {\bibinfo {author} {\bibfnamefont {E.}~\bibnamefont
  {{Rowell}}}, \bibinfo {author} {\bibfnamefont {R.}~\bibnamefont {{Stong}}}, \
  and\ \bibinfo {author} {\bibfnamefont {Z.}~\bibnamefont {{Wang}}},\
  }\href@noop {} {\bibfield  {journal} {\bibinfo  {journal} {Comm. Math.
  Phys.}\ }\textbf {\bibinfo {volume} {292}},\ \bibinfo {pages} {343} (\bibinfo
  {year} {2009})},\ \Eprint {http://arxiv.org/abs/arXiv:0712.1377}
  {arXiv:0712.1377} \BibitemShut {NoStop}%
\bibitem [{\citenamefont {Wen}(2015{\natexlab{a}})}]{W150605768}%
  \BibitemOpen
  \bibfield  {author} {\bibinfo {author} {\bibfnamefont {X.-G.}\ \bibnamefont
  {Wen}},\ }\href {\doibase 10.1093/nsr/nwv077} {\bibfield  {journal} {\bibinfo
   {journal} {Nat. Sci. Rev.}\ }\textbf {\bibinfo {volume} {3}},\ \bibinfo
  {pages} {68} (\bibinfo {year} {2015}{\natexlab{a}})},\ \Eprint
  {http://arxiv.org/abs/arXiv:1506.05768} {arXiv:1506.05768} \BibitemShut
  {NoStop}%
\bibitem [{\citenamefont {Lan}\ \emph {et~al.}(2016)\citenamefont {Lan},
  \citenamefont {Kong},\ and\ \citenamefont {Wen}}]{LW150704673}%
  \BibitemOpen
  \bibfield  {author} {\bibinfo {author} {\bibfnamefont {T.}~\bibnamefont
  {Lan}}, \bibinfo {author} {\bibfnamefont {L.}~\bibnamefont {Kong}}, \ and\
  \bibinfo {author} {\bibfnamefont {X.-G.}\ \bibnamefont {Wen}},\ }\href
  {\doibase 10.1103/physrevb.94.155113} {\bibfield  {journal} {\bibinfo
  {journal} {Phys. Rev. B}\ }\textbf {\bibinfo {volume} {94}},\ \bibinfo
  {pages} {155113} (\bibinfo {year} {2016})},\ \Eprint
  {http://arxiv.org/abs/arXiv:1507.04673} {arXiv:1507.04673} \BibitemShut
  {NoStop}%
\bibitem [{\citenamefont {Lan}\ \emph {et~al.}(2018)\citenamefont {Lan},
  \citenamefont {Kong},\ and\ \citenamefont {Wen}}]{LW170404221}%
  \BibitemOpen
  \bibfield  {author} {\bibinfo {author} {\bibfnamefont {T.}~\bibnamefont
  {Lan}}, \bibinfo {author} {\bibfnamefont {L.}~\bibnamefont {Kong}}, \ and\
  \bibinfo {author} {\bibfnamefont {X.-G.}\ \bibnamefont {Wen}},\ }\href
  {\doibase 10.1103/physrevx.8.021074} {\bibfield  {journal} {\bibinfo
  {journal} {Phys. Rev. X}\ }\textbf {\bibinfo {volume} {8}},\ \bibinfo {pages}
  {021074} (\bibinfo {year} {2018})},\ \Eprint
  {http://arxiv.org/abs/1704.04221} {arXiv:1704.04221} \BibitemShut {NoStop}%
\bibitem [{\citenamefont {{Lan}}\ and\ \citenamefont
  {{Wen}}(2019)}]{LW180108530}%
  \BibitemOpen
  \bibfield  {author} {\bibinfo {author} {\bibfnamefont {T.}~\bibnamefont
  {{Lan}}}\ and\ \bibinfo {author} {\bibfnamefont {X.-G.}\ \bibnamefont
  {{Wen}}},\ }\href {\doibase 10.1103/PhysRevX.9.021005} {\bibfield  {journal}
  {\bibinfo  {journal} {Phys. Rev. X}\ }\textbf {\bibinfo {volume} {9}},\
  \bibinfo {pages} {021005} (\bibinfo {year} {2019})},\ \Eprint
  {http://arxiv.org/abs/1801.08530} {arXiv:1801.08530} \BibitemShut {NoStop}%
\bibitem [{\citenamefont {Gu}\ \emph {et~al.}(2015)\citenamefont {Gu},
  \citenamefont {Wang},\ and\ \citenamefont {Wen}}]{GWW1017}%
  \BibitemOpen
  \bibfield  {author} {\bibinfo {author} {\bibfnamefont {Z.-C.}\ \bibnamefont
  {Gu}}, \bibinfo {author} {\bibfnamefont {Z.}~\bibnamefont {Wang}}, \ and\
  \bibinfo {author} {\bibfnamefont {X.-G.}\ \bibnamefont {Wen}},\ }\href
  {\doibase 10.1103/physrevb.91.125149} {\bibfield  {journal} {\bibinfo
  {journal} {Phys. Rev. B}\ }\textbf {\bibinfo {volume} {91}},\ \bibinfo
  {pages} {125149} (\bibinfo {year} {2015})},\ \Eprint
  {http://arxiv.org/abs/arXiv:1010.1517} {arXiv:1010.1517} \BibitemShut
  {NoStop}%
\bibitem [{\citenamefont {Kitaev}(2001)}]{K0131}%
  \BibitemOpen
  \bibfield  {author} {\bibinfo {author} {\bibfnamefont {A.~Y.}\ \bibnamefont
  {Kitaev}},\ }\href {\doibase 10.1070/1063-7869/44/10S/S29} {\bibfield
  {journal} {\bibinfo  {journal} {Phys.-Usp.}\ }\textbf {\bibinfo {volume}
  {44}},\ \bibinfo {pages} {131} (\bibinfo {year} {2001})},\ \Eprint
  {http://arxiv.org/abs/cond-mat/0010440} {cond-mat/0010440} \BibitemShut
  {NoStop}%
\bibitem [{\citenamefont {{Levitov}}\ \emph {et~al.}(2001)\citenamefont
  {{Levitov}}, \citenamefont {{Orlando}}, \citenamefont {{Majer}},\ and\
  \citenamefont {{Mooij}}}]{LMc0108266}%
  \BibitemOpen
  \bibfield  {author} {\bibinfo {author} {\bibfnamefont {L.~S.}\ \bibnamefont
  {{Levitov}}}, \bibinfo {author} {\bibfnamefont {T.~P.}\ \bibnamefont
  {{Orlando}}}, \bibinfo {author} {\bibfnamefont {J.~B.}\ \bibnamefont
  {{Majer}}}, \ and\ \bibinfo {author} {\bibfnamefont {J.~E.}\ \bibnamefont
  {{Mooij}}},\ }\href@noop {} {\  (\bibinfo {year} {2001})},\ \Eprint
  {http://arxiv.org/abs/cond-mat/0108266} {cond-mat/0108266} \BibitemShut
  {NoStop}%
\bibitem [{\citenamefont {Chen}\ \emph {et~al.}(2013)\citenamefont {Chen},
  \citenamefont {Gu}, \citenamefont {Liu},\ and\ \citenamefont
  {Wen}}]{CGL1314}%
  \BibitemOpen
  \bibfield  {author} {\bibinfo {author} {\bibfnamefont {X.}~\bibnamefont
  {Chen}}, \bibinfo {author} {\bibfnamefont {Z.-C.}\ \bibnamefont {Gu}},
  \bibinfo {author} {\bibfnamefont {Z.-X.}\ \bibnamefont {Liu}}, \ and\
  \bibinfo {author} {\bibfnamefont {X.-G.}\ \bibnamefont {Wen}},\ }\href
  {\doibase 10.1103/physrevb.87.155114} {\bibfield  {journal} {\bibinfo
  {journal} {Phys. Rev. B}\ }\textbf {\bibinfo {volume} {87}},\ \bibinfo
  {pages} {155114} (\bibinfo {year} {2013})},\ \Eprint
  {http://arxiv.org/abs/arXiv:1106.4772} {arXiv:1106.4772} \BibitemShut
  {NoStop}%
\bibitem [{\citenamefont {Wen}(2013)}]{W1313}%
  \BibitemOpen
  \bibfield  {author} {\bibinfo {author} {\bibfnamefont {X.-G.}\ \bibnamefont
  {Wen}},\ }\href {\doibase 10.1103/physrevd.88.045013} {\bibfield  {journal}
  {\bibinfo  {journal} {Phys. Rev. D}\ }\textbf {\bibinfo {volume} {88}},\
  \bibinfo {pages} {045013} (\bibinfo {year} {2013})},\ \Eprint
  {http://arxiv.org/abs/arXiv:1303.1803} {arXiv:1303.1803} \BibitemShut
  {NoStop}%
\bibitem [{\citenamefont {Gu}\ and\ \citenamefont {Wen}(2009)}]{GW0931}%
  \BibitemOpen
  \bibfield  {author} {\bibinfo {author} {\bibfnamefont {Z.-C.}\ \bibnamefont
  {Gu}}\ and\ \bibinfo {author} {\bibfnamefont {X.-G.}\ \bibnamefont {Wen}},\
  }\href@noop {} {\bibfield  {journal} {\bibinfo  {journal} {Phys. Rev. B}\
  }\textbf {\bibinfo {volume} {80}},\ \bibinfo {pages} {155131} (\bibinfo
  {year} {2009})},\ \Eprint {http://arxiv.org/abs/arXiv:0903.1069}
  {arXiv:0903.1069} \BibitemShut {NoStop}%
\bibitem [{\citenamefont {Chen}\ \emph
  {et~al.}(2011{\natexlab{a}})\citenamefont {Chen}, \citenamefont {Liu},\ and\
  \citenamefont {Wen}}]{CLW1141}%
  \BibitemOpen
  \bibfield  {author} {\bibinfo {author} {\bibfnamefont {X.}~\bibnamefont
  {Chen}}, \bibinfo {author} {\bibfnamefont {Z.-X.}\ \bibnamefont {Liu}}, \
  and\ \bibinfo {author} {\bibfnamefont {X.-G.}\ \bibnamefont {Wen}},\ }\href
  {\doibase 10.1103/physrevb.84.235141} {\bibfield  {journal} {\bibinfo
  {journal} {Phys. Rev. B}\ }\textbf {\bibinfo {volume} {84}},\ \bibinfo
  {pages} {235141} (\bibinfo {year} {2011}{\natexlab{a}})},\ \Eprint
  {http://arxiv.org/abs/arXiv:1106.4752} {arXiv:1106.4752} \BibitemShut
  {NoStop}%
\bibitem [{\citenamefont {Wen}(2015{\natexlab{b}})}]{W1477}%
  \BibitemOpen
  \bibfield  {author} {\bibinfo {author} {\bibfnamefont {X.-G.}\ \bibnamefont
  {Wen}},\ }\href {\doibase 10.1103/physrevb.91.205101} {\bibfield  {journal}
  {\bibinfo  {journal} {Phys. Rev. B}\ }\textbf {\bibinfo {volume} {91}},\
  \bibinfo {pages} {205101} (\bibinfo {year} {2015}{\natexlab{b}})},\ \Eprint
  {http://arxiv.org/abs/arXiv:1410.8477} {arXiv:1410.8477} \BibitemShut
  {NoStop}%
\bibitem [{\citenamefont {Wen}(2017{\natexlab{a}})}]{W161003911}%
  \BibitemOpen
  \bibfield  {author} {\bibinfo {author} {\bibfnamefont {X.-G.}\ \bibnamefont
  {Wen}},\ }\href {\doibase 10.1103/revmodphys.89.041004} {\bibfield  {journal}
  {\bibinfo  {journal} {Rev. Mod. Phys.}\ }\textbf {\bibinfo {volume} {89}},\
  \bibinfo {pages} {041004} (\bibinfo {year} {2017}{\natexlab{a}})},\ \Eprint
  {http://arxiv.org/abs/arXiv:1610.03911} {arXiv:1610.03911} \BibitemShut
  {NoStop}%
\bibitem [{\citenamefont {Chen}\ \emph {et~al.}(2010)\citenamefont {Chen},
  \citenamefont {Gu},\ and\ \citenamefont {Wen}}]{CGW1038}%
  \BibitemOpen
  \bibfield  {author} {\bibinfo {author} {\bibfnamefont {X.}~\bibnamefont
  {Chen}}, \bibinfo {author} {\bibfnamefont {Z.-C.}\ \bibnamefont {Gu}}, \ and\
  \bibinfo {author} {\bibfnamefont {X.-G.}\ \bibnamefont {Wen}},\ }\href
  {\doibase 10.1103/physrevb.82.155138} {\bibfield  {journal} {\bibinfo
  {journal} {Phys. Rev. B}\ }\textbf {\bibinfo {volume} {82}},\ \bibinfo
  {pages} {155138} (\bibinfo {year} {2010})},\ \Eprint
  {http://arxiv.org/abs/arXiv:1004.3835} {arXiv:1004.3835} \BibitemShut
  {NoStop}%
\bibitem [{\citenamefont {Lu}\ and\ \citenamefont {Vishwanath}(2013)}]{LV1334}%
  \BibitemOpen
  \bibfield  {author} {\bibinfo {author} {\bibfnamefont {Y.-M.}\ \bibnamefont
  {Lu}}\ and\ \bibinfo {author} {\bibfnamefont {A.}~\bibnamefont
  {Vishwanath}},\ }\href@noop {} {\  (\bibinfo {year} {2013})},\ \Eprint
  {http://arxiv.org/abs/arXiv:1302.2634} {arXiv:1302.2634} \BibitemShut
  {NoStop}%
\bibitem [{\citenamefont {Chang}\ \emph {et~al.}(2015)\citenamefont {Chang},
  \citenamefont {Cheng}, \citenamefont {Cui}, \citenamefont {Hu}, \citenamefont
  {Jin}, \citenamefont {Movassagh}, \citenamefont {Naaijkens}, \citenamefont
  {Wang},\ and\ \citenamefont {Young}}]{CY14126589}%
  \BibitemOpen
  \bibfield  {author} {\bibinfo {author} {\bibfnamefont {L.}~\bibnamefont
  {Chang}}, \bibinfo {author} {\bibfnamefont {M.}~\bibnamefont {Cheng}},
  \bibinfo {author} {\bibfnamefont {S.~X.}\ \bibnamefont {Cui}}, \bibinfo
  {author} {\bibfnamefont {Y.}~\bibnamefont {Hu}}, \bibinfo {author}
  {\bibfnamefont {W.}~\bibnamefont {Jin}}, \bibinfo {author} {\bibfnamefont
  {R.}~\bibnamefont {Movassagh}}, \bibinfo {author} {\bibfnamefont
  {P.}~\bibnamefont {Naaijkens}}, \bibinfo {author} {\bibfnamefont
  {Z.}~\bibnamefont {Wang}}, \ and\ \bibinfo {author} {\bibfnamefont
  {A.}~\bibnamefont {Young}},\ }\href@noop {} {\bibfield  {journal} {\bibinfo
  {journal} {Journal of Physics A: Mathematical and Theoretical}\ }\textbf
  {\bibinfo {volume} {48}},\ \bibinfo {pages} {12FT01} (\bibinfo {year}
  {2015})},\ \Eprint {http://arxiv.org/abs/arXiv:1412.6589} {arXiv:1412.6589}
  \BibitemShut {NoStop}%
\bibitem [{\citenamefont {{Ye}}(2018)}]{Y161008645}%
  \BibitemOpen
  \bibfield  {author} {\bibinfo {author} {\bibfnamefont {P.}~\bibnamefont
  {{Ye}}},\ }\href {\doibase 10.1103/PhysRevB.97.125127} {\bibfield  {journal}
  {\bibinfo  {journal} {Phys. Rev. B}\ }\textbf {\bibinfo {volume} {97}},\
  \bibinfo {pages} {125127} (\bibinfo {year} {2018})},\ \Eprint
  {http://arxiv.org/abs/arXiv:1610.08645} {arXiv:1610.08645} \BibitemShut
  {NoStop}%
\bibitem [{\citenamefont {Chen}\ \emph
  {et~al.}(2011{\natexlab{b}})\citenamefont {Chen}, \citenamefont {Gu},\ and\
  \citenamefont {Wen}}]{CGW1107}%
  \BibitemOpen
  \bibfield  {author} {\bibinfo {author} {\bibfnamefont {X.}~\bibnamefont
  {Chen}}, \bibinfo {author} {\bibfnamefont {Z.-C.}\ \bibnamefont {Gu}}, \ and\
  \bibinfo {author} {\bibfnamefont {X.-G.}\ \bibnamefont {Wen}},\ }\href@noop
  {} {\bibfield  {journal} {\bibinfo  {journal} {Phys. Rev. B}\ }\textbf
  {\bibinfo {volume} {83}},\ \bibinfo {pages} {035107} (\bibinfo {year}
  {2011}{\natexlab{b}})},\ \Eprint {http://arxiv.org/abs/arXiv:1008.3745}
  {arXiv:1008.3745} \BibitemShut {NoStop}%
\bibitem [{\citenamefont {Schuch}\ \emph {et~al.}(2011)\citenamefont {Schuch},
  \citenamefont {Perez-Garcia},\ and\ \citenamefont {Cirac}}]{SPC1139}%
  \BibitemOpen
  \bibfield  {author} {\bibinfo {author} {\bibfnamefont {N.}~\bibnamefont
  {Schuch}}, \bibinfo {author} {\bibfnamefont {D.}~\bibnamefont
  {Perez-Garcia}}, \ and\ \bibinfo {author} {\bibfnamefont {I.}~\bibnamefont
  {Cirac}},\ }\href@noop {} {\bibfield  {journal} {\bibinfo  {journal} {Phys.
  Rev. B}\ }\textbf {\bibinfo {volume} {84}},\ \bibinfo {pages} {165139}
  (\bibinfo {year} {2011})},\ \Eprint {http://arxiv.org/abs/arXiv:1010.3732}
  {arXiv:1010.3732} \BibitemShut {NoStop}%
\bibitem [{\citenamefont {Lan}\ \emph {et~al.}(2017)\citenamefont {Lan},
  \citenamefont {Kong},\ and\ \citenamefont {Wen}}]{LW160205946}%
  \BibitemOpen
  \bibfield  {author} {\bibinfo {author} {\bibfnamefont {T.}~\bibnamefont
  {Lan}}, \bibinfo {author} {\bibfnamefont {L.}~\bibnamefont {Kong}}, \ and\
  \bibinfo {author} {\bibfnamefont {X.-G.}\ \bibnamefont {Wen}},\ }\href
  {\doibase 10.1103/physrevb.95.235140} {\bibfield  {journal} {\bibinfo
  {journal} {Phys. Rev. B}\ }\textbf {\bibinfo {volume} {95}},\ \bibinfo
  {pages} {235140} (\bibinfo {year} {2017})},\ \Eprint
  {http://arxiv.org/abs/arXiv:1602.05946} {arXiv:1602.05946} \BibitemShut
  {NoStop}%
\bibitem [{\citenamefont {{Kapustin}}\ and\ \citenamefont
  {{Thorngren}}(2017)}]{KT170108264}%
  \BibitemOpen
  \bibfield  {author} {\bibinfo {author} {\bibfnamefont {A.}~\bibnamefont
  {{Kapustin}}}\ and\ \bibinfo {author} {\bibfnamefont {R.}~\bibnamefont
  {{Thorngren}}},\ }\href {\doibase 10.1007/JHEP10(2017)080} {\bibfield
  {journal} {\bibinfo  {journal} {Journal of High Energy Physics}\ }\textbf
  {\bibinfo {volume} {10}},\ \bibinfo {pages} {80} (\bibinfo {year} {2017})},\
  \Eprint {http://arxiv.org/abs/1701.08264} {arXiv:1701.08264} \BibitemShut
  {NoStop}%
\bibitem [{\citenamefont {Wang}\ and\ \citenamefont {Gu}(2018)}]{WG170310937}%
  \BibitemOpen
  \bibfield  {author} {\bibinfo {author} {\bibfnamefont {Q.-R.}\ \bibnamefont
  {Wang}}\ and\ \bibinfo {author} {\bibfnamefont {Z.-C.}\ \bibnamefont {Gu}},\
  }\href {\doibase 10.1103/PhysRevX.8.011055} {\bibfield  {journal} {\bibinfo
  {journal} {Phys. Rev.}\ }\textbf {\bibinfo {volume} {X8}},\ \bibinfo {pages}
  {011055} (\bibinfo {year} {2018})},\ \Eprint
  {http://arxiv.org/abs/1703.10937} {arXiv:1703.10937} \BibitemShut {NoStop}%
\bibitem [{\citenamefont {{Wang}}\ and\ \citenamefont
  {{Gu}}(2018)}]{WG181100536}%
  \BibitemOpen
  \bibfield  {author} {\bibinfo {author} {\bibfnamefont {Q.-R.}\ \bibnamefont
  {{Wang}}}\ and\ \bibinfo {author} {\bibfnamefont {Z.-C.}\ \bibnamefont
  {{Gu}}},\ }\href@noop {} {\  (\bibinfo {year} {2018})},\ \Eprint
  {http://arxiv.org/abs/1811.00536} {arXiv:1811.00536} \BibitemShut {NoStop}%
\bibitem [{\citenamefont {Kapustin}\ and\ \citenamefont
  {Thorngren}(2014)}]{KT1430}%
  \BibitemOpen
  \bibfield  {author} {\bibinfo {author} {\bibfnamefont {A.}~\bibnamefont
  {Kapustin}}\ and\ \bibinfo {author} {\bibfnamefont {R.}~\bibnamefont
  {Thorngren}},\ }\href@noop {} {\  (\bibinfo {year} {2014})},\ \Eprint
  {http://arxiv.org/abs/arXiv:1404.3230} {arXiv:1404.3230} \BibitemShut
  {NoStop}%
\bibitem [{\citenamefont {’t Hooft}(1980)}]{H8035}%
  \BibitemOpen
  \bibfield  {author} {\bibinfo {author} {\bibfnamefont {G.}~\bibnamefont {’t
  Hooft}},\ }\href@noop {} {\bibfield  {journal} {\bibinfo  {journal} {NATO
  Adv. Study Inst. Ser. B Phys.}\ }\textbf {\bibinfo {volume} {59}},\ \bibinfo
  {pages} {135} (\bibinfo {year} {1980})}\BibitemShut {NoStop}%
\bibitem [{\citenamefont {Kong}\ \emph {et~al.}(2019)\citenamefont {Kong},
  \citenamefont {Lan}, \citenamefont {Tian},\ and\ \citenamefont {Wen}}]{KLTW}%
  \BibitemOpen
  \bibfield  {author} {\bibinfo {author} {\bibfnamefont {L.}~\bibnamefont
  {Kong}}, \bibinfo {author} {\bibfnamefont {T.}~\bibnamefont {Lan}}, \bibinfo
  {author} {\bibfnamefont {Y.}~\bibnamefont {Tian}}, \ and\ \bibinfo {author}
  {\bibfnamefont {X.-G.}\ \bibnamefont {Wen}},\ }\href@noop {} {\bibfield
  {journal} {\bibinfo  {journal} {to appear}\ } (\bibinfo {year}
  {2019})}\BibitemShut {NoStop}%
\bibitem [{\citenamefont {Levin}\ and\ \citenamefont {Wen}(2005)}]{LW0510}%
  \BibitemOpen
  \bibfield  {author} {\bibinfo {author} {\bibfnamefont {M.~A.}\ \bibnamefont
  {Levin}}\ and\ \bibinfo {author} {\bibfnamefont {X.-G.}\ \bibnamefont
  {Wen}},\ }\href {\doibase 10.1103/physrevb.71.045110} {\bibfield  {journal}
  {\bibinfo  {journal} {Phys. Rev. B}\ }\textbf {\bibinfo {volume} {71}},\
  \bibinfo {pages} {045110} (\bibinfo {year} {2005})},\ \Eprint
  {http://arxiv.org/abs/cond-mat/0404617} {cond-mat/0404617} \BibitemShut
  {NoStop}%
\bibitem [{\citenamefont {Walker}\ and\ \citenamefont {Wang}(2012)}]{WW1132}%
  \BibitemOpen
  \bibfield  {author} {\bibinfo {author} {\bibfnamefont {K.}~\bibnamefont
  {Walker}}\ and\ \bibinfo {author} {\bibfnamefont {Z.}~\bibnamefont {Wang}},\
  }\href {\doibase 10.1007/s11467-011-0194-z} {\bibfield  {journal} {\bibinfo
  {journal} {Frontiers of Physics}\ }\textbf {\bibinfo {volume} {7}},\ \bibinfo
  {pages} {150} (\bibinfo {year} {2012})},\ \Eprint
  {http://arxiv.org/abs/arXiv:1104.2632} {arXiv:1104.2632} \BibitemShut
  {NoStop}%
\bibitem [{\citenamefont {{Bhardwaj}}\ \emph {et~al.}(2017)\citenamefont
  {{Bhardwaj}}, \citenamefont {{Gaiotto}},\ and\ \citenamefont
  {{Kapustin}}}]{BK160501640}%
  \BibitemOpen
  \bibfield  {author} {\bibinfo {author} {\bibfnamefont {L.}~\bibnamefont
  {{Bhardwaj}}}, \bibinfo {author} {\bibfnamefont {D.}~\bibnamefont
  {{Gaiotto}}}, \ and\ \bibinfo {author} {\bibfnamefont {A.}~\bibnamefont
  {{Kapustin}}},\ }\href {\doibase 10.1007/JHEP04(2017)096} {\bibfield
  {journal} {\bibinfo  {journal} {Journal of High Energy Physics}\ }\textbf
  {\bibinfo {volume} {2017}},\ \bibinfo {pages} {96} (\bibinfo {year}
  {2017})},\ \Eprint {http://arxiv.org/abs/1605.01640} {arXiv:1605.01640}
  \BibitemShut {NoStop}%
\bibitem [{\citenamefont {{Williamson}}\ and\ \citenamefont
  {{Wang}}(2017)}]{WW160607144}%
  \BibitemOpen
  \bibfield  {author} {\bibinfo {author} {\bibfnamefont {D.~J.}\ \bibnamefont
  {{Williamson}}}\ and\ \bibinfo {author} {\bibfnamefont {Z.}~\bibnamefont
  {{Wang}}},\ }\href {\doibase 10.1016/j.aop.2016.12.018} {\bibfield  {journal}
  {\bibinfo  {journal} {Annals of Physics}\ }\textbf {\bibinfo {volume}
  {377}},\ \bibinfo {pages} {311} (\bibinfo {year} {2017})},\ \Eprint
  {http://arxiv.org/abs/1606.07144} {arXiv:1606.07144} \BibitemShut {NoStop}%
\bibitem [{\citenamefont {Kong}\ and\ \citenamefont {Wen}(2014)}]{KW1458}%
  \BibitemOpen
  \bibfield  {author} {\bibinfo {author} {\bibfnamefont {L.}~\bibnamefont
  {Kong}}\ and\ \bibinfo {author} {\bibfnamefont {X.-G.}\ \bibnamefont {Wen}},\
  }\href@noop {} {\  (\bibinfo {year} {2014})},\ \Eprint
  {http://arxiv.org/abs/arXiv:1405.5858} {arXiv:1405.5858} \BibitemShut
  {NoStop}%
\bibitem [{\citenamefont {Wen}(2017{\natexlab{b}})}]{W161201418}%
  \BibitemOpen
  \bibfield  {author} {\bibinfo {author} {\bibfnamefont {X.-G.}\ \bibnamefont
  {Wen}},\ }\href {\doibase 10.1103/physrevb.95.205142} {\bibfield  {journal}
  {\bibinfo  {journal} {Phys. Rev. B}\ }\textbf {\bibinfo {volume} {95}},\
  \bibinfo {pages} {205142} (\bibinfo {year} {2017}{\natexlab{b}})},\ \Eprint
  {http://arxiv.org/abs/arXiv:1612.01418} {arXiv:1612.01418} \BibitemShut
  {NoStop}%
\bibitem [{\citenamefont {{Lan}}\ \emph {et~al.}(2018)\citenamefont {{Lan}},
  \citenamefont {{Zhu}},\ and\ \citenamefont {{Wen}}}]{LW180901112}%
  \BibitemOpen
  \bibfield  {author} {\bibinfo {author} {\bibfnamefont {T.}~\bibnamefont
  {{Lan}}}, \bibinfo {author} {\bibfnamefont {C.}~\bibnamefont {{Zhu}}}, \ and\
  \bibinfo {author} {\bibfnamefont {X.-G.}\ \bibnamefont {{Wen}}},\ }\href@noop
  {} {\  (\bibinfo {year} {2018})},\ \Eprint {http://arxiv.org/abs/1809.01112}
  {arXiv:1809.01112} \BibitemShut {NoStop}%
\bibitem [{\citenamefont {{Wen}}\ and\ \citenamefont
  {{Wang}}(2018)}]{WW180109938}%
  \BibitemOpen
  \bibfield  {author} {\bibinfo {author} {\bibfnamefont {X.-G.}\ \bibnamefont
  {{Wen}}}\ and\ \bibinfo {author} {\bibfnamefont {Z.}~\bibnamefont {{Wang}}},\
  }\href@noop {} {\  (\bibinfo {year} {2018})},\ \Eprint
  {http://arxiv.org/abs/1801.09938} {arXiv:1801.09938} \BibitemShut {NoStop}%
\bibitem [{\citenamefont {Blok}\ and\ \citenamefont {Wen}(1990)}]{BW9045}%
  \BibitemOpen
  \bibfield  {author} {\bibinfo {author} {\bibfnamefont {B.}~\bibnamefont
  {Blok}}\ and\ \bibinfo {author} {\bibfnamefont {X.-G.}\ \bibnamefont {Wen}},\
  }\href@noop {} {\bibfield  {journal} {\bibinfo  {journal} {Phys. Rev. B}\
  }\textbf {\bibinfo {volume} {42}},\ \bibinfo {pages} {8145} (\bibinfo {year}
  {1990})}\BibitemShut {NoStop}%
\bibitem [{\citenamefont {Fr{\"o}hlich}\ and\ \citenamefont
  {Kerler}(1991)}]{FK9169}%
  \BibitemOpen
  \bibfield  {author} {\bibinfo {author} {\bibfnamefont {J.}~\bibnamefont
  {Fr{\"o}hlich}}\ and\ \bibinfo {author} {\bibfnamefont {T.}~\bibnamefont
  {Kerler}},\ }\href@noop {} {\bibfield  {journal} {\bibinfo  {journal} {Nucl.
  Phys. B}\ }\textbf {\bibinfo {volume} {354}},\ \bibinfo {pages} {369}
  (\bibinfo {year} {1991})}\BibitemShut {NoStop}%
\bibitem [{\citenamefont {Wen}\ and\ \citenamefont {Zee}(1992)}]{WZ9290}%
  \BibitemOpen
  \bibfield  {author} {\bibinfo {author} {\bibfnamefont {X.~G.}\ \bibnamefont
  {Wen}}\ and\ \bibinfo {author} {\bibfnamefont {A.}~\bibnamefont {Zee}},\
  }\href {\doibase 10.1103/physrevb.46.2290} {\bibfield  {journal} {\bibinfo
  {journal} {Phys. Rev. B}\ }\textbf {\bibinfo {volume} {46}},\ \bibinfo
  {pages} {2290} (\bibinfo {year} {1992})}\BibitemShut {NoStop}%
\bibitem [{\citenamefont {DeMarco}\ and\ \citenamefont {Wen}(2019)}]{DW}%
  \BibitemOpen
  \bibfield  {author} {\bibinfo {author} {\bibfnamefont {M.}~\bibnamefont
  {DeMarco}}\ and\ \bibinfo {author} {\bibfnamefont {X.-G.}\ \bibnamefont
  {Wen}},\ }\href@noop {} {\bibfield  {journal} {\bibinfo  {journal} {to
  appear}\ } (\bibinfo {year} {2019})}\BibitemShut {NoStop}%
\bibitem [{\citenamefont {{Beckman}}\ \emph {et~al.}(2002)\citenamefont
  {{Beckman}}, \citenamefont {{Gottesman}}, \citenamefont {{Kitaev}},\ and\
  \citenamefont {{Preskill}}}]{BPh0110205}%
  \BibitemOpen
  \bibfield  {author} {\bibinfo {author} {\bibfnamefont {D.}~\bibnamefont
  {{Beckman}}}, \bibinfo {author} {\bibfnamefont {D.}~\bibnamefont
  {{Gottesman}}}, \bibinfo {author} {\bibfnamefont {A.}~\bibnamefont
  {{Kitaev}}}, \ and\ \bibinfo {author} {\bibfnamefont {J.}~\bibnamefont
  {{Preskill}}},\ }\href {\doibase 10.1103/PhysRevD.65.065022} {\bibfield
  {journal} {\bibinfo  {journal} {\prd}\ }\textbf {\bibinfo {volume} {65}},\
  \bibinfo {pages} {065022} (\bibinfo {year} {2002})},\ \Eprint
  {http://arxiv.org/abs/hep-th/0110205} {arXiv:hep-th/0110205} \BibitemShut
  {NoStop}%
\bibitem [{\citenamefont {{Shi}}(2018)}]{S181001986}%
  \BibitemOpen
  \bibfield  {author} {\bibinfo {author} {\bibfnamefont {B.}~\bibnamefont
  {{Shi}}},\ }\href@noop {} {\bibfield  {journal} {\bibinfo  {journal} {arXiv
  e-prints}\ ,\ \bibinfo {pages} {arXiv:1810.01986}} (\bibinfo {year}
  {2018})},\ \Eprint {http://arxiv.org/abs/1810.01986} {arXiv:1810.01986}
  \BibitemShut {NoStop}%
\bibitem [{\citenamefont {{Huse}}\ \emph {et~al.}(2013)\citenamefont {{Huse}},
  \citenamefont {{Nandkishore}}, \citenamefont {{Oganesyan}}, \citenamefont
  {{Pal}},\ and\ \citenamefont {{Sondhi}}}]{HS13041158}%
  \BibitemOpen
  \bibfield  {author} {\bibinfo {author} {\bibfnamefont {D.~A.}\ \bibnamefont
  {{Huse}}}, \bibinfo {author} {\bibfnamefont {R.}~\bibnamefont
  {{Nandkishore}}}, \bibinfo {author} {\bibfnamefont {V.}~\bibnamefont
  {{Oganesyan}}}, \bibinfo {author} {\bibfnamefont {A.}~\bibnamefont {{Pal}}},
  \ and\ \bibinfo {author} {\bibfnamefont {S.~L.}\ \bibnamefont {{Sondhi}}},\
  }\href {\doibase 10.1103/PhysRevB.88.014206} {\bibfield  {journal} {\bibinfo
  {journal} {Physical Review B}\ }\textbf {\bibinfo {volume} {88}},\ \bibinfo
  {eid} {014206} (\bibinfo {year} {2013})},\ \Eprint
  {http://arxiv.org/abs/1304.1158} {arXiv:1304.1158} \BibitemShut {NoStop}%
\bibitem [{\citenamefont {{Bauer}}\ and\ \citenamefont
  {{Nayak}}(2013)}]{BN13065753}%
  \BibitemOpen
  \bibfield  {author} {\bibinfo {author} {\bibfnamefont {B.}~\bibnamefont
  {{Bauer}}}\ and\ \bibinfo {author} {\bibfnamefont {C.}~\bibnamefont
  {{Nayak}}},\ }\href {\doibase 10.1088/1742-5468/2013/09/P09005} {\bibfield
  {journal} {\bibinfo  {journal} {Journal of Statistical Mechanics: Theory and
  Experiment}\ }\textbf {\bibinfo {volume} {2013}},\ \bibinfo {eid} {09005}
  (\bibinfo {year} {2013})},\ \Eprint {http://arxiv.org/abs/1306.5753}
  {arXiv:1306.5753} \BibitemShut {NoStop}%
\bibitem [{\citenamefont {{Chandran}}\ \emph {et~al.}(2014)\citenamefont
  {{Chandran}}, \citenamefont {{Khemani}}, \citenamefont {{Laumann}},\ and\
  \citenamefont {{Sondhi}}}]{CS13101096}%
  \BibitemOpen
  \bibfield  {author} {\bibinfo {author} {\bibfnamefont {A.}~\bibnamefont
  {{Chandran}}}, \bibinfo {author} {\bibfnamefont {V.}~\bibnamefont
  {{Khemani}}}, \bibinfo {author} {\bibfnamefont {C.~R.}\ \bibnamefont
  {{Laumann}}}, \ and\ \bibinfo {author} {\bibfnamefont {S.~L.}\ \bibnamefont
  {{Sondhi}}},\ }\href {\doibase 10.1103/PhysRevB.89.144201} {\bibfield
  {journal} {\bibinfo  {journal} {Physical Review B}\ }\textbf {\bibinfo
  {volume} {89}},\ \bibinfo {eid} {144201} (\bibinfo {year} {2014})},\ \Eprint
  {http://arxiv.org/abs/1310.1096} {arXiv:1310.1096} \BibitemShut {NoStop}%
\bibitem [{\citenamefont {{Grozdanov}}\ \emph {et~al.}(2017)\citenamefont
  {{Grozdanov}}, \citenamefont {{Hofman}},\ and\ \citenamefont
  {{Iqbal}}}]{GI161007392}%
  \BibitemOpen
  \bibfield  {author} {\bibinfo {author} {\bibfnamefont {S.}~\bibnamefont
  {{Grozdanov}}}, \bibinfo {author} {\bibfnamefont {D.~M.}\ \bibnamefont
  {{Hofman}}}, \ and\ \bibinfo {author} {\bibfnamefont {N.}~\bibnamefont
  {{Iqbal}}},\ }\href {\doibase 10.1103/PhysRevD.95.096003} {\bibfield
  {journal} {\bibinfo  {journal} {\prd}\ }\textbf {\bibinfo {volume} {95}},\
  \bibinfo {pages} {096003} (\bibinfo {year} {2017})},\ \Eprint
  {http://arxiv.org/abs/1610.07392} {arXiv:1610.07392} \BibitemShut {NoStop}%
\bibitem [{\citenamefont {Glorioso}\ and\ \citenamefont
  {Son}(2018)}]{GS181104879}%
  \BibitemOpen
  \bibfield  {author} {\bibinfo {author} {\bibfnamefont {P.}~\bibnamefont
  {Glorioso}}\ and\ \bibinfo {author} {\bibfnamefont {D.~T.}\ \bibnamefont
  {Son}},\ }\href@noop {} {\  (\bibinfo {year} {2018})},\ \Eprint
  {http://arxiv.org/abs/1811.04879} {arXiv:1811.04879} \BibitemShut {NoStop}%
\bibitem [{\citenamefont {Motrunich}\ and\ \citenamefont
  {Senthil}(2002)}]{MS0204}%
  \BibitemOpen
  \bibfield  {author} {\bibinfo {author} {\bibfnamefont {O.~I.}\ \bibnamefont
  {Motrunich}}\ and\ \bibinfo {author} {\bibfnamefont {T.}~\bibnamefont
  {Senthil}},\ }\href@noop {} {\bibfield  {journal} {\bibinfo  {journal} {Phys.
  Rev. Lett.}\ }\textbf {\bibinfo {volume} {89}},\ \bibinfo {pages} {277004}
  (\bibinfo {year} {2002})}\BibitemShut {NoStop}%
\bibitem [{\citenamefont {Wen}(2003{\natexlab{b}})}]{W0313}%
  \BibitemOpen
  \bibfield  {author} {\bibinfo {author} {\bibfnamefont {X.-G.}\ \bibnamefont
  {Wen}},\ }\href {\doibase 10.1103/physrevb.68.115413} {\bibfield  {journal}
  {\bibinfo  {journal} {Phys. Rev. B}\ }\textbf {\bibinfo {volume} {68}},\
  \bibinfo {pages} {115413} (\bibinfo {year} {2003}{\natexlab{b}})},\ \Eprint
  {http://arxiv.org/abs/cond-mat/0210040} {cond-mat/0210040} \BibitemShut
  {NoStop}%
\bibitem [{\citenamefont {Levin}\ and\ \citenamefont {Wen}(2006)}]{LW0622}%
  \BibitemOpen
  \bibfield  {author} {\bibinfo {author} {\bibfnamefont {M.}~\bibnamefont
  {Levin}}\ and\ \bibinfo {author} {\bibfnamefont {X.-G.}\ \bibnamefont
  {Wen}},\ }\href {\doibase 10.1103/physrevb.73.035122} {\bibfield  {journal}
  {\bibinfo  {journal} {Phys. Rev. B}\ }\textbf {\bibinfo {volume} {73}},\
  \bibinfo {pages} {035122} (\bibinfo {year} {2006})},\ \Eprint
  {http://arxiv.org/abs/hep-th/0507118} {hep-th/0507118} \BibitemShut {NoStop}%
\bibitem [{\citenamefont {Wess}\ and\ \citenamefont {Zumino}(1971)}]{WZ7195}%
  \BibitemOpen
  \bibfield  {author} {\bibinfo {author} {\bibfnamefont {J.}~\bibnamefont
  {Wess}}\ and\ \bibinfo {author} {\bibfnamefont {B.}~\bibnamefont {Zumino}},\
  }\href {\doibase 10.1016/0370-2693(71)90582-X} {\bibfield  {journal}
  {\bibinfo  {journal} {Phys. Lett. B}\ }\textbf {\bibinfo {volume} {37}},\
  \bibinfo {pages} {95} (\bibinfo {year} {1971})}\BibitemShut {NoStop}%
\bibitem [{\citenamefont {Witten}(1983)}]{W8322}%
  \BibitemOpen
  \bibfield  {author} {\bibinfo {author} {\bibfnamefont {E.}~\bibnamefont
  {Witten}},\ }\href {\doibase 10.1016/0550-3213(83)90063-9} {\bibfield
  {journal} {\bibinfo  {journal} {Nucl. Phys. B}\ }\textbf {\bibinfo {volume}
  {223}},\ \bibinfo {pages} {422 } (\bibinfo {year} {1983})}\BibitemShut
  {NoStop}%
\bibitem [{\citenamefont {Costantino}(2005)}]{C0527}%
  \BibitemOpen
  \bibfield  {author} {\bibinfo {author} {\bibfnamefont {F.}~\bibnamefont
  {Costantino}},\ }\href@noop {} {\bibfield  {journal} {\bibinfo  {journal}
  {Math. Z.}\ }\textbf {\bibinfo {volume} {251}},\ \bibinfo {pages} {427}
  (\bibinfo {year} {2005})},\ \Eprint {http://arxiv.org/abs/math/0403014}
  {math/0403014} \BibitemShut {NoStop}%
\bibitem [{\citenamefont {Chen}\ \emph {et~al.}(2012)\citenamefont {Chen},
  \citenamefont {Gu}, \citenamefont {Liu},\ and\ \citenamefont
  {Wen}}]{CGL1204}%
  \BibitemOpen
  \bibfield  {author} {\bibinfo {author} {\bibfnamefont {X.}~\bibnamefont
  {Chen}}, \bibinfo {author} {\bibfnamefont {Z.-C.}\ \bibnamefont {Gu}},
  \bibinfo {author} {\bibfnamefont {Z.-X.}\ \bibnamefont {Liu}}, \ and\
  \bibinfo {author} {\bibfnamefont {X.-G.}\ \bibnamefont {Wen}},\ }\href
  {\doibase 10.1126/science.1227224} {\bibfield  {journal} {\bibinfo  {journal}
  {Science}\ }\textbf {\bibinfo {volume} {338}},\ \bibinfo {pages} {1604}
  (\bibinfo {year} {2012})},\ \Eprint {http://arxiv.org/abs/arXiv:1301.0861}
  {arXiv:1301.0861} \BibitemShut {NoStop}%
\bibitem [{\citenamefont {Steenrod}(1947)}]{S4790}%
  \BibitemOpen
  \bibfield  {author} {\bibinfo {author} {\bibfnamefont {N.}~\bibnamefont
  {Steenrod}},\ }\href {\doibase 10.2307/1969172} {\bibfield  {journal}
  {\bibinfo  {journal} {Annals of Mathematics}\ }\textbf {\bibinfo {volume}
  {48}},\ \bibinfo {pages} {290} (\bibinfo {year} {1947})}\BibitemShut
  {NoStop}%
\end{thebibliography}%

\end{document}